%
%

\input epsf.tex
\input phyzzx.tex
\newcount\figno
\figno=0
\def\fig#1#2#3{
\par\begingroup\parindent=0pt\leftskip=1cm\rightskip=1cm\parindent=0pt
\baselineskip=20pt
\global\advance\figno by 1
\midinsert
\epsfxsize=#3
\centerline{\epsfbox{#2}}
\vskip 12pt
{\bf Fig.\ \the\figno: } #1\par
\endinsert\endgroup\par
}
\def\figlabel#1{\xdef#1{\the\figno}}
\def\encadremath#1{\vbox{\hrule\hbox{\vrule\kern8pt\vbox{\kern8pt
\hbox{$\displaystyle #1$}\kern8pt}
\kern8pt\vrule}\hrule}}
\overfullrule=0pt

\def\Title#1#2{\rightline{#1}
\ifx\answ\bigans\nopagenumbers\pageno0\vskip1in%
\else
\def\listrefs{\footatend\vskip 1in\immediate\closeout\rfile\writestoppt
\baselineskip=20pt\centerline{{\bf References}}\bigskip{\frenchspacing%
\parindent=20pt\escapechar=` \input
refs.tmp\vfill\eject}\nonfrenchspacing}
\pageno1\vskip.8in\fi \centerline{\titlefont #2}\vskip .5in}

\ifx\answ\bigans\def\tcbreak#1{}\else\def\tcbreak#1{\cr&{#1}}\fi
%
\message{If you do not have msbm (blackboard bold) fonts,}
\message{change the option at the top of the tex file.}
\font\blackboard=msbm10 
\font\blackboards=msbm7
\font\blackboardss=msbm5
%

\def\half{{1\over 2}}



\input epsf

\def\SUSY#1{{{\cal N}= {#1}}}                   
\def\lbr{{\lbrack}}                             
\def\rbr{{\rbrack}}                             

\def\wdg{{\wedge}}                              
\def\MR#1{{{\BR}^{#1}}}               	
\def\MC#1{{{\BC}^{#1}}}               	
\def\MR#1{{{\BR}^{#1}}}               
\def\MC#1{{{\BC}^{#1}}}               
\def\MS#1{{{\bf S}^{#1}}}               
\def\MT#1{{{\bf T}^{#1}}}              	 
\def\px#1{{\partial_{#1}}}              	
\def\trp#1{{{\rm tr}\{ {#1} \} }}            		
\def\rep#1{{{\bf {#1}}}}                      		
\def\hepth#1{{\it hep-\-th/{#1}}}
\def\frac#1#2{{{{#1}}\over {{#2}}}}           	
\def\e8{E_8 \times E_8}                       	

\def\lam{{\lambda}}

\def\Title#1#2{\rightline{#1}
\ifx\answ\bigans\nopagenumbers\pageno0\vskip1in%
\baselineskip 15pt plus 1pt minus 1pt
\else
\def\listrefs{\footatend\vskip 1in\immediate\closeout\rfile\writestoppt
\baselineskip=20pt\centerline{{\bf References}}\bigskip{\frenchspacing%
\parindent=20pt\escapechar=` \input
refs.tmp\vfill\eject}\nonfrenchspacing}
\pageno1\vskip.8in\fi \centerline{\titlefont #2}\vskip .5in}

\ifx\answ\bigans\def\tcbreak#1{}\else\def\tcbreak#1{\cr&{#1}}\fi
%
\message{If you do not have msbm (blackboard bold) fonts,}
\message{change the option at the top of the tex file.}
\font\blackboard=msbm10 
\font\blackboards=msbm7
\font\blackboardss=msbm5
\def\Bbb#1{{\fam\black\relax#1}}
%
\def\yboxit#1#2{\vbox{\hrule height #1 \hbox{\vrule width #1
\vbox{#2}\vrule width #1 }\hrule height #1 }}
\def\fillbox#1{\hbox to #1{\vbox to #1{\vfil}\hfil}}
\def\ybox{{\lower 1.3pt \yboxit{0.4pt}{\fillbox{8pt}}\hskip-0.2pt}}
\def\np#1#2#3{Nucl. Phys. {\bf B#1} (#2) #3}
\def\pl#1#2#3{Phys. Lett. {\bf #1B} (#2) #3}

\def\physrev#1#2#3{Phys. Rev. {\bf D#1} (#2) #3}

\def\cmp#1#2#3{Comm. Math. Phys. {\bf #1} (#2) #3}

\def\comments#1{}

\def\half{{1\over 2}}

\def\a{\alpha}

\def\II{\relax{I\kern-.07em I}}

\def\hk{{hyperk\"ahler}}

\def\IZ{\relax\ifmmode\mathchoice
{\hbox{\cmss Z\kern-.4em Z}}{\hbox{\cmss Z\kern-.4em Z}}
{\lower.9pt\hbox{\cmsss Z\kern-.4em Z}}
{\lower1.2pt\hbox{\cmsss Z\kern-.4em Z}}\else{\cmss Z\kern-.4em
Z}\fi}
\def\IB{\relax{\rm I\kern-.18em B}}
\def\IC{\bf C}
\def\ID{\relax{\rm I\kern-.18em D}}
\def\IE{\relax{\rm I\kern-.18em E}}
\def\IF{\relax{\rm I\kern-.18em F}}
\def\IG{\relax\hbox{$\inbar\kern-.3em{\rm G}$}}
\def\IGa{\relax\hbox{${\rm I}\kern-.18em\Gamma$}}
\def\IH{\relax{\rm I\kern-.18em H}}
\def\II{\relax{\rm I\kern-.18em I}}
\def\IK{\relax{\rm I\kern-.18em K}}
\def\IP{\relax{\rm I\kern-.18em P}}


\font\cmss=cmss10 \font\cmsss=cmss10 at 7pt
\def\IR{\relax{\rm I\kern-.18em R}}

\def\BR{\IR}
\def\BZ{\IZ}
\def\BR{\IR}
\def\BC{\IC}

\def\tilde{\widetilde}



\newif\iffigs\figstrue

%
\let\useblackboard=\iftrue
%
%
\newfam\black

\input harvmac.tex

\input epsf

\newcount\figno
\figno=0
\def\fig#1#2#3{
\par\begingroup\parindent=0pt\leftskip=1cm\rightskip=1cm\parindent=0pt
\baselineskip=11pt
\global\advance\figno by 1
\midinsert
\epsfxsize=#3
\centerline{\epsfbox{#2}}
\vskip 12pt
{\bf Fig.\ \the\figno: } #1\par
\endinsert\endgroup\par
}
\def\figlabel#1{\xdef#1{\the\figno}}
\def\encadremath#1{\vbox{\hrule\hbox{\vrule\kern8pt\vbox{\kern8pt
\hbox{$\displaystyle #1$}\kern8pt}
\kern8pt\vrule}\hrule}}
\overfullrule=0pt

\def\Title#1#2{\rightline{#1}
\ifx\answ\bigans\nopagenumbers\pageno0\vskip1in%
\baselineskip 15pt plus 1pt minus 1pt
\else
\def\listrefs{\footatend\vskip 1in\immediate\closeout\rfile\writestoppt
\baselineskip=20pt\centerline{{\bf References}}\bigskip{\frenchspacing%
\parindent=20pt\escapechar=` \input
refs.tmp\vfill\eject}\nonfrenchspacing}
\pageno1\vskip.8in\fi \centerline{\titlefont #2}\vskip .5in}

\ifx\answ\bigans\def\tcbreak#1{}\else\def\tcbreak#1{\cr&{#1}}\fi
\useblackboard
\message{If you do not have msbm (blackboard bold) fonts,}
\message{change the option at the top of the tex file.}
\font\blackboard=msbm10 
\font\blackboards=msbm7
\font\blackboardss=msbm5
\textfont\black=\blackboard
\scriptfont\black=\blackboards
\scriptscriptfont\black=\blackboardss
\def\Bbb#1{{\fam\black\relax#1}}
\else
\def\Bbb#1{{\bf #1}}
\fi
%
\def\yboxit#1#2{\vbox{\hrule height #1 \hbox{\vrule width #1
\vbox{#2}\vrule width #1 }\hrule height #1 }}
\def\fillbox#1{\hbox to #1{\vbox to #1{\vfil}\hfil}}
\def\ybox{{\lower 1.3pt \yboxit{0.4pt}{\fillbox{8pt}}\hskip-0.2pt}}
\def\np#1#2#3{Nucl. Phys. {\bf B#1} (#2) #3}
\def\pl#1#2#3{Phys. Lett. {\bf #1B} (#2) #3}

\def\physrev#1#2#3{Phys. Rev. {\bf D#1} (#2) #3}

\def\cmp#1#2#3{Comm. Math. Phys. {\bf #1} (#2) #3}

\def\comments#1{}

\def\half{{1\over 2}}

\def\a{\alpha}

\def\II{\relax{I\kern-.07em I}}

\def\hk{{hyperk\"ahler}}

\def\IZ{\relax\ifmmode\mathchoice
{\hbox{\cmss Z\kern-.4em Z}}{\hbox{\cmss Z\kern-.4em Z}}
{\lower.9pt\hbox{\cmsss Z\kern-.4em Z}}
{\lower1.2pt\hbox{\cmsss Z\kern-.4em Z}}\else{\cmss Z\kern-.4em
Z}\fi}
\def\IB{\relax{\rm I\kern-.18em B}}
\def\IC{\bf C}
\def\ID{\relax{\rm I\kern-.18em D}}
\def\IE{\relax{\rm I\kern-.18em E}}
\def\IF{\relax{\rm I\kern-.18em F}}
\def\IG{\relax\hbox{$\inbar\kern-.3em{\rm G}$}}
\def\IGa{\relax\hbox{${\rm I}\kern-.18em\Gamma$}}
\def\IH{\relax{\rm I\kern-.18em H}}
\def\II{\relax{\rm I\kern-.18em I}}
\def\IK{\relax{\rm I\kern-.18em K}}
\def\IP{\relax{\rm I\kern-.18em P}}

\useblackboard
\def\IZ{\relax\Bbb{Z}}
\fi

\font\cmss=cmss10 \font\cmsss=cmss10 at 7pt
\def\IR{\relax{\rm I\kern-.18em R}}

\def\BR{\IR}
\def\BZ{\IZ}
\def\BR{\IR}
\def\BC{\IC}

\def\tilde{\widetilde}



\def\lim{{lim}}

\input epsf

\def\SUSY#1{{{\cal N}= {#1}}}                   
\def\lbr{{\lbrack}}                             
\def\rbr{{\rbrack}}                             

\def\wdg{{\wedge}}                              



\def\MR#1{{{\BR}^{#1}}}               
\def\MC#1{{{\BC}^{#1}}}               

\def\MR#1{{{\BR}^{#1}}}               
\def\MC#1{{{\BC}^{#1}}}               
\def\MS#1{{{\bf S}^{#1}}}               
\def\MT#1{{{\bf T}^{#1}}}               
\def\MHT#1{{{\bf \widetilde{T}}^{#1}}}               

\def\px#1{{\partial_{#1}}}              





\def\trp#1{{{\rm tr}\{ {#1} \} }}            

\def\rep#1{{{\bf {#1}}}}                      



\def\hepth#1{{\it hep-th/{#1}}}

\def\frac#1#2{{{{#1}}\over {{#2}}}}           

\def\u{{\mu}}
\def\v{{\nu}}
\def\b{{\beta}}

\def\lam{{\lambda}}



\def\Modsp{{\cal M}}     

\def\Vol#1{{{V\!\!ol\left({#1}\right)}}}    
\def\tw{{\theta}}         
\def\btw{{\eta}}         

\def\jhep#1#2#3{{{{\bf JHEP {#1}}({#2}){#3} }}}



\newif\iffigs\figstrue

%
\let\useblackboard=\iftrue
%
%
\newfam\black

\input harvmac.tex


\def\Title#1#2{\rightline{#1}
\ifx\answ\bigans\nopagenumbers\pageno0\vskip1in%
\baselineskip 15pt plus 1pt minus 1pt
\else
\def\listrefs {
\footatend\vskip 1in\immediate\closeout\rfile\writestoppt
\baselineskip=20pt
\centerline{{ \bf References}}
\bigskip{\frenchspacing%
\parindent=20pt\escapechar=` \input
refs.tmp\vfill\eject}\nonfrenchspacing}
\pageno1\vskip.8in\fi \centerline{\titlefont #2}\vskip .5in}

\ifx\answ\bigans\def\tcbreak#1{}
\else\def\tcbreak#1{\cr&{#1}}\fi
\useblackboard
\message{If you do not have msbm (blackboard bold) fonts,}
\message{change the option at the top of the tex file.}
\font\blackboard=msbm10 
\font\blackboards=msbm7
\font\blackboardss=msbm5
\textfont\black=\blackboard
\scriptfont\black=\blackboards
\scriptscriptfont\black=\blackboardss
\def\Bbb#1{{\fam\black\relax#1}}
\else
\def\Bbb#1{{\bf #1}}
\fi
\def\yboxit#1#2{\vbox{\hrule height #1 \hbox{\vrule width #1
\vbox{#2}\vrule width #1 }\hrule height #1 }}
\def\fillbox#1{\hbox to #1{\vbox to #1{\vfil}\hfil}}
\def\ybox{{\lower 1.3pt \yboxit{0.4pt}{\fillbox{8pt}}\hskip-0.2pt}}
\def\np#1#2#3{Nucl. Phys. {\bf B#1} (#2) #3}
\def\pl#1#2#3{Phys. Lett. {\bf #1B} (#2) #3}

\def\physrev#1#2#3{Phys. Rev. {\bf D#1} (#2) #3}

\def\cmp#1#2#3{Comm. Math. Phys. {\bf #1} (#2) #3}

\def\jhep#1#2#3{{{JHEP.{\bf {#1}}({#2}){#3}}}}
\def\atmp#1#2#3{Adv. Theor. Math. Phys. {\bf #1}(#2) #3}

\def\comments#1{}

\def\half{{1\over 2}}

\def\a{\alpha}

\def\II{\relax{I\kern-.07em I}}

\def\hk{{hyperk\"ahler}}

\def\IZ{\relax\ifmmode\mathchoice
{\hbox{\cmss Z\kern-.4em Z}}{\hbox{\cmss Z\kern-.4em Z}}
{\lower.9pt\hbox{\cmsss Z\kern-.4em Z}}
{\lower1.2pt\hbox{\cmsss Z\kern-.4em Z}}\else{\cmss Z\kern-.4em
Z}\fi}
\def\IB{\relax{\rm I\kern-.18em B}}
\def\IC{\bf C}
\def\ID{\relax{\rm I\kern-.18em D}}
\def\IE{\relax{\rm I\kern-.18em E}}
\def\IF{\relax{\rm I\kern-.18em F}}
\def\IG{\relax\hbox{$\inbar\kern-.3em{\rm G}$}}
\def\IGa{\relax\hbox{${\rm I}\kern-.18em\Gamma$}}
\def\IH{\relax{\rm I\kern-.18em H}}
\def\II{\relax{\rm I\kern-.18em I}}
\def\IK{\relax{\rm I\kern-.18em K}}
\def\IP{\relax{\rm I\kern-.18em P}}

\useblackboard
\def\IZ{\relax\Bbb{Z}}
\fi

\font\cmss=cmss10 \font\cmsss=cmss10 at 7pt
\def\IR{\relax{\rm I\kern-.18em R}}

\def\BR{\IR}
\def\BZ{\IZ}
\def\BR{\IR}
\def\BC{\IC}

\def\tilde{\widetilde}

%
%

\def\lim{{lim}}


\def\SUSY#1{{{\cal N}= {#1}}}                   
\def\lbr{{\lbrack}}                             
\def\rbr{{\rbrack}}                             

\def\wdg{{\wedge}}                              
\def\MR#1{{{\BR}^{#1}}}               
\def\MC#1{{{\BC}^{#1}}}               
\def\MR#1{{{\BR}^{#1}}}               
\def\MC#1{{{\BC}^{#1}}}               
\def\MS#1{{{\bf S}^{#1}}}               
\def\MT#1{{{\bf T}^{#1}}}               
\def\MHT#1{{{\bf \widetilde{T}}^{#1}}}               

\def\px#1{{\partial_{#1}}}              




\def\trp#1{{{\rm tr}\{ {#1} \} }}            

\def\rep#1{{{\bf {#1}}}}                      



\def\hepth#1{{\it hep-th/{#1}}}

\def\frac#1#2{{{{#1}}\over {{#2}}}}           


\def\u{{\mu}}
\def\v{{\nu}}
\def\b{{\beta}}

\def\lam{{\lambda}}




\def\Modsp{{\cal M}}     

\def\Vol#1{{{V\!\!ol\left({#1}\right)}}}    
\def\tw{{\alpha}}         
\def\etw{{\eta}}         

\def\bZ{{\overline{Z}}}
\def\bW{{\overline{W}}}
\def\qx#1{{\partial^{{#1}}}}

\def\ve{{\hat{e}}}
\def\TNS{{\rho}}        
\def\TBNT{{{\bf TN}}}   

\def\Ur#1{U_{\left[#1\right]}(|\vec{r}|)}
\def\tp{\tilde{\psi}}

\def\vr{|\vec{r}|}


\nopagenumbers

\vskip 7.5truein
\centerline{\titlefont Noncommutative Geometry and }
\centerline{\titlefont Twisted Little-String Theories}
\vskip 1.0truein
\centerline{\bf Morten Krogh}
\vskip 1.5 truein

\centerline{\bf A DISSERTATION}
\centerline{\bf PRESENTED TO THE FACULTY} 
\centerline{\bf OF PRINCETON UNIVERSITY}
\centerline{\bf IN CANDIDACY FOR THE DEGREE}
\centerline{\bf OF DOCTOR OF PHILOSOPHY}
\vskip .5truein
\centerline{\bf RECOMMENDED FOR ACCEPTANCE}
\centerline{\bf BY THE DEPARTMENT OF PHYSICS}
\vskip .5truein
\centerline{\bf June 1999}

\vfill\eject 
\centerline{} 
\vskip 1.0truein
\centerline{\bf \copyright\ Copyright by Morten Krogh, 1999}
\centerline{\bf All rights reserved.}

\vfill\eject

\baselineskip=10pt plus 2pt minus 1pt

\noblackbox
\footline={\hss\tenrm\folio\hss}
\pageno=-3
\centerline{\bf ABSTRACT}
\vskip 1.0 truein

In this thesis we will discuss various aspects of 
noncommutative geometry and compactified Little-String theories. 

First we will give an introduction to the use of 
noncommutative geometry in string theory. Thereafter we will 
present a proof of the connection between D-brane dynamics and 
noncommutative geometry. This proof was made in collaboration 
with Edna Cheung. Then we will explain the concept of instantons 
in noncommutative gauge theories which will be relevant for the 
last chapters. 

	The last chapters shift the focus to Little-String- 
and $(2,0)$-theories. We study compactifications of these 
theories on tori with twists. First we study the case of 
two coinciding branes in detail. This is based on work with 
Edna Cheung and Ori Ganor. Finally we study the case of an 
arbitrary number of coinciding branes. The main result here 
is that the moduli spaces of vacua for the twisted compactifications 
are equal to moduli spaces of instantons on a noncommutative torus. 
A special case of this is that a large class of gauge theories with 
$\SUSY{2}$ supersymmetry in $D=4$ or $\SUSY{4}$ in $D=3$ has  
moduli spaces which are moduli spaces of instantons on 
noncommutative tori.  
This work was done in collaboration with Edna Cheung, Ori Ganor and 
Andrei Mikhailov.

\vfill\eject



\centerline{\bf TABLE OF CONTENTS} 
\settabs\+3...&aaa&3.3..&aaa&3.3.3..&aaaaaaaaaaaaaaaaaaaaaaaaaaaaaaaaaaa
aaaaaaaaaaaaaaa&\cr
\bigskip\bigskip\bigskip 
\+Acknowledgements&&&&&&vi\cr 
\bigskip\medskip 
\+1.&Introduction to Noncommutative Geometry in string theory&&&&&1\cr 
\bigskip\medskip
\+2. &Worldsheet derivation of Noncommutative Geometry from &&&&&\cr 
\+  &D0-branes in a Background B-field&&&&&16\cr
\medskip
\+&&2.1&Introduction&&&16\cr
\smallskip
\+&&2.2&Zero-branes on $T^2$ with background B-field&&&17\cr
\smallskip
\+&&2.3&Non-trvial Gauge Bundles&&&26\cr
\smallskip
\+&&2.4&Incorporating 2-branes&&&28\cr
\smallskip
\+&&2.5&Conclusion&&&31\cr
\bigskip\medskip
\+3.&Instantons in noncommutative gauge theories&&&&&&33\cr 
\bigskip\medskip 
\+4.&Twisted Little-String Theories&&&&&&40\cr
\medskip
\+&&4.1&The problem &&&41\cr
\medskip
\+&&4.2&Solution &&&45\cr
\medskip
\+&&4.3&Limits &&&49\cr
\medskip
\+&&4.4&Reduction of the twisted $(2,0)$theory to 4+1DT&&&60\cr
\medskip
\+&&4.5&R-symmetry twists in the little-string theories&&&63\cr
\medskip
\+&&4.6&Discussion&&&67\cr
\bigskip\medskip
\+5.&Instantons on a Non-commutative $T^4$ from Twisted 
      $(2,0)$&&&&& \cr
\+ &   and Little-String Theories&&&&&&68\cr
\medskip
\+&&5.1&The Solution &&&72\cr
\medskip
\+&&5.2&Review of Noncommutative Gauge Theories&&&80\cr
\medskip
\+&&5.3&Noncommutative Instantons as the Moduli-space&&&83\cr
\medskip
\+&&5.4&The 3+1D limit &&&91\cr
\medskip
\+&&5.5&Another Look at the $\eta$-twists &&&94\cr
\medskip
\+&&5.6&Conclusion &&&106\cr
\bigskip\medskip
\+References&&&&&&109\cr

\vfill\eject 

\centerline{\bf ACKNOWLEDGEMENTS}
\bigskip
I would like to thank my advisor, Prof. Ori Ganor, for 
many interesting and educating discussions, 
our collaborations and his 
encouragement of my research. 

I would also like to thank Edna Cheung, Sangmin Lee and 
Andrei Mikhailov for our collaborations on joint papers.    

I am grateful to many people at Princeton University from whom 
I have enjoyed conversations about physics and other matters. 
Among these people I would particularly like to thank: 
Edna, Andrei, Shiraz, Mark, Paul, Don, Judith, Lorenzo, Steve, 
 Vamsi, Ashvin, Mukund, Chang, Oyvind, Alberto, Kyunghwa, 
Sangmin, Anastasia, Tamar, Sanjaye, Zachary, Savdeep, Yuji, 
Michael, Frank, Prof. Klebanov, Prof. Callan, Prof. Verlinde, 
Prof. Periwal, Prof. Thorlacius and Prof. Taylor. 

I would also like to thank many people from outside the 
Physics Dept. of Princeton University who have been important 
for me. This includes my family: Leila, Torben, Mikala, Merete, 
 Terese, Helle, Tobias, Marie, Simon and many others. It includes 
friends from Denmark: Dan, Jens, Marie, Bodil, Michael, Niels, Ernst, 
 Joakim, Soren and many others. It also includes friends from 
Princeton outside physics: Jon, Jan, Peter, Zhimin, Inge, Attila 
and many others.

I am sorry for not being able to mention everybody. 

I am grateful to the Danish Research Academy for a 
fellowship that gave me the opportunity to engage in research 
without having to worry about my financial situation.

\vfill\eject


\lref\Sen{Ashoke Sen,
  {\it ``D0 Branes on $T^n$ and Matrix Theory,''} \hepth{9709220.}}


\lref\Seiwhy{N. Seiberg,
  {\it ``Why is the Matrix Model Correct?'',
  Phys.Rev.Lett. 79 (1997) 3577-3580} \hepth{9710009.}}

\lref\WOS{Ori J. Ganor, Sanjaye Ramgoolam and Washington Taylor,
  {\it ``Branes, Fluxes and Duality in M(atrix)-Theory'',
   Nucl. Phys. B492 (1997) 191-204} \hepth{9611202.}}

\lref\Li{Miao Li,
  {\it ``Comments on Supersymmetric Yang-Mills Theory on
         a Noncommutative Torus'',} \hepth{9802052.}}

\lref\HWW{Pei-Ming Ho,Yi-Yen Wu and Yong-Shi Wu,
  {\it ``Towards a Noncommutative Geometric Approach to 
         Matrix Compactification'',} \hepth{9712201.}}

\lref\Ming{Pei-Ming Ho and Yong-Shi Wu,
  {\it ``Noncommutative Gauge Theories in Matrix Theory,''}
   \hepth{9801147.}}

\lref\Casal{R Casalbuoni,
  {\it ``Algebraic treatment of compactification on 
         noncommutative tori'',} \hepth{9801170.}}

\lref\Halpern{M. Claudson and M. Halpern,
  {\it Nucl.Phys. B250 (1985) 689.}}

\lref\Flume{R. Flume,
  {\it Ann. of Phys. 164 (1985) 189.}}

\lref\BRR{M. Baake, P. Reinicke and V. Rittenberg,
  {\it J.Math. Phys. 26 (1985) 1070.}}

\lref\Witp{Edward Witten,
  {\it ``Bound States of Strings and p-Branes'',
  Nucl.Phys.B460 (1996) 335} \hepth{9510135.}}

\lref\Giveon{A. Giveon, M. Porrati and E. Rabinovici,
  {\it ``Target Space Duality in String Theory'',
  Phys.Rept. 244 (1994) 77-202} \hepth{9401139.}}


\lref\rWAdSII{E. Witten, 
   {``Anti-de Sitter Space, Thermal Phase Transition, and Confinement
    in Gauge Theories,''} \hepth{9803131}}

\lref\rG{O.J. Ganor,
  {``Toroidal Compactification of Heterotic 6D  
  Non-Critical Strings Down to Four Dimensions,''}
  \np{488}(1997){223}, \hepth{9608109}}

\lref\rGMS{O.J. Ganor, D.R. Morrison and N. Seiberg,
  {``Branes, Calabi-Yau Spaces, and Toroidal Compactification
  of the $N=1$ Six-Dimensional $E_8$ Theory,''} 
  \np{487}{1997}{93}, \hepth{9610251}}

\lref\rIMS{K. Intriligator, D.R. Morrison and N. Seiberg,
  {``Five-Dimen\-sional Supersymmetric Gauge Theories and Degenerations
  of Calabi-Yau Spaces,''} \np{497}{97}{56-100}, \hepth{9702198}}

\lref\rWitCOM{ E. Witten,
  {``Some Comments on String Dynamics,''}
  \hepth{9507121},
  published in {\it ``Future perspectives in string theory,''} 501-523.}

\lref\rStrOPN{A. Strominger,
  {``Open p-Branes,''} \pl{383}{1996}{44-47}, \hepth{9512059}}

\lref\rWitNGT{E. Witten,
  {``New ``Gauge'' Theories In Six Dimensions,''}
  \hepth{9710065}}

\lref\rSeiVBR{N. Seiberg,
  {``New Theories in Six-Dimensions and
  Matrix Description of M-theory on $T^5$ and $T^5/Z_2$,''}
  \hepth{9705221}, \pl{408}{97}{98}}

\lref\rBFSS{T. Banks, W. Fischler, S.H. Shenker and L. Susskind,
  {``M Theory As A Matrix Model: A Conjecture,''}
  \hepth{9610043}, \physrev{55}{1997}{5112-5128}}

\lref\rDVVQ{R. Dijkgraaf, E. Verlinde, H. Verlinde,
  {``BPS Quantization Of The 5-Brane,''}
  \np{486}{97}{77}, \hepth{9604055}}
\lref\rDVVS{R. Dijkgraaf, E. Verlinde, H. Verlinde,
  {``BPS Spectrum Of The 5-Brane And Black-Hole Entropy,''}
  \np{486}{97}{89}, \hepth{9604055}}

\lref\rSWSIXD{N. Seiberg and E. Witten,
  {``Comments On String Dynamics In Six-Dimensions,''}
  \hepth{9603003}, \np{471}{1996}{121}}

\lref\rKS{A. Kapustin and S. Sethi,
  {``The Higgs Branch of Impurity Theories,''} \hepth{9804027}}

\lref\rSeiSTN{N. Seiberg,
  {\it ``Notes on Theories with 16 Supercharges,''}
  \hepth{9705117}}

\lref\rKV{S. Kachru and C. Vafa,
  {``Exact Results For $N=2$ Compactifications
  Of Heterotic Strings,''} \np{450}{95}{69}, \hepth{9505105}}

\lref\rWitFBR{E. Witten,
  {``Solutions Of Four-Dimensional Field Theories Via M Theory,''}
  \np{500}{1997}{3--42},\hepth{9703166}}

\lref\rSenFO{A. Sen,
  {``F-theory and Orientifolds,''}
  \np{475}{1996}{562-578}, \hepth{9605150}}

\lref\rSeiIRD{N. Seiberg,
  {``IR Dynamics on Branes and Space-Time Geometry,''}
  \pl{384}{1996}{81--85}, \hepth{9606017}}

\lref\rSav{S. Sethi,
  {``The Matrix Formulation of Type IIB Five-Branes,''}
  \hepth{9710005}}

\lref\rGS{O.J. Ganor and S. Sethi,
  {``New Perspectives on Yang-Mills Theories With Sixteen Supersymmetries,''}
  \hepth{9712071}}

\lref\rGK{S.S. Gubser and I.R. Klebanov,
  {``Absorption by Branes and Schwinger Terms in the World Volume Theory,''}
  \pl{413}{1997}{41--48}, \hepth{9708005}}

\lref\rMS{J. Maldacena and A. Strominger,
  {``Semiclassical Decay of Near Extremal 5-Branes,''}
  \jhep{12}{97}{008}, \hepth{9710014}}

\lref\rSWGDC{N. Seiberg and E. Witten,
  {``Gauge Dynamics And Compactification To Three Dimensions,''}
  \hepth{9607163}}

\lref\rSeiHOL{N. Seiberg, {``Naturalness Versus Supersymmetric 
  Non-Renor\-malization Theorems,''}
 \pl{318}{1993}{469--475}, {\tt hep-ph/9309335}.}

\lref\rAspin{P. Aspinwall,
  {``K3 Surfaces and String Duality,''} \hepth{9611137}} 

\lref\rSeiFIV{N. Seiberg,
  {``Five Dimensional SUSY Field Theories, Non-trivial Fixed Points
  and String Dynamics,''} \pl{388}{1996}{753-760},\hepth{9608111}}

\lref\rSWII{N. Seiberg and E. Witten,
  {``Monopoles, Duality and Chiral Symmetry Breaking in $N=2$
  Supersymmetric QCD,''} \np{431}{1994}{484--550}, \hepth{9408099}}

\lref\rBarak{B. Kol,
  {``On 6d ``Gauge'' Theories with Irrational Theta Angle''},
  \hepth{9711017}}

\lref\rDH{M.R. Douglas and C. Hull,
  {``D-branes and the Noncommutative Torus,''}
  \hepth{9711165}}

\lref\rSavLen{S. Sethi and L. Susskind,
  {``Rotational Invariance in the M(atrix) Formulation
  of Type IIB Theory,''} \pl{400}{1997}{265--268}, \hepth{9702101}}

\lref\rTomNat{T. Banks and N. Seiberg,
  {``Strings from Matrices,''} \np{497}{1997}{41--55}, \hepth{9702187}}

\lref\rPolWit{J. Polchinski and E. Witten,
  {``Evidence For Heterotic - Type I String Duality,''}
  \np{460}{1996}{525}, \hepth{9510169}}

\lref\rBerDou{M. Berkooz and M.R. Douglas,
  {``Five-branes in M(atrix) Theory,''} \hepth{9610236}}

\lref\rSeiWHY{N. Seiberg,
  {``Why is the Matrix Model Correct?,''} 
  \physrev{79}{1997}{3577--3580}, \hepth{9710009}}

\lref\rSenTD{A. Sen,
  {``D0 Branes on $T^n$ and Matrix Theory,''}
  \hepth{9709220}}

\lref\rWati{W. Taylor,
  {``D-brane field theory on compact spaces,''}
  \hepth{9611042}, \pl{394}{1997}{283}.}

\lref\rCDS{A. Connes, M.R. Douglas and A. Schwarz,
  {``Noncommutative Geometry and Matrix Theory: Compactification on Tori,''}
  \hepth{9711162}}

\lref\rCK{Y.-K. E. Cheung and M. Krogh,
  {``Noncommutative Geometry from 0-branes in a Background B-field,''}
  \hepth{9803031}}

\lref\rProg{Work in progress.}


\lref\rMicha{M. Berkooz,
  {``Non-local Field Theories and the Non-commu\-tative Torus,''}
  \hepth{9802069}}

\lref\Nahm{W. Nahm, 
  {``Self-dual monopoles and calorons,''}
  Lecture Notes in Physics, vol 201, Springer 1984}

\lref\rGTest{O.J. Ganor,
  {``A Test Of The Chiral E8 Current Algebra On A 6D
  Non-Critical String,''} \np{479}{1996}{197--217}, \hepth{9607020}}

\lref\rKMV{A. Klemm, P. Mayr and C. Vafa, 
  {``BPS States of Exceptional Non-Critical Strings,''} \hepth{9607139}}  

\lref\rMNWI{J.A. Minahan, D. Nemeschansky and N.P. Warner,
  {``Investigating the BPS Spectrum of Non-Critical $E_n$ Strings,''}
  \np{508}{1997}{64--106}, \hepth{9705237}} 

\lref\rMNWII{J. A. Minahan, D. Nemeschansky and N. P. Warner,
  {``Partition Functions for BPS States of the Non-Critical $E_8$ String,''}
  Adv.Theor.Math.Phys. 1 (1998) 167-183, \hepth{9707149}}

\lref\rMNVW{J. A. Minahan, D. Nemeschansky, C. Vafa and N. P. Warner,
  {``E-Strings and $N=4$ Topological Yang-Mills Theories,''}
  \hepth{9802168}}


\lref\rKumVaf{A. Kumar and C. Vafa,
  {``U-Manifolds,''} \pl{396}{1997}{85-90}, \hepth{9611007}}


\lref\rSWI{N. Seiberg and E. Witten,
  {``Electric-Magnetic Duality, Mono\-pole Condensation,
  and Confinement in $N=2$ Super\-sym\-metric Yang-Mills Theory,''}
  \np{426}{1994}{19}, \hepth{9407087}.}

\lref\rKV{S. Kachru and C. Vafa,
  {``Exact Results For $N=2$ Compactifications
  Of Heterotic Strings,''} \np{450}{1995}{69}, \hepth{9505105}.}

\lref\rWitCOM{E. Witten,
  {``Some Comments on String Dynamics,''}
  \hepth{9507121},
  published in {\it ``Future perspectives in string theory,''} 501-523.}

\lref\rSeiVBR{N. Seiberg,
  {``New Theories in Six-Dimensions and
  Matrix Description of M-theory on $T^5$ and $T^5/Z_2$,''}
  \hepth{9705221}, \pl{408}{1997}{98}}

\lref\rDVVQ{R. Dijkgraaf, E. Verlinde, H. Verlinde,       
 {``BPS Quantization Of The 5-Brane,''}
  \np{486}{1997}{77}, \hepth{9604055}}
\lref\rDVVS{R. Dijkgraaf, E. Verlinde and H. Verlinde,
  {``BPS Spectrum Of The 5-Brane And Black-Hole Entropy,''}
  \np{486}{1997}{89}, \hepth{9604055}}

\lref\rWitFBR{E. Witten,
  {``Solutions Of Four-Dimensional Field Theories Via M Theory,''}
  \np{500}{1997}{3--42}, \hepth{9703166}}

\lref\rCGK{Y.-K.E. Cheung, O.J. Ganor and M. Krogh,
  {``On the twisted $(2,0)$ and Little-String Theories,''}
  \np{536}{1998}{175}, \hepth{9805045}}

\lref\rBluInt{J.D. Blum and K. Intriligator,
  {``New Phases of String Theory and 6d RG Fixed Points via Branes at
  Orbifold Singularities,''} \np{506}{1997}{199}, \hepth{9705044}}

\lref\rIntNEW{K. Intriligator,
  {``New String Theories in Six Dimensions via Branes at Orbifold
  Singularities,''} \atmp{1}{1998}{271}, \hepth{9708117}}

\lref\rDM{M.R. Douglas and G. Moore, 
         {``D-branes, Quivers, and ALE Instantons,''}
         \hepth{9603167}}

\lref\rGanSet{O.J. Ganor and S. Sethi,
   {``New Perspectives On Yang-Mills Theories With 16 Supersymmetries,''}
     \jhep{01}{1998}{007}, \hepth{9712071}}

\lref\rKapSet{A. Kapustin and S. Sethi,
  {``The Higgs Branch of Impurity Theories,''}
  \atmp{2}{1998}{571}, \hepth{9804027}}

\lref\rWitNGT{E. Witten,
  {``New ``Gauge'' Theories In Six Dimensions,''}
  \atmp{2}{1998}{61}, \hepth{9710065}}

\lref\rBDS{T. Banks, M.R. Douglas and N. Seiberg,
  {``Probing F-theory With Branes,''}
  \hepth{9605199}, \pl{387}{1996}{278--281} }

\lref\rSeiIRD{N. Seiberg,
  {``IR Dynamics on Branes and Space-Time Geometry,''}
  \pl{384}{1996}{81--85}, \hepth{9606017}}

\lref\rSWGDC{N. Seiberg and E. Witten,
  {``Gauge Dynamics And Compactification To Three Dimensions,''}
  \hepth{9607163}}

\lref\rDouglas{ M.R.~Douglas, ``Branes within Branes'', \hepth{9512077}}

\lref\rCDS{A. Connes, M.R. Douglas and A. Schwarz,
  {``Noncommutative Geometry and Matrix Theory: Compactification on Tori,''}
  \jhep{02}{1998}{003}, \hepth{9711162}}

\lref\rDH{M.R. Douglas and C. Hull,
  {``D-branes and the Noncommutative Torus,''}\break
  \jhep{02}{1998}{008}, \hepth{9711165}}

\lref\rANS{A. Astashkevich, N. Nekrasov and A. Schwarz,
     {``On noncommutative Nahm transform,''}
   \hepth{9810147}}


\lref\rConnesbog{A. Connes, 
      {``Noncommutative Geometry,''}
      Academic Press, 1994.}

\lref\rSchwarzdual{Albert Schwarz, 
      {``Morita equivalence and duality,''}
    \np{B534}{1998}{289-300} \hepth{9805034}}

\lref\rHoWuWu{P.-M. Ho, Y.-Y. Wu and Y.-S. Wu, 
    {``Towards a Noncommutative Geometric Approach to 
       Matrix Compactification,''}
    \physrev{58}{1998}{026006}, \hepth{9712201}} 

\lref\rHoWu{P.-M. Ho, Y.-S. Wu,
{``Noncommutative Gauge Theories in Matrix Theory,''}
  \physrev{58}{1998}{066003}, \hepth{9801147}}

\lref\rNekSch{N. Nekrasov and A. Schwarz,
  {``Instantons on noncommutative $R^4$ and (2,0) superconformal       
  six dimensional theory,''} \cmp{198}{1998}{689}, \hepth{9802068}}

\lref\rMicha{M. Berkooz,
  {``Non-local Field Theories and the Non-commu\-tative Torus,''}
  \pl{430}{1998}{237}, \hepth{9802069}}

\lref\rBig{D. Bigatti, {``Non commutative geometry for outsiders,''}
  \hepth{9802129}} 

\lref\rCK{Y.-K. E. Cheung and M. Krogh,
  {``Noncommutative Geometry from 0-branes in a Background B-field,''}
  \np{528}{1998}{185-196},    \hepth{9803031}}

\lref\rKO{T. Kawano and K. Okuyama,
  {``Matrix Theory on Noncommutative Torus,'' }
 \pl{433}{1998}{29}, \hepth{9803044}} 

\lref\rAAS{ F. Ardalan, H. Arfaei and M.M. Sheikh-Jabbari,
{``Mixed Branes and M(atrix) Theory on Noncommutative Torus,''}
 \hepth{9803067}}

\lref\rHo{P.-M. Ho,
   {``Twisted Bundle on Quantum Torus and
     BPS States in Matrix Theory,''}
  \pl{434}{1998}{41}, \hepth{9803166}}

\lref\rMZ{B. Morariu, B. Zumino,
  {``Super Yang-Mills on the Noncommutative Torus,''}
  \pl{433}{1998}{279}, \hepth{9807198}}

\lref\rAASJ {F. Ardalan, H. Arfaei and M.M. Sheikh-Jabbari,
  {``Noncommutative Geometry form Strings and Branes,''}
 \hepth{9810072}}

\lref\rHVer{C. Hofman and E. Verlinde,
    {``U-duality of Born-Infeld on the Noncommutative  Two Torus,''}
 \hepth{9810116}}

\lref\rBraceMorariu{Daniel Brace and Bogdan Morariu, 
    {``A Note on the BPS Spectrum of the Matrix Model,''}
  \hepth{9810185}}

\lref\rHofVer{C. Hofman and E. Verlinde, 
    {``Gauge bundles and Born-Infeld on the Noncommutative Torus,''}
   \hepth{9810219}}

\lref\rSchwarzKon{A. Konechny and A. Schwarz,
    {``BPS states on noncommutative tori and duality,''}
   \hepth{9811159}}

\lref\rWati{W. Taylor,
  {``D-brane field theory on compact spaces,''}
  \pl{394}{1997}{283}, \hepth{9611042}}


\lref\rNek{N. Nekrasov,
  {``Five-Dimensional Gauge Theories
   and Relativistic Integrable Systems,''}
  \np{531}{1998}{323}, \hepth{9609219}}

\lref\rFMW{R. Friedman, J. Morgan, and E. Witten,
  {``Vector Bundles And F Theory,''}
  \cmp{187}{1997}{679}, \hepth{9701162}}

\lref\rBJPS{M. Bershadsky, A. Johansen, T. Pantev and V. Sadov,
  {``Four-Dimensional Compactifications of F-theory,''}
  \np{505}{1997}{165}, \hepth{9701165}}

\lref\rGriHar{Griffiths and Harris,
   {\it ``Principles of Algebraic Geometry''},
    Wi\-ley-Interscience, New-York, 1978}

\lref\rABS{O. Aharony, M. Berkooz and N. Seiberg,
  {``Linear Dilatons, NS Five-Branes and Holography,''}
  \jhep{9810}{1998}{004}, \hepth{9808149}}



\hoffset=.6truein\hsbody=5.8truein\hsize=\hsbody 
\hstitle=6.5truein
\voffset=.2truein
\vsize = 8truein

\pageno=1

\chapter{Introduction to Noncommutative geometry in string theory}

\vskip .2truein

Noncommutative geometry was invented by mathematicians as a 
generalization of the notion of a topological space. Alain 
Connes' book \rConnesbog\ is an exposition to these ideas. Here we 
will focus on its implications in string and gauge theory.
Noncommutative geometry entered string theory in the fall of 1997 in 
the paper by Connes,Douglas and Schwarz \rCDS. Here matrix theory 
for M-theory on $T^d$ was considered. One can put a background value 
for the 3-form potential of M-theory with one index along the lightlike 
circle and 2 indices along $T^d$. We see that we need $d \geq 2$ to do 
this. In that paper it was argued that the matrix model is a gauge 
theory on a noncommutative torus. Following Sen and Seiberg \refs
{\rSenTD , \rSeiWHY} the matrix model for M-theory in a certain background
 is given by the dynamics of D0-branes in the same background in a 
certain limit. The limit is such that the resulting theory becomes 
a gauge theory when $d \leq 3$ or another 
kind of field theory without gravity for higher $d$. The 
3-form potential with one lightlike index becomes an NS-NS B-field.

We see that the question of the role of noncommutative geometry in string 
theory can be phrased solely in terms of D-branes in a background 
$B^{NS}$-field. The connection to Matrix theory then follows trivially.

We thus consider the following question. Consider Type IIA compactified 
 on $T^2$ of radii $R_1 , R_2$ with a constant $B^{NS}$-field along 
$T^2$. Higher dimensional tori can be dealt with in a similar way. 
Let the string mass and coupling be $m_s$ and $g$. 
Define $\theta^{NS} 
= {1 \over 2\pi} \int_{T^2} B^{NS}$. $\theta^{NS}$ is a pure number. 
The gauge invariance of $B^{NS}$ makes $\theta^{NS}$ periodic with period 
1. Let us put $N_0$ D0-branes and $N_2$ D2-branes in this background. 
We will take the limit 
\eqn\granse{\eqalign{
m_s \rightarrow& \; \infty \cr
g \rightarrow& \; 0 \cr
R_1 , R_2 \rightarrow & \; 0 \cr
m_s^2 R_1 , m_s^2R_2  & \; \;\;\; {\rm   fixed} \cr
g m_s^3 & \;\;\;\; {\rm  fixed}
}}
This limit is the usual matrix theory limit, which gives a gauge 
theory decoupled from gravity. We want to give a description of this 
system.   

A method for dealing with D0-branes on a torus was developed by 
Taylor \rWati\ and Ganor,Ramgoolam and Taylor \WOS. This was taken 
further by Connes,Douglas and Schwarz  \rCDS. The method is to 
consider D0-branes described the quantum mechanics Lagrangian
\eqn\nulaktion{
L = {(2\pi)^2 \over 4gm_s^3 } Tr ( 
 \sum_i 2(D_0 X^i)^2 + \sum_{i,j} [X^i ,X^j]^2 
  + 2 i \bar{\Psi}\Gamma^0 D_0 \Psi + \bar{\Psi} \Gamma^i 
  [X^i, \Psi])
}
Here $X^i$, $i=1,\ldots,9$ are Hermitian matrices, $\Psi_{\alpha}$ 
are Majorana-Weyl spinors of SO(1,9) and 
\eqn\koder{
D_0 X^i = {\partial X^i \over \partial t} + i [A_0 , X^i]
}
and similarly for $D_0 \Psi$.
This action has a supersymmetry
\eqn\susyt{\eqalign{
\delta X^i =& {i \over 2} \bar{\epsilon} \Gamma^i \Psi \cr
\delta \Psi =& - {1 \over 4}( 2 D_0 X^i \Gamma^{0i} + 
  [X^i , X^j] \Gamma^{ij}) \epsilon
}}
with $\Gamma^{ij}= [\Gamma^i , \Gamma^j]$
There is also a nonlinearly realised supersymmetry 
\eqn\ikkelin{\eqalign{
\delta X^i =& 0 \cr
\delta \Psi_{\alpha} =& \zeta_{\alpha} 
}}
Rermark that the last symmetry only acts in the $U(1)$ part 
of $U(N)$. To deal with D0-branes on a $T^2$ we should make 
$X^1,X^2$ periodic. Since field configurations that differ by 
a $U(N)$ gauge transformation should be identified the periodicity 
of $X^1,X^2$ means that there exist unitary operators $U_1,U_2$ 
such that
\eqn\perlign{\eqalign{
U_1 X^1 U_1^{-1} =& X^1 + R_1 m_s^2 \cr
U_2 X^2 U_2^{-1} =& X^2 + R_2 m_s^2 \cr
U_i X^j U_i^{-1} =& X^j  \qquad i \neq j \cr
U_i \Psi_{\alpha} U_i^{-1} =& \Psi_{\alpha}
}}
We will take solutions of these equations to describe a certain 
D0-brane configuration on $T^2$. Remark that $N_0$,$N_2$ and 
$\theta^{NS}$ appear nowhere. We then expect that there are several 
classes of solutions to these equations, one class for each choice 
of $N_0$,$N_2$ and $\theta^{NS}$.

By taking the trace of the above equations it is readily seen 
that these equations do not admit finite dimensional matrix 
solutions. The natural 
thing is to look for solutions where $U_i$,$X^i$ and $\Psi_{\alpha}$ 
are operators in an infinite dimensional Hilbert space.
This infinite dimensionality just reflects that there are infinitely 
many D0-branes on the covering space of the $T^2$.

Let us now discuss the solutions of these equations. It is easily 
seen that $U_1 U_2 U_1^{-1} U_2^{-1}$ commutes with all 
$X^i$ and $\Psi_{\alpha}$. Since we are interested in an irreducible 
representation of $X^i$ and $\Psi_{\alpha}$ Schur's lemma tells us that 
 $U_1 U_2 U_1^{-1} U_2^{-1}$ must be a constant. Since $U_1$,$U_2$ 
are unitary the constant must be on the unit circle in the complex 
plane. We thus have
\eqn\ikkom{
U_1 U_2 = e^{2\pi i  \theta} U_2 U_1
}
We will later see that the interpretation of $\theta$ is that 
$\theta = \theta^{NS} = {1 \over 2\pi} \int_{T^2} B^{NS}$.
 
$U_1$ and $U_2$ generate an algebra. If $\theta =0$ this is the 
algebra of functions on a $T^2$: For instance we could take 
$U_1 = e^{ix_1}$,$U_2 = e^{ix_2}$. For nonzero $\theta$ this algebra 
is known as the noncommutative torus, $T^2_{\theta}$. The point of 
noncommutative geometry is that often one does not need the actual 
space but only the algebra of functions on the space. There is a 
one-to-one correspondence between 
topological spaces and commutative $C^{*}$-algebras.
Whereas  it is hard to generalize the space to a ``noncommutative'' 
space it is straightforward to drop the requirement that the 
$C^{*}$-algebra be commutative. The algebra $T^2_{\theta}$ is a natural 
generalisation of $T^2$, and is henceforth called the noncommutative 
torus. Remark that $\theta$ is periodic with period 1.

Solving eq.\perlign\ involves finding a Hilbert space and the set 
of operators $X$,$\Psi$ obeying eq.\perlign . Another way of saying 
this is to say that the Hilbert space is a module over $T^2_{\theta}$, 
and $X^1$,$X^2$ are connections in this module and $X^i \; i >2$ and 
$\Psi_{\alpha}$ are endomorphisms of the module. We will not explain 
the mathematics here. An exposition is given in \refs{\rCDS, \rSchwarzdual} 
among others. 
For $\theta =0$ the solution is given by bundles over 
$T^2$ \refs{\WOS}. The modules are the sections of these bundles. 
In the noncommutative case there is no space anymore and we can not 
define bundles in the usual way. The modules are the right generalisation. 

Now we will explicitly describe these modules following 
\refs{\rHo , \rMZ}. Let $\sigma_1$,$\sigma_2$ be coordinates 
periodic with period $2\pi \over m_s^2R_1$,$2\pi \over m_s^2R_2$ 
and obeying 
\eqn\rela{
[ \sigma_1 ,\sigma_2 ] = 2\pi i \theta {1 \over m_s^4R_1R_2}
}
Let $\partial_1$,$\partial_2$ be derivatives with respect to $\sigma_1$, 
$\sigma_2$. They obey 
\eqn\relato{\eqalign{
[ \partial_i , \sigma_j] =& \delta_{ij} \cr
[ \partial_i , \partial_j] =& 0
}} 
Define
\eqn\defu{\eqalign{
U_1 =& e^{i\sigma_1 m_s^2R_1} e^{2\pi \theta {1 \over m_s^2R_2} 
\partial_2}  \cr
U_2 =& e^{i\sigma_2 m_s^2R_2} e^{-2\pi \theta {1 \over m_s^2R_1} 
\partial_1} 
}}
Then 
\eqn\ikkomto{
U_1 U_2 = e^{i 2\pi \theta} U_2 U_1
}
The Hilbert space consists of n-component vector functions 
$\Phi(\sigma_1 , \sigma_2)$ obeying the boundary conditions
\eqn\randb{\eqalign{
\Phi(\sigma_1 + {2\pi \over m_s^2R_1} , \sigma_2) =&
\Omega_1(\sigma_1 ,\sigma_2) \Phi(\sigma_1 , \sigma_2)  \cr
\Phi(\sigma_1 , \sigma_2 + {2\pi \over m_s^2R_2}) =&
\Omega_2(\sigma_1 ,\sigma_2)  \Phi(\sigma_1 , \sigma_2)
}}
Here 
\eqn\overg{\eqalign{
\Omega_1(\sigma_1 ,\sigma_2)=& e^{i {m \over n} \sigma_2 m_s^2 R_2} U \cr
\Omega_2(\sigma_1, \sigma_2)=& V        
}}
where m is an integer. Here $U$, $V$ are $n \times n$ unitary matrices 
satisfying 
\eqn\klokke{
UV = e^{-2\pi i {m \over n}} VU 
}
Such matrices are readily found, for example as clock and shift 
matrices. m and n are integers characterizing the module.
$\Omega_1$,$\Omega_2$ satisfy the cocycle condition
\eqn\kocykel{
\Omega_1(\sigma_1 , \sigma_2 + {2\pi \over m_s^2R_2})\Omega_2(\sigma_1 ,
\sigma_2) =
\Omega_2(\sigma_1 + {2\pi \over m_s^2R_1}, \sigma_2) \Omega_1(\sigma_1 ,
\sigma_2) 
}
which is necessary for eq.\randb to make sense. Sections of the adjoint
bundle, or endomorphisms of the module, satisfy
\eqn\adjbundt{\eqalign{
\psi(\sigma_1 + {2\pi \over m_s^2 R_1}, \sigma_2) =& 
\Omega_1(\sigma_1 ,\sigma_2) \psi(\sigma_1 , \sigma_2) \Omega_1^{-1}( 
\sigma_1 ,\sigma_2) \cr
\psi(\sigma_1 , \sigma_2 + {2\pi \over m_s^2 R_2}) =& 
\Omega_2(\sigma_1 ,\sigma_2) \psi(\sigma_1 , \sigma_2) \Omega_2^{-1}( 
\sigma_1 ,\sigma_2) 
 }}
Connections, or covariant derivatives, satisfy the same equation
\eqn\conbun{\eqalign{
 D_i(\sigma_1 + {2\pi \over m_s^2 R_1}, \sigma_2) =& 
\Omega_1(\sigma_1 ,\sigma_2) D_i(\sigma_1 , \sigma_2) \Omega_1^{-1}( 
\sigma_1 ,\sigma_2) \cr
 D_i(\sigma_1 , \sigma_2 + {2\pi \over m_s^2 R_2}) =& 
\Omega_2(\sigma_1 ,\sigma_2) D_i(\sigma_1 , \sigma_2) \Omega_2^{-1}( 
\sigma_1 ,\sigma_2) 
 }}
One connection is 
\eqn\conn{\eqalign{
D_1 =& \partial_1 \cr
D_2 =& \partial_2 - if \sigma_1 
}}
where f is a constant. By plugging $D_1$,$D_2$ into eq.\conbun we see 
that 
\eqn\fligmed{
2 \pi f = {m \over n-m\theta} m_s^4 R_1 R_2 
}
The most general connection is this one plus an arbitrary adjoint section: 
\eqn\gencon{
\nabla_i = D_i + A_i
}

Now let us solve our original equations, eq.\perlign . $X^i$, $i \neq 1,2$ 
and $\Psi_{\alpha}$ commute with $U_1$ and $U_2$. $U_1$ and $U_2$ are 
constructed such that any function of $\sigma_1$ and $\sigma_2$ commute 
with it. We thus conclude that $X^i$, $i \neq 1,2$,$\Psi_{\alpha}$ can 
be any adjoint section, i.e. it solves eq.\adjbundt 
and is a function of 
$\sigma_1$,$\sigma_2$, but must not contain derivatives. In mathematical 
terms they are endomorphisms of the module over $T^2_{\theta}$. $X^1$,$X^2$ 
are solved by 
\eqn\xetto{\eqalign{
X^1 =& i \nabla_1 \cr
X^2 =& i \nabla_2 
}}
where $\nabla_1$,$\nabla_2$ are arbitrary connections of the module. 
The fields $X^i$,$\Psi_{\alpha}$ should also be Hermitian as always for 
D0-branes. Now we have the fields of the theory. The action is given by 
the time integral of eq.\nulaktion . The only thing we still need to 
specify is the definition of the trace. The trace is always taken of an 
endomorphism.
 Let $\Psi$ be an endomorphism, 
i.e. it obeys eq.\adjbundt . First we can take the trace of $\Psi$ as 
a $n \times n$ matrix. The trace of eq.\adjbundt shows that 
$tr \Psi(\sigma_1 , \sigma_2)$ is a periodic function of $\sigma_1$, 
$\sigma_2$. It makes sense to integrate $tr \Psi (\sigma_1 , \sigma_2)$ 
over $\sigma_1$, $\sigma_2$. We thus define 
\eqn\sporet{
Tr \Psi(\sigma_1 , \sigma_2) = C \int d\sigma_1 d\sigma_2 tr \Psi( 
\sigma_1 , \sigma_2) 
}
We only need to fix the constant,$C$. A natural normalisation, which 
is consistent with the commutative case $\theta =0$, is the following.
Let $F_{12}$ be the field strength of a connection $\nabla$
\eqn\styrke{
F_{12} = [ \nabla_1 , \nabla_2 ]
}
Then $Tr(F_{12})$, which is a topological invariant, should 
satisfy
\eqn\topnummer{
{(2\pi)^2 \over m_s^4 R_1 R_2} {1 \over 2\pi i} Tr F_{12} = -m
}
where $m$ is the integer from above. We remember that 
${(2\pi)^2 \over m_s^4 R_1 R_2}$ is the volume of the torus spanned by 
$\sigma_1$,$\sigma_2$. Let us use $D_i$ from above
\eqn\doven{
F_{12} = [D_1 ,D_2] = -i f 
}
We have 
\eqn\kvantis{\eqalign{
{m_s^4 R_1 R_2 \over (2\pi)^2} (-m) =& {1 \over 2\pi i} Tr F_{12} \cr 
 =& {1 \over 2\pi i} C \int d\sigma_1 d\sigma_2 tr(-i f) \cr 
 =& {-1 \over (2 \pi)^2} C {2\pi \over m_s^2 R_1}{2 \pi \over m_s^2R_2} 
     n {m \over n-m\theta} m_s^4 R_1 R_2 
}}
Here we used the value for $f$ given in eq.\fligmed . This fixes
\eqn\fixc{
C = {n-m\theta \over n} {m_s^4R_1 R_2 \over (2\pi)^2} 
}
In this discussion we need $n - m\theta > 0$. The trace is then defined 
as 
\eqn\spordef{Tr \Psi(\sigma_1 , \sigma_2) = 
 {m_s^4R_1 R_2 \over (2\pi)^2} {n-m\theta \over n}  
 \int d\sigma_1 d\sigma_2 tr \Psi( \sigma_1 , \sigma_2) 
}
Here $\Psi(\sigma_1 , \sigma_2)$ is any endomorphism. Notice that when 
$\theta = 0$ we get the usual definition of the trace. Notice also 
that $Tr 1 = n- m \theta$.

Let us shortly recapitulate. We have solved the equations for the 
compactification of the D0-brane quantum mechanics. The most general 
solution depended on integers $m,n$ and an angle $\theta$. We got an 
action for the fields. The only ambiguity was the definition 
of the trace. What we will do now is to identify the parameters in 
terms of the physical parameters $N_0$,$N_2$ and $\theta^{NS}$. We will 
also calculate the mass of some BPS-states and compare with the 
correct result, which is known from type IIA string theory. The 
agreement of these BPS masses will prove the correctness of the definition 
of the trace and give some evidence for the relevance of modules over 
noncommutative tori to D-brane dynamics. In chapter 2 we will provide 
a proof that this noncommutative gauge theory is indeed the correct answer. 

Let us first match the parameters $n,m,\theta$ with the number of 
D0-branes,$N_0$, the number of D2-branes,$N_2$ and $\theta^{NS} = 
{1 \over 2\pi} \int_{T^2} B^{NS}$. For $\theta^{NS} =0$ we understand 
the situation very well. The system is described by a gauge theory 
obtained after T-duality on both directions of $T^2$. After T-duality 
there are $N_0$ D2-branes and $N_2$ D0-branes giving a $U(N_0)$ gauge 
theory with first chern class, $c_1 =N_2$. Comparing with the module 
above we get 
\eqn\parmatch{\eqalign{
n =& N_0 \cr
m =& N_2 
}}
and $\theta =0$, since there is no noncommutativity. For general $\theta^{NS}$
 the relation \parmatch\ must still hold since $n,m,N_0,N_2$ are integers and 
cannot change continously. We know from string theory that in the presence 
of a $B^{NS}$-field the ``effective'' number of D0-branes is 
$N_0 - \theta^{NS} N_2$. Since the combination $n- \theta m$ appears 
all over in the above construction we are led to $\theta^{NS} = \theta$. 
In other words the noncommutativity parameter is set exactly by the 
$B^{NS}$-flux. 

Let us now calculate the energy of some BPS states. First we take 
the lowest energy state which just corresponds to the pure D0,D2- 
brane system with no excitations. This state preserves 16 supercharges. 
In the noncommutative gauge theory this is given by setting 
\eqn\rentsys{\eqalign{
\Psi_{\alpha} =& 0 \cr
X^i =& 0 \qquad i=3,\ldots,9 \cr
X^1 =& i D_1 \cr
X^2 =& i D_2 
}}
The energy, which can be calculated classically since it is a BPS 
state, is 
\eqn\renenergi{\eqalign{
E =& -{(2\pi)^2 \over 4gm_s^3} Tr 2 [X^1,X^2]^2 \cr
  =& - {(2\pi)^2 \over 4gm_s^3} 2  Tr (-f^2) \cr
  =& {2 (2\pi)^2 \over 4gm_s^3} f^2 Tr 1 \cr
  =& {1 \over 2gm_S^3} ({m \over n-m\theta})^2 (m_s^4R_1R_2)^2 
      (n-m\theta) \cr
  =& {m_s^5 (R_1 R_2)^2 \over 2g} {m^2 \over n-m\theta} 
}}
Let us compare this with the correct result known from string theory. 
 $N_2$ D2-branes wrapped on a $T^2$ with a $B^{NS}$ has an effective 
D0-brane number, $-\theta^{NS}N_2$. If there are $N_0$ D0-branes in 
addition, the effective D0-brane number is $N_0 - \theta^{NS} N_2 
= n-m\theta$. The energy of D0- and D2-branes add in quadrature. 
One way of understanding this is to T-dualize along a single direction 
and get a diagonal wound D-string. This T-dual picture also explains the 
shift $N_0 \rightarrow N_0 - \theta^{NS} N_2$, since $\theta^{NS}$ 
makes the T-dual torus non-rectangular. The energy is thus 
\eqn\stringrenenergi{\eqalign{ 
E =& \sqrt{((n-m\theta){m_s \over g})^2 + (m{R_1R_2m_s^3 \over g})^2}\cr 
  =& (n-m\theta){m_s \over g} + \half {m^2 \over n-m\theta}(R_1R_2)^2
  m_s^5 {1 \over g} + \ldots 
}}
where we have expanded the square root. In the limit 
\eqn\graense{\eqalign{
m_s \rightarrow& \; \infty \cr
g \rightarrow& \; 0 \cr
R_1 , R_2 \rightarrow& \; 0 \cr
m_s^2 R_1 , m_s^2 R_2 & \;\;\;\; {\rm fixed} \cr
g m_s^3 & \;\;\;\; {\rm fixed}
}}
which we consider, the first term goes to infinity, the second 
term stays finite and all higher terms vanish. The first term is 
well known from matrix theory. It is equivalent to the term $N \over R$ 
which is always subtracted. The second term is the interesting one 
which should be reproduced by the SYM action. This story is also 
well known from expanding the Born-Infeld action, where the first 
term is the background energy of the brane, the second term is the 
SYM action and higher terms are suppressed at low energy.

By comparing the second term in eq.\stringrenenergi with 
the energy obtained from the noncommutative SYM, eq.\renenergi , 
we find agreement. We also obtained a further understanding for 
the ubiquity of the expression $n- \theta m$.

Let us now consider the D0,D2-brane system with momentum and 
fundamental string winding along the cycles of the $T^2$. 
For generic momenta and winding this state preserves 8 supercharges, 
but for special values 16 supercharges are preserved. Let us just look 
at a particular example of a 16 supercharge state. The others can 
be done similarly but with a bit more work. We consider a state 
with $N_0$ D0-branes, $N_2$ D2-branes, k units of momentum along the 
first axis and w units of fundamental string winding along the 
second axis. By performing a T-duality along the first axis this becomes 
a wrapped D-string and a wrapped fundamental string. They combine 
into a maximally supersymmetric state when they are parallel. The 
condition for that is 
\eqn\parkon{
wN_0 = k N_2 
}
If they are not parallel they will form a string web. We will 
only consider the parallel case. The energy of the state is readily 
calculated since the different contributions add in quadrature. 
The fact that they add in quadrature is particularly clear from 
the T-dual picture. 
\eqn\avanceret{\eqalign{
E^2 =& ((N_0 - \theta N_2){m_s \over g})^2 + 
        ({N_2 m_s^3 R_1 R_2 \over g})^2 \cr
    +& ({k - \theta w \over R_1})^2 + (w m_s^2R_2)^2 
}}
Taking into account the limit, eq.\graense , the energy becomes 
\eqn\avan{\eqalign{ 
E =& (N_0 - \theta N_2) {m_s \over g} \cr 
  +& \half {m_s^5 (R_1 R_2)^2 \over g} {N_2^2 \over N_0 -\theta N_2} 
  + \half {g \over m_s R_1^2}{(k- \theta w)^2 \over (N_0 - \theta N_2)}\cr 
  +& \;  {\rm vanishing \;\; terms}
}}
Again it is the finite terms which should be compared with the gauge theory.
  
Let us now turn to the noncommutative gauge theory and reproduce this 
state. Let us consider the fields of the form
\eqn\formaffelter{\eqalign{
X^1 =& iD_1 + A(t) 1 \cr
X^2 =& iD_2 
}}
where $D_1$,$D_2$ are the connections given in eq.\conn . $A(t)$ is 
a function of time. Remark that $A(t)$ multiplies the unit element 
in $End_{T^2_{\theta}}(E)$. We are going to take $A(t)$ to be 
linear in time. The corresponding field strength is then constant 
and we see from the supersymmetry variations, eq.\susyt ,eq.\ikkelin , 
that the corresponding state preserves 16 supercharges. This especially 
means that the  energy receives no corrections. Let us write down 
the lagrangian for these fields, obtained by plugging into eq.\nulaktion .
\eqn\specaktion{\eqalign{
L =& {(2\pi)^2 \over 4gm_s^3} Tr (2 \dot{A}^2 +2 (if)^2) \cr
  =& \half {(2\pi)^2 \over gm_s^3}(n-m\theta) \dot{A}^2 
  - \half m_s^5 (R_1R_2)^2 {1 \over g} {m^2 \over n-m\theta}
}}
This action is very simple and we can easily find the energy levels. 
The only thing we need to know is the period of $A$. $A$ is periodic, 
because there are gauge transformations that shift $A$. A similar 
story holds, of course, in the commutative case, $\theta=0$,
 where it is more well known. Let $D=gcd(n,m)$ then the period of $A$ 
is 
\eqn\aper{
A \rightarrow A + m_s^2 R_1 {D \over n-\theta m} 
}
This is not hard to show, but we will not present the calculation here. 
The energies are now readily found 
\eqn\energiniv{
E = \half m_s^5 (R_1R_2)^2{1 \over g} {m^2 \over n-m\theta}
  + \half {g \over m_sR_1^2}{n-\theta m \over D^2} l^2 
}
where $l$ is the integer quantum number for the conjugate 
momentum to $A$. Using eq.\parkon which implies 
\eqn\forhold{ 
l = {k- \theta w \over n-\theta m}D
}
we now see that there is perfect agreement with the string 
theory result, eq.\avan .

This finishes our treatment of comparing BPS masses. We will 
not discuss the less supersymmetric states. This is done in detail 
 in the literature, \refs{\rBraceMorariu, \rHofVer, \rSchwarzKon}.

In this chapter we have argued that the D-brane system with $B^{NS}$ 
flux is desrcribed by gauge theory on a noncommutative torus. Firstly 
we obtained it as a solution to the quotient conditions eq.\perlign , 
secondly BPS masses agree exactly. In the next chapter we will provide 
a proof of this connection from a worldsheet point of view. Before 
we do this we need to explain another equivalent
 formulation of gauge theory on 
a noncommutative torus.

We saw that the fields of the gauge theory were connections and 
endomorphisms over the noncommutative torus. This is beautiful 
from a mathematical point of view, but from a practical point of
 view it would be preferable to work with ordinary functions or 
sections of a bundle. Indeed there is a way of reformulating 
the theory to a theory of ordinary functions or sections. Let us 
here explain the case of $N_2=0$. In this case the fields could be 
considered as functions, $\Phi(\sigma_1 ,\sigma_2)$, of noncommuting 
variables $\sigma_1$,$\sigma_2$
\eqn\hvordanikkekom{
[\sigma_1 , \sigma_2 ] = 2\pi i \theta {1 \over m_s^4 R_1 R_2}
}
$\sigma_1$ and $\sigma_2$ are periodic with period
\eqn\deresper{\eqalign{
\sigma_1 \rightarrow & \sigma_1 + {2\pi \over m_s^2R_1} \cr
\sigma_2 \rightarrow & \sigma_2 + {2\pi \over m_s^2R_2}
}}
Any field can be written as 
\eqn\feltform{
\Phi(\sigma_1 ,\sigma_2)= \sum_{n_1 ,n_2} C_{n_1 ,n_2} 
 e^{in_1 \sigma_1 m_s^2 R_1} e^{i n_2 \sigma_2 m_s^2 R_2}
}
Let us now consider ordinary commuting variables, $s_1$,$s_2$,
 of the same period as $\sigma_1$,$\sigma_2$. We can define 
a map from $\Phi(\sigma_1 , \sigma_2)$ to a function $F(s_1, s_2)$ 
as follows 
\eqn\moduleltilfun{
F(s_1 ,s_2) =  \sum_{n_1 ,n_2} C_{n_1 ,n_2} 
 e^{in_1 s_1 m_s^2 R_1} e^{i n_2 s_2 m_s^2 R_2}
 e^{-i\pi\theta n_1 n_2}
}
The product of fields $\Phi(\sigma_1 , \sigma_2)$ map to the following 
$*$-product
\eqn\stjernemultiplikation{\eqalign{
&(F_1 * F_2)(s_1 ,s_2) \cr
&= 
e^{i\pi \theta {1 \over m_s^4 R_1 R_2}
({\partial \over \partial a_1}{\partial \over \partial b_2}
 -{\partial \over \partial a_2}{\partial \over \partial b_1})}
F_1(a_1 ,a_2) F_2(b_1 ,b_2) |_{a_i =b_i = s_i}
}}
The gauge theory action can thus be formulated in terms of 
ordinary functions but with the $*$-product replacing any 
normal product. This presentation is more suitable for 
standard quantum field theory treatment. We remark 
that the action is nonlocal. The effect of the exponential 
is to introduce nonlocal interactions along the first axis 
for processes which have a momentum transfer in the 2. direction 
or vice versa.

\vfill\eject


\chapter{Worldsheet derivation of 
       Noncommutative Geometry from D0-branes in a Background B-field}

In this chapter 
we continue the 
study D0-branes in type IIA on $T^2$ with a background B-field turned on.  
We calculate explicitly how the background B-field modifies the D0-brane action.   
The effect of the B-field is to replace 
ordinary multiplication with a noncommutative $*$-product.   
This theory is exactly
 the non-local 2+1 dim SYM theory on a dual $T^2$ 
 proposed by Connes, Douglas and Schwarz, which was discussed in chapter 1. 
We calculate the radii and the gauge coupling for the SYM on the dual $T^2$
 for all 
choices of longitudinal momentum  and membrane wrapping number on the 
$T^2$.

\section{Introduction}

Last fall Connes, Douglas and Schwarz \rCDS\ made a very interesting 
proposal relating the matrix theory of M-theory 
on $T^2$ with a background three form potential, $C_{-12}$, to 
a gauge theory on a noncommutative torus. Shortly 
after Douglas and Hull justified this claim by relating 
these theories to a theory on a D-string \rDH. 
They also mentioned that this could be seen in the 
original 0-brane picture.
	
The purpose of the present paper is to precisely 
incorporate the background B-field in the dynamics 
of 0-branes. In this way we will obtain the 
matrix model for M-theory on $T^2$ with $C_{-12} 
\neq 0$. It confirms the claims made by Connes,Douglas 
and Schwarz. The theory is a SYM theory on 
a dual torus with a modified interaction 
depending on the B-field. The theory 
contains higher derivative terms of arbitrarily 
high power and is thus non-local. We 
calculate the radii of the dual torus and the 
gauge coupling constant. We get a noncommutative 
gauge theory for any choice of longitudinal momentum 
and number of membranes wrapped 
around the $T^2$. The radii and gauge coupling   
depend on these numbers.

Aspects of the connection between compactifications 
of M-theory and noncommutative 
geometry has, among others, been worked out in 
\refs{\HWW, \Ming, \Casal}.

\def\fig#1#2#3{
\par\begingroup\parindent=0pt\leftskip=1cm\rightskip=1cm\parindent=0pt
\baselineskip=20pt
\global\advance\figno by 1
\midinsert
\epsfxsize=#3
\centerline{\epsfbox{#2}}
\vskip 12pt
{\bf Fig.\ \the\figno: } #1\par
\endinsert\endgroup\par
}
\def\figlabel#1{\xdef#1{\the\figno}}
\def\encadremath#1{\vbox{\hrule\hbox{\vrule\kern8pt\vbox{\kern8pt
\hbox{$\displaystyle #1$}\kern8pt}
\kern8pt\vrule}\hrule}}
\section{Zero-branes on $T^2$ with background B-field}

Let us consider M-theory on $T^2$ with radii, $R_1$, $R_2$. 
The torus is taken to be rectangular for simplicity. 
Making the torus oblique does not introduce anything interesting.
 We are interested in the matrix 
description of this
theory with a background $C$-field, ${C_{-12}\neq 0}$.  Here - denotes the lightlike circle
and $1, 2$ the directions along the torus.  Let the Plank mass be M and the radius 
of the lightlike circle be $R$.  Following \refs{\Sen,\Seiwhy}, we take this to mean a limit of 
spatial compactifications.  We also perform a rescaling to keep the interesting energies
finite.  The upshot is that we consider M theory on $T^2 \times S^1$ with Plank mass $\tilde M$
and radii $\tilde R_1$, $\tilde R_2$, $\tilde R$ in the limit ${\tilde R \rightarrow 0}$ with
\eqn\rescaling{\eqalign{
\tilde M^2 \tilde R &= M^2 R \cr
\tilde M \tilde R_i &= MR_i \qquad i=1,2. 
}}
This turns into type IIA on $T^2$ with string mass $m_s$, coupling $\lambda$ and radii
$r_1$, $r_2$ given by
\eqn\IIAscaling{\eqalign{
{m_s}^2 &= \tilde M^3 \tilde R \cr
\lambda &=(\tilde M\tilde R)^{3 \over 2}\cr
r_i&=\tilde R_i \qquad i=1,2. 
}}
Furthermore there is a flux of the $B_{ij}$ field through the torus.  We call the flux $B$:
\eqn\Bflux{
	B=R R_1 R_2 C_{-12}. 
}
We are interested in the sector of theory labelled by two integers, namely the number of 
D0-branes, $N_0$, and the number of D2-branes, $N_2$, wrapped on $T^2$.  In this section we
will solely be interested in the case $N_2=0$.  In the next section we will treat the general 
case.  Let us for the moment set $N_0$=1.  This makes us avoid some essentially irrevalent 
indices.  The generalization to general $N_0$ is straightforward.

The method for dealing with this situation has been developed in \refs{\rWati, \WOS}.  We work with 
the covering space of $T^2$, namely ${\bf R^2}$ and place 0-branes in a lattice (Figure 1).

\fig{0-branes on the covering space,$\bf R^2$ }{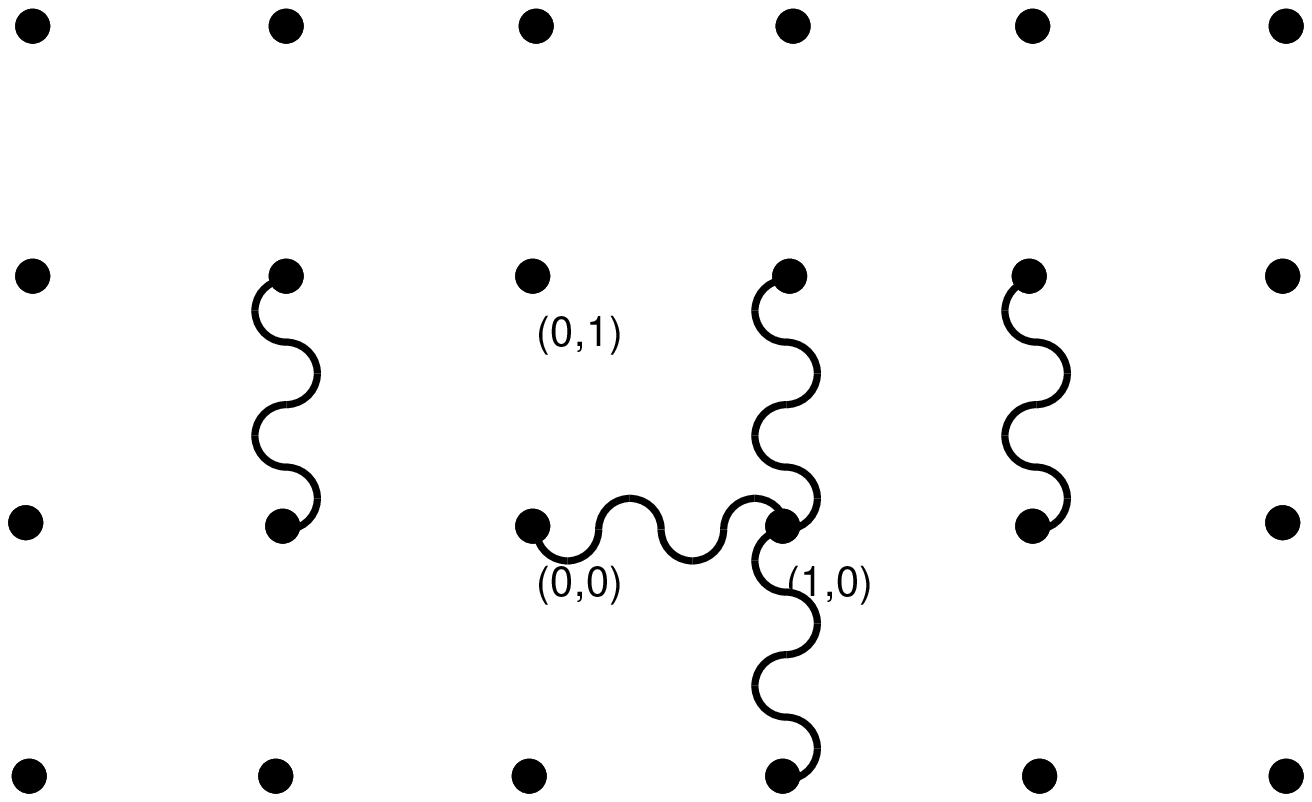}{4truein}

Let us label the 0-branes $(a, b)$  where $a$, $b$ are integers.  The open strings obey Dirichlet boundary
condition on the 0-branes.  This is the point where we need $N_2=0$.  If there had been D2-branes,
 the 0-branes would have been dispersed as fluxes inside the 2-branes and the open strings would
have Dirichlet boundary conditions on the 2-branes and the above picture does not apply.

The fields in the theory come from quantizing the open strings and calculating their interactions.
In the limit we are taking, ${m_s\rightarrow \infty}$, only the lowest modes survive and when 
$B=0$ the theory becomes a $SYM$ quantum mechanics \refs{\Halpern, \Flume, \BRR, \Witp, \rBFSS}.  
The gauge group is $"U(\infty)"$ since there are infinitely many 0-branes.  To be more precise let us
define a Hilbert space on which the fields will be operators.  There is a basis vector for each 
0-brane, i.e. the basis is $|a,b>$ where $a, b \in {\bf Z}$.  Let $\phi$ be any field in theory, then 
the matrix element $\phi_{a_1b_1,a_2b_2}$ has the interpretation as a field which annihilates a 
state of an open string starting at $a_1b_1$ and ending at $a_2b_2$, see 
figure 2.
\fig{String starting at 0-brane $(a_1,b_1)$ and ending at 
     0-brane $(a_2,b_2)$} {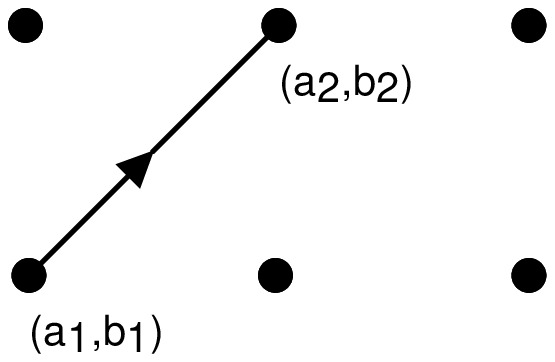} {2truein}
The fields of the theory are
\item{1.} bosons:$ \; X^i \qquad i=1,\ldots,8$ 
\item{2.} fermions:$ \; \Psi_{\alpha} \qquad \alpha=1,\ldots,16$ 
\item{3.} The gauge field: $A_0$.

The fields are constrained to obey the following conditions:
\eqn\constraints{\eqalign{
U_i^{-1}\;X^a\; U_i\; &= X^a + 2\pi r_a \delta^{ai}\;\; i=1,2; \cr
U_i^{-1}\;\Psi^{\alpha}\;U_i\;&=\; \Psi^{\alpha}, \cr
U_i^{-1}A_0 U_i &= A_0;
}}
where $U_i$ are translation operators on the states in the Hilbert space:
\eqn\defineu{\eqalign{
U_1|a,b>&=\;|a+1, b> \cr
U_2|a,b>&=\;|a,b+1>.
}}
The gauge field $A^0$ can be gauged away, and we will work in the gauge $A^0=0$. 
When $B=0$ the action is
\eqn\Obraneaction{\eqalign{
L= {m_s\over 2\lambda} \bf{Tr} \lbr
& \dot{X^a} \dot{X^a} + {m_s^4 \over (2\pi)^2} \sum_{a<b} {\lbr{X^a,X^b}\rbr}^2 \cr
 +& {m_s^2 \over {2\pi}} \Psi^T i \dot{\Psi} - {m_s^4 \over (2\pi)^2} \Psi^T
  \Gamma_a \lbr {X^a,\Psi} \rbr \rbr .
}}
What about $B\neq0$?  We will now show how to incorporate $B$-dependence in the action.
The two-form $B_{ij}$ couples to the worldsheet through the interaction $\int_{W.S.} B_{ij}$,
i.e. the $B$-field is pulled back to the worldsheet and integrated.  Let us look at the example shown 
in Figure 3 below.
\fig{The interaction between these three strings give rise 
      to a cubic vertex.} {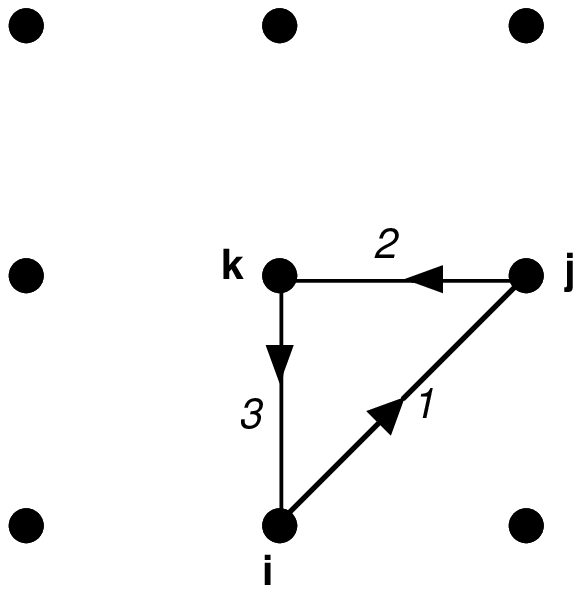} {2truein}
This interaction is represented, in the case $B=0$, by a term:
\eqn\threephi{ \kappa
\phi^{(3)}_{ik} \phi^{(2)}_{kj} \phi^{(1)}_{ji}
}
where $\kappa$ is a constant.  This term could, for instance, annihilate string 1 and 2
and create 3 with opposite orientation.  The worldsheet would look 
as shown in figure 4.
\fig{The worldsheet for a cubic vertex}{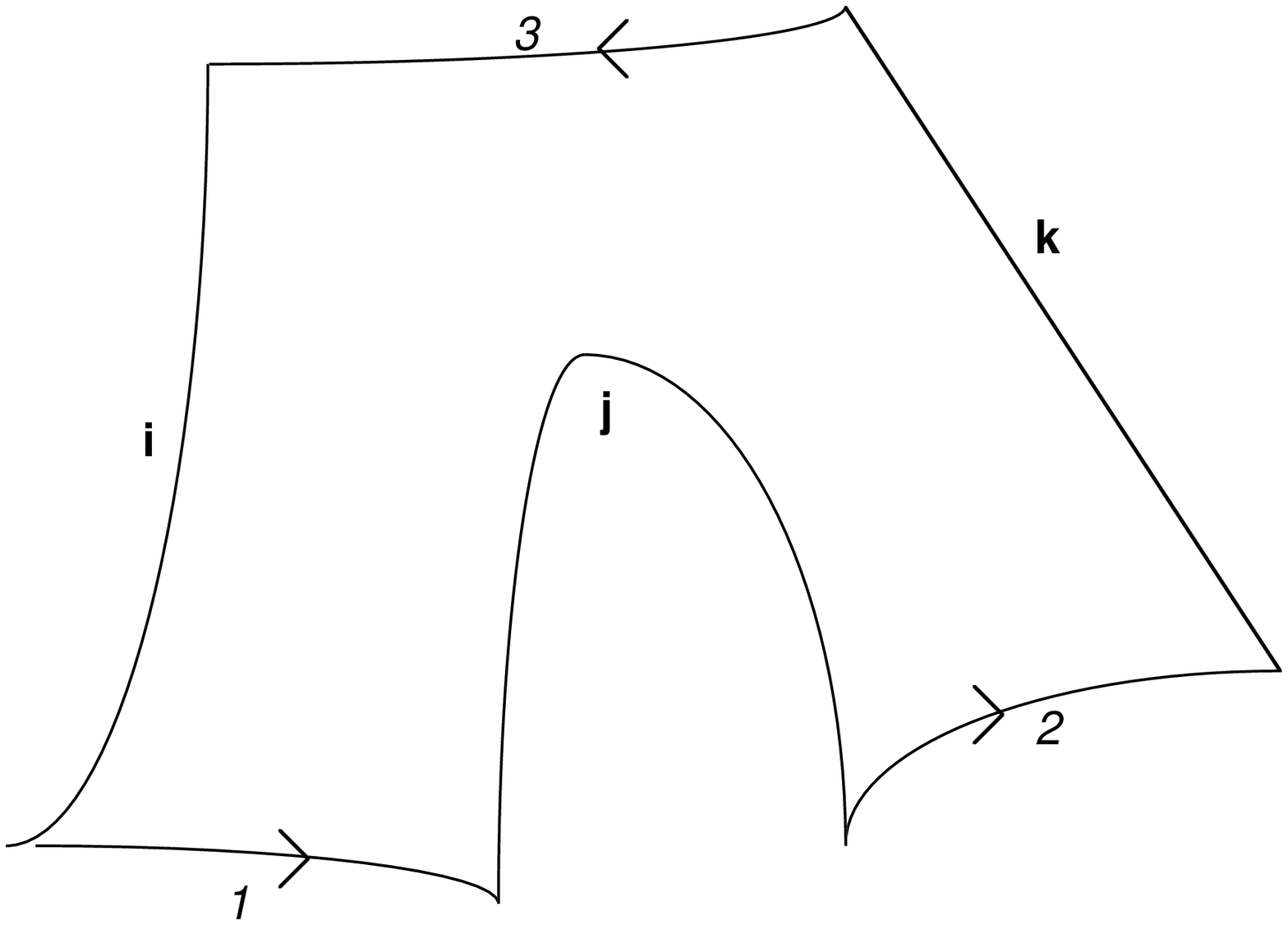}{4truein}
To calculate $\int_{W.S.} B_{ij}$ we just need the projection into the plane of the torus, since 
this is the only direction in which $B_{ij}\neq0$.  This projection is exactly given by the area 
between the three strings in Figure 3.  The important point is that $B_{ij}$ is closed so
$\int B_{ij}$ only depends on the homotopy type of the worldsheet imbedding and is 
insensitive to the finer details of how the interaction takes place.  For the example in Figure 3,
$\int_{W.S.} B_{ij} =\half B$, where we remark that $B$ was defined to be the flux through the 
torus.  This means that the interaction eq.\threephi\ now is replaced with
\eqn\Bthreephi{
e^{i\half B} \kappa \phi^{(3)}_{ik} \phi^{(2)}_{kj} \phi^{(1)}_{ji}.
}
It is now a straightforward exercise to figure out what happens to a general interaction 
between the fields:
\eqn\kphi{
	\phi^{(k)}_{a_1b_1,a_kb_k} \ldots \phi^{(2)}_{a_3b_3,a_2b_2} 
\phi^{(1)}_{a_2b_2,a_1b_1}.
}
We have to find the integral of the $B$-field through the polygon shown in Figure 5.
\fig{Generic form of a vertex with k strings} {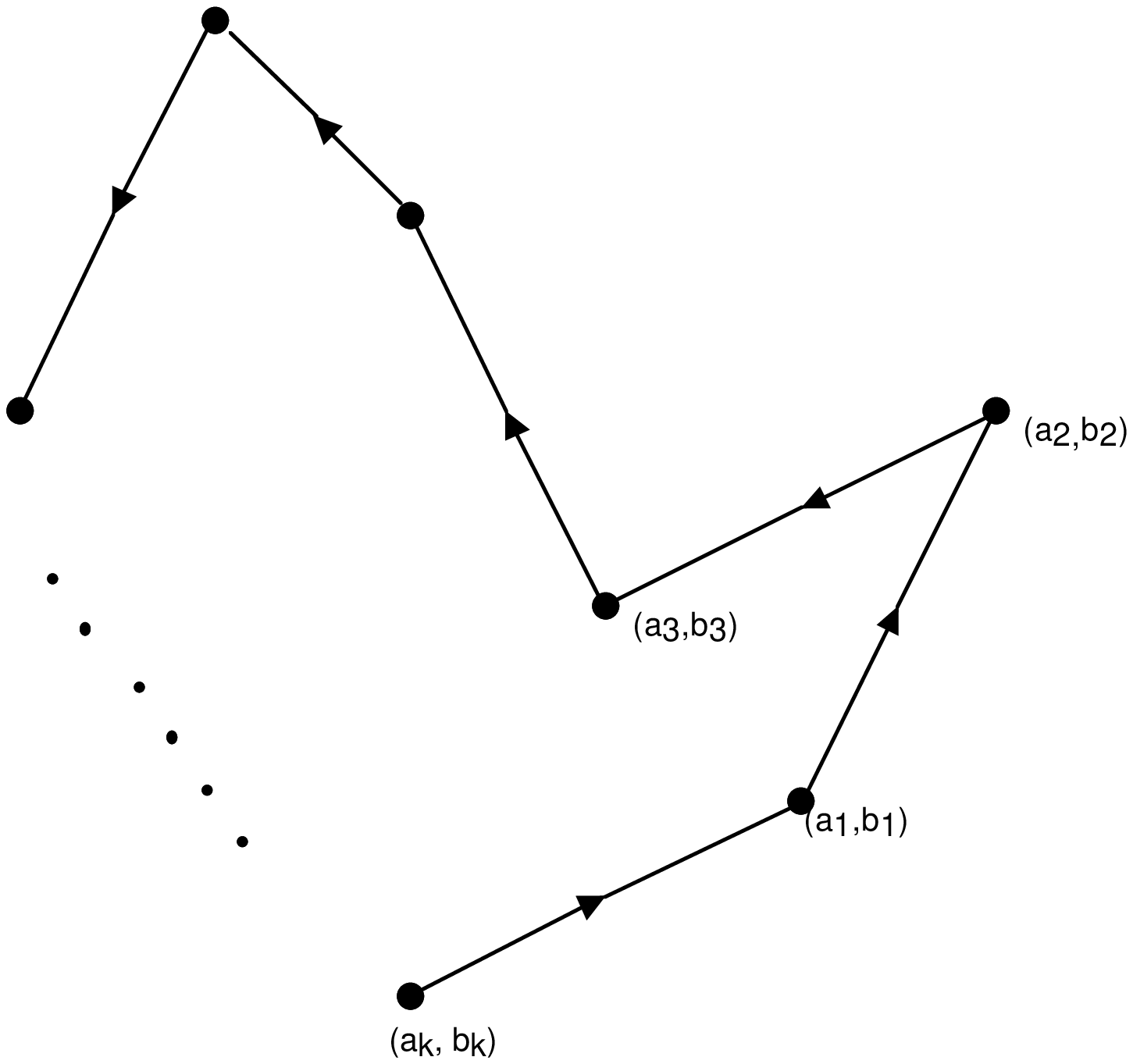} {3truein} 

We should 
remember to count with sign.  Orienting a polygon oppositely would change the sign of
$\int_{W.S.} B_{ij}$.  It is easily deduced that the result is 
\eqn\Bkphi{
\int_{W.S.} B_{ij}= \half B \sum _{i=2}^{k-1} \biggl|{\matrix{{a_{i+1} -a_i} &{a_i-a_1}\cr
	 {b_{i+1}-b_i}  & {b_i-b_1}\cr}\biggr|}
}
where $\Bigl|\matrix{a &b\cr c&d}\Bigr| = ad-bc$.  This means that the interaction eq.\kphi\ now becomes 
\eqn\intB{\eqalign{&
\phi^{(k)}_{a_1b_1,a_kb_k} \phi^{(k-1)}_{a_kb_k,a_{k-1}b_{k-1}}
e^{i \half  B  \biggl| \matrix {{a_k -a_{k-1}} &{a_{k-1}-a_1}\cr
		      {b_k -b_{k-1}}  & {b_{k-1}-b_1} \cr} \biggr| } \ldots \ldots  \cr                  
& \phi^{(3)}_{a_4b_4,a_3b_3} 
 e^{i \half  B  \biggl| \matrix{{a_4 -a_3} &{a_3-a_1}\cr
		                  {b_4 -b_3}  & {b_3-b_1}\cr}  \biggr| }
\phi^{(2)}_{a_3b_3,a_2b_2} \cr
 & e^{i \half  B  \biggl| \matrix{{a_3 -a_2} &{a_2-a_1}\cr
		                  {b_3 -b_2}  & {b_2-b_1} \cr}  \biggr|}
\phi^{(1)}_{a_2b_2,a_1b_1}
}}
The reason for distributing the exponentials in this way will become clear in a moment. 
One could put an exponential between $\phi^{(k)}$ and $\phi^{(k-1)}$, but it would be 
identically 1, so we omitted it. We can introduce a notation which will make this look 
simpler. First note that the interactions always appear with a sum over indices.
\eqn\bigsum{
\sum_{a1b1,a_kb_k}{ \phi^{(k)}_{a_1b_1,a_kb_k} \phi^{(k-1)}_{a_kb_k,a_{k-1}b_{k-1}}
\ldots \phi^{(3)}_{a_4b_4,a_3b_3} \phi^{(2)}_{a_3b_3,a_2b_2} \phi^{(1)}_{a_2b_2,a_1b_1}}}
If we think of the fields as matrices this is just 
\eqn\trace{
Tr{ (\phi^{(k)}\cdot \phi^{(k-1)} \ldots \phi^{(2)} \cdot \phi^{(1)})}.
}
Now we define a product, called $*$, by
\eqn\star{
(\phi^{(2)}*\phi^{(1)})_{a_3b_3,a_1b_1}=
\sum_{a_2b_2}{e^ {i\half B  \biggl| {\matrix{{a_3 -a_2} &{a_2-a_1}\cr
				{b_3-b_2}  & {b_2-b_1}\cr}} \biggr|}}  
	\phi^{(2)}_{a_3b_3,a_2b_2} \phi^{(1)}_{a_2b_2,a_1b_1}.
}	
Now the interaction with a $B$-field, eq.\intB, can be written
\eqn\startrace{
Tr(\phi^{(k)} *\phi^{(k-1)}* \ldots *\phi^{(2)} *\phi^{(1)}).
}
This is really nice.  It shows that to generalize the action eq\Obraneaction\ to include
a $B$-field we just need to replace ordinary matrix product with $*$-product.  For $B=0$
the $*$-product coincides with the ordinary product.  This point of view, that the fields take
value in another algebra, was of course the main point of \rCDS.

In the case $B=0$ the fields with the constraints eq.\constraints\ and action can be 
rewritten to a $SYM$ theory on a dual $T^2$  \refs{\rWati,\WOS}.  Let us briefly repeat 
that construction for the case, $B\neq 0$.  Let us first express the basis of the Hilbert 
space in another form.  We want to think of the Hilbert space as $L_2$ functions on a dual torus
with radii, $1\over m_s^2 r_1$, $1\over m_s^2r_2$. Let the basis vector $|ab>$ correspond to 
$e^{iaxm_s^2r_1} e^{ibym_s^2r_2}$, then the operators $U_1$, $U_2$ become multiplication
operators
\eqn\Uonetwo{
	U_1=e^{ixm_s^2r_1}, \qquad U_2=e^{iym_s^2r_2}.
}
It is now easy to solve the constraints for the fields eq.\constraints
\eqn\solvecont{\eqalign{
X^1 &= -i2\pi{1 \over m_s^2} {\partial \over {\partial x}} + \sum_{a,b} X^1_{ab,00}
e^{iam_s^2r_1x}e^{ibm_s^2r_2y}\cr
X^2 &= -i2\pi {1\over m_s^2} {\partial \over {\partial y}} + \sum_{a,b} X^2_{ab,00}
e^{iam_s^2r_1x}e^{ibm_s^2r_2y}\cr
X^j & = \sum_{a,b} X^j_{ab,00} e^{iam_s^2r_1x} e^{ibm_s^2r_2y}\cr
\Psi & =\sum_{a,b}  \Psi_{ab,00} e^{iam_s^2r_1x} e^{ibm_s^2r_2y}.
}}
We see that $X^1$, $X^2$ become covariant derivatives and all other fields are multiplication operators.
These are exactly the fields of 2+1 dim $SYM$ on a torus of radii ${1\over m_s^2r_1}$, ${1\over m_s^2r_2}$.

All this is independent of $B$.  We saw that the only B-dependence was to change products of fields
to the $*$-product.  Let us see how the $*$-product looks in this basis.  We only need to consider the types
of operators which appear in the action.  We see from eq.\solvecont\ that these are the differential
operators, ${\partial\over{\partial x}}$, ${\partial\over{\partial y}}$, and multiplication operators.
${\partial \over{\partial x}}$ and $\partial \over{\partial y}$ are diagonal and it is easily seen from
eq.\star\ that for diagonal operators the $*$-product is equal to the usual product.  Let us now look at
two multiplication operators $\phi^{(1)}(x,y)$ and $\phi^{(2)}(x,y)$.  We have 
\eqn\fourieret{
\phi^{(i)}(x,y) = \sum_{a,b} \phi^{i}_{ab,00}e^{iam_s^2r_1x}e^{ibm_s^2r_2y}
}
with $\phi^{i}_{a_2b_2,a_1b_1} =  \phi^{i}_{(a_2-a_1)(b_2-b_1),00}$.
Plugging into eq.\star\ it is seen that 
\eqn\stjerne{\eqalign{
&(\phi^{(2)}*\phi^{(1)})(x,y) = \cr 
& e^{-{i \over 2} {B \over m_s^4r_1r_2}( {\partial \over \partial x_2} {\partial \over \partial y_1}
- {\partial \over \partial y_2} {\partial \over \partial x_1})}
\phi^{(2)}(x_2,y_2) \phi^{(1)}(x_1,y_1)|_{x_2=x_1=x,y_2=y_1=y}.
}}
Let us recapitulate what we have obtained so far. We consider one 0-brane 
on a $T^2$, $N_0=1$, and no membranes, $N_2=0$. The flux of the $B_{ij}$-field 
through the torus is $B$. We consider the limit coming from matrix theory. 
If $B=0$ the resulting theory is a 2+1 dim. SYM on a dual $T^2$. 
In terms of the M-theory variables the radii of the $T^2$ are
\eqn\param{\eqalign{
r_1^{'}= {1 \over m_s^{2}r_1}= {1 \over M^3RR_1} \cr 
r_2^{'}= {1 \over m_s^{2}r_2}= {1 \over M^3RR_2} 
}}
and the gauge coupling is 
\eqn\koblinggauge{
{1 \over g^2}= {m_s r_1 r_2 \over \lambda} = 
{R_1R_2 \over R}.
}
The gauge bundle 
on $T^2$ is trivial. This is a consequence of the fact that all the fields in 
eq.\solvecont\ are functions on $T^2$ instead of sections of a non-trivial bundle. 
Equivalently $c_1 = {1 \over 2\pi}\int trF = 0$. For any $B$ the only difference 
is that every time two fields are being multiplied in the action one should 
instead use the $*$-product.

When $B=0$ the $*$-product, of course, coincides with the usual product. 
Looking at eq.\star\ we see that the product has a periodicity in 
$B$ of $4\pi$. For $B=2\pi$ it is different from $B=0$. At first sight 
this is problematic, since the theory is known to have a periodicity in 
$B$ of $2\pi$. The puzzle is resolved by noting that there is a 
field redefinition which takes the theory at $B=2\pi$ into 
$B=0$. The field redefinition is 
\eqn\feltre{
\phi_{a_2b_2,a_1b_1} \rightarrow (-1)^{(a_2-a_1)(b_2-b_1)}
\phi_{a_2b_2,a_1b_1}.				
}
Thus the gauge theories actually have the correct $2\pi$ 
periodicity in $B$.

So far we have only discussed the case with $N_0=1$ and $N_2=0$, i.e. one 
0-brane and no 2-branes. The case with any $N_0$ goes in exactly 
the same way. It is now a $U(N_0)$ theory instead of a $U(1)$ 
theory. Nothing else is changed. Especially the same form of the 
$*$-product should be used, except that now the fields are 
$N_0 \times N_0$ matrices.

\section{Non-trivial gauge bundles}

In the previous section we only considered cases with no membranes, 
$N_2 =0$. What about $N_2 \neq 0$? In the case $B =0$ we know 
the answer. Here the final theory is obtained by T-duality on both 
circles. After T-duality we get the decoupled theory of 
$N_0$ D2-branes with $N_2$ D0-branes dispersed in the 2-branes. 
0-branes in 2-branes is just magnetic flux. In other words now 
it is a $U(N_0)$ theory with a non-trivial bundle on $T^2$. The 
first Chern class is $c_1= {1 \over 2\pi}\int trF = N_2 $. In the 
previous section we saw that for $B \neq 0$ and $N_2=0$ the theory 
became a $U(N_0)$ theory with $c_1=0$ and deformed by the 
$*$-product. All these theories have radii and coupling given by 
eq.\param. The obvious guess is now that the case with $B \neq 0$
and $N_2 \neq 0$ was described by a $U(N_0)$ theory with 
$c_1 = N_2$ and an action deformed by the $*$-product. However this 
cannot be true, at least not in this naive sense. The reason is 
that if the bundle is non-trivial we really need to define the 
fields in coordinate patches. The $*$-product does not transform 
correctly under change of patch. To make it do so we would have 
to replace $\partial \over \partial x$, $\partial \over \partial y$ 
by covariant derivatives. Even if this is possible there are other 
reasons to doubt that this is correct. Firstly $B \rightarrow 
B + 2\pi$ is not a symmetry in the presence of 2-branes. It is 
only a symmetry if one changes the number of 0-branes: 
$N_0 \rightarrow N_0 - N_2$. For $N_2=1$ one could always 
change to $N_0 =0$. If the above guess was correct this would 
lead to strange connections between ``$U(0)$'' and $U(N)$
theories. 

Another reason to doubt this naive guess is the following. 
When one uses Sen's and Seiberg's prescription \refs{\Sen,\Seiwhy} to derive 
a matrix model the energy has the form
\eqn\matenergi{
E = \sqrt{({N \over R})^2 + P_{\perp}^2 + m^2} = {N \over R} 
+ \half {R \over N}(P_{\perp}^2 + m^2) + \ldots .
}
Here it is written for uncompactified M-theory, but a similar 
expression is valid for all compactifications. The point is 
that in the limit $R \rightarrow 0$, the first term goes to 
infinity, the second term stays finite(after rescaling of all 
energies) and the terms indicated by dots vanish. The first 
term goes to infinity but is fixed and independent of 
any dynamics. Therefore we can ignore it and just keep the 
second term. A matrix theory Hamiltonian always gives the 
second term. When we change N the theory changes drastically. 
For instance the gauge group changes. In other words when the 
infinite term is changed we expect the finite term to change 
drastically. Let us now look at our situation. With $N_0$ 
0-branes, $N_2$ 2-branes and a $B_{ij}$-field flux $B$. Here 
the energy takes the form
\eqn\energib{
E = {N_0 + BN_2 \over R} + finite.
}
For $B \neq 0$ the infinite term changes when $N_2$ is changed. 
So following the remarks above we expect the theory to 
change drastically. Specifically it is probably not enough 
to change the bundle, but also radii and gauge coupling 
changes. 

Whether or not the case of $N_2 \neq 0$ is solved by just 
changing the first Chern class, there is another way of 
solving it. This is the subject of next section.


\section{Incorporating 2-branes}

In this section we will obtain the matrix model for the general case, 
with $N_0$ 0-branes,$N_2$ 2-branes and a flux $B$. We will do that 
by performing a T-duality to transform to the case $N_2=0$

For a review of T-duality, see \Giveon. The T-duality group for 
IIA on $T^2$ contains an $SL(2,Z)$ subgroup which acts as follows. 
It leaves the complex structure of $T^2$ invariant. Define a 
complex number in the upper halfplane, $\rho = B +iV$. Here V is 
the volume of the torus measured in string units and $B$ is the 
flux of $B_{ij}$ through the torus. In our case $V=m_s^2 r_1r_2$. An 
element $\pmatrix{a&b \cr c&d \cr} \in SL(2,Z)$ acts as follows
\eqn\slaction{\eqalign{
\rho^{'}&= {a\rho +b \over c\rho +d} \cr
\pmatrix{N_0^{'} \cr -N_2^{'}\cr} &= \pmatrix{a&b \cr c&d \cr} 
\pmatrix{N_0 \cr -N_2 \cr}.
}}
Let us use this in our case. Let $Q$ be the greatest common 
divisor of $N_0$ and $N_2$. Write 
\eqn\stoerste{\eqalign{
N_2 &= Q \tilde{N_2} \cr
N_0 &= Q \tilde{N_0}.
}}
Since $\tilde N_0$,$\tilde N_2$ are relatively prime we can choose 
$a,b$ such that $a \tilde{N_0} - b \tilde{N_2}=1$. Let us now 
perform a T-duality transformation with the matrix 
\eqn\tmatrix{
\pmatrix{a&b \cr \tilde N_2 & \tilde N_0 \cr }.
}
An easy calculation gives the new radii, $B_{ij}$ flux, 0-brane and 
2-brane numbers
\eqn\nyparam{\eqalign{
r_1^{'} &= {r_1 \over \tilde N_0 + \tilde N_2 B} \cr
r_2^{'} &= {r_2 \over \tilde N_0 + \tilde N_2 B} \cr
B^{'} &= {aB+b \over \tilde N_0 + \tilde N_2 B} \cr
N_0^{'} &=Q \cr
N_2^{'} &=0. 
}}
The string mass is invariant
\eqn\strengmasse{
m_s^{'} = m_s 
}
and the new coupling is 
\eqn\nykobling{
\lambda^{'}= {\lambda \over \tilde N_0 + \tilde N_2 B}.
}
In these formulas we have taken the matrix theory limit in the 
quantities which have a non-zero limit. We remark that the 
denominator $\tilde N_0 + \tilde N_2 B$ is positive since 
\eqn\naevner{
P_{-} = {\tilde N_0 + \tilde N_2 B \over R}
}
and $P_-$ is positive as always in matrix theory. We now see 
that the parameters of the theory go to zero and infinity 
in exactly the same way as in last section. This means that 
we are in exactly the same situation as in last section.

In other words the matrix theory is a 2+1 dim. SYM on a $T^2$ 
with gauge group $U(Q)$ where $Q$ is the greatest common divisor of 
$N_0$ and $N_2$. The action is deformed with the $*$-product 
with a value of $B$ equal to 
\eqn\vaerdi{
B^{'} = {aB+b \over \tilde N_0 + \tilde N_2 B}.
}
The $T^2$ has radii
\eqn\stor{\eqalign{
r_1^{''}= {1 \over m_s^{'2}r_1^{'}}= {\tilde N_0 + \tilde N_2 B \over M^3RR_1} \cr 
r_2^{''}= {1 \over m_s^{'2}r_2^{'}}= {\tilde N_0 + \tilde N_2 B \over M^3RR_2} 
}}
and the gauge coupling is 
\eqn\kobgauge{
{1 \over g^2}= {m_s^{'}r_1^{'}r_2^{'} \over \lambda^{'}} = 
{R_1R_2 \over R(\tilde N_0 + \tilde N_2 B)}.
}
The SL(2,Z) duality employed has a very easy geometric 
interpretation if one performs a T-duality on a single 
circle as in \rDH. Here $N_0,N_2$ parametrize which
homology cycle the D-strings wrap. The factor $\tilde N_0 + \tilde N_2 B$ 
is just the length of the D-string. The T-duality transformation is just 
a geometric change of $\tau$-parameter of the torus.


\section{Conclusion}

 It was explained in \refs{\rWati,\WOS} how to describe 0-branes on $T^2$ 
by working on the covering space ${\bf R}^2$ and modding out by 
translations. We did this in the presence of a background 
B-field. This enabled us to get a matrix theory of M-theory on 
$T^2$ with a background $C_{-12}$. The result agrees with 
\refs{\rCDS, \rDH} and is a gauge theory on a noncommutative 
torus. 
	
There are some interesting aspects of this. In the case $B=0$ this 
procedure leads to a 2+1 dim SYM which is exactly the same as the 
theory on the D2-brane in the T-dual picture. In other words the 
procedure of compactifying the 0-branes agrees with 
T-duality. For $B \neq 0$ this is not the case. T-duality 
 does not give a theory on a finite torus when $B \neq 0$. 
This is the whole reason for all this interest in $B \neq 0$. 
This means that working with 0-branes on the covering space is 
not the same as T-duality. We believe, of course, that T-duality 
still is true. The point is just that the T-dual description 
is not simpler. The T-dual description is the theory on D2-branes 
wrapped on a dual $T^2$ which is again shrinking. To extract a 
well defined action out of that one has to expand the full 
Born-Infeld action as advocated in {\Li}. It would indeed 
be interesting to use the noncommutative theory to put 
constraints on the full Born-Infeld action. All the higher 
derivative terms should come out of this.

Originally it was thought that compactifications of 
M-theory could be gotten by compactifying the 
0-brane quantum mechanics. That was indeed the case 
for toroidal compactifications up to $T^3$. For other 
compactifications certain degrees of freedom were missing. 
It was later realized that the correct way of obtaining 
the matrix model was to use string dualities in order 
to realise the theory as a theory living on a brane decoupled 
from gravity. In the case of $C_{-12} \neq 0$ we are in some 
sense back to the first philosophy. We can obtain the 
final theory starting with the 0-brane theory but 
we do not know how to realise it as a sensible limit of 
a theory on a brane.

It is an interesting question whether these new theories 
make sense as renormalizable quantum theories. In the case 
of $B =0$ we know that the procedure of putting 0-branes 
on the covering space gives a renormalizable theory 
up to $T^3$ and not for higher tori. So certainly arguing 
that this procedure should give a well defined theory 
is wrong. However, one might hope that the question of 
renormalisability is related to the ``number of degrees of 
freedom''. In that sense the theory on $T^d$ with $B \neq 0$ 
behaves like the theory on $T^d$ with $B=0$ and we might 
expect that the noncommutative theories are well defined up 
to $T^3$. Realizing these theories as theories on branes 
would resolve this issue, but as discussed above this 
might require knowing the full Born-Infeld action.

It will be very interesting to see what the methods 
of noncommutative geometry can teach us about string 
theory and the other way round.

\vfill\eject


\chapter{Instantons in noncommutative gauge theories}

In this chapter we will discuss certain aspects of 
the D0-D4 brane system because that 
will be relevant to us in chapter 5. 

Consider type IIA on $T^4$ with radii, $R_i$,$i=1,\ldots,4$, 
with $N_4$ D4-branes wrapped on $T^4$. There is no $B^{NS}$-field 
at the moment. Let the string mass be $m_s$ and the coupling be $\lambda$. 
 If $m_s R_i \gg 1$ it is well known that this system is described by 
 4+1 dimensional maximally supersymmetric Yang-Mills theory on a 
$T^4$. A $U(N_4)$ bundle on $T^4$ is characterised by the chern numbers, 
which are integers. In physical terms the first chern class 
 corresponds to the number of 
D2-branes wrapped on the 6 two-cycles on $T^4$ and the
second chern class to the number of D0-branes. 
The moduli space of lowest energy configurations are the 
instanton configurations where 
\eqn\inst{
F^+ + \omega^+ =0 
}
Here $F^+$ means the selfdual part of $F$ and $\omega^+$ 
is a selfdual constant in the $U(1)$ part of $U(N)$. 
Especially the $SU(N)$ part is antiselfdual. Of course 
one could exchange selfdual with antiselfdual above. This depends on 
whether there is a positive or negative number of D0-branes.

Let the number of D0-branes be $N_0$. Suppose that $m_s R_i \ll 1$. 
then we can obtain a description of the same system by a T-duality on 
all 4 circles. This will give us a $U(N_0)$ gauge theory with 
chern numbers determined by the D2-branes and $N_4$. The lowest 
energy configurations are the instanton configurations, 
\eqn\instto{
F^+ + \omega^+ =0 
}
now with the $U(N_0)$ gauge field.
We will now make the following claim. Whereas the gauge theory is only 
a good description when the radii are big the moduli space of lowest 
energy configurations is always equivalent to 
an instanton moduli space. To our knowledge there is no rigorous proof 
of this but it is very plausible for the following reasons. The 
moduli space is a \hk\ manifold because of the amount of supersymmetry. 
An instanton moduli space is also \hk. They are equal at large radii 
as explained above. Furthermore for very small radii we can perform 
T-duality to another instanton moduli space as discussed above. 
However these two instanton moduli spaces are known to be equal 
by the so called Nahm transformation \Nahm. These facts taken together 
strongly suggests that the space of lowest energy configurations 
is given exactly by the instanton moduli space for all radii, not 
just large one. This holds irrespective of what the coupling is.

Let us now include the $B^{NS}$-field. We consider type IIA on 
$T^4$ of radii $R_i$ as before. There is a flux of the NS-NS 
B-field on $T^4$. Define 
\eqn\mangeik{
\theta_{ij} = {1 \over 2\pi} \int_{T^2_{ij}} B^{NS}
}
$\theta_{ij}$ is antisymmetric, so there are 6 independent 
numbers. There are D0,D2,D4-branes wrapped on cycles of 
$T^4$. In the previous chapters only a two-torus was 
discussed. However the discussion can be repeated in any dimension. 
This time the low energy physics is described by a gauge theory 
on a noncommutative $T^4$. The noncommutative $T^4$ is defined as 
the algebra generated by $U_1$,$U_2$,$U_3$ and $U_4$ satisfying
\eqn\mangerelation{
U_i U_j = e^{2\pi i \theta_{ij}}U_j U_i 
}
Bundles, or finitely genrated projective modules, over the 
noncommutative $T^4$ are classified by 8 integers, exactly 
corresponding to the numbers of D0,D2 and D4 branes wrapped on 
the original $T^4$. For a certain regime of radii,$m_s$ and $\lam$ 
the low energy physics of this system is described by a 4+1 dimensional 
SYM on a noncommutative $T^4$. 
This is not a renormalizable theory, so new degrees of freedom is needed 
at high energy. However the lowest energy configurations, which are BPS, 
can be obtained from the gauge theory. Let us recall the supersymmetry 
variation from eq.\susyt ,eq.\ikkelin in the case of 
static bosonic configurations 
\eqn\statbos{
\delta \Psi = -{1 \over 4}[X^i, X^j] \Gamma^{ij} \epsilon 
  + \zeta
}
We are interested in configurations with $X^i \neq 0$ only 
 for $i=1,2,3,4$. These $X^i$ are connections in the module 
over the noncommutative torus as explained in 
chapter 1. $[X^i ,X^j]$ is the field strength $F^{ij}$. Let us 
look for solutions to 
\eqn\varipsi{
\delta \Psi =0
}
We can try to take $\epsilon$ chiral 
\eqn\endnuetnavn{
\Gamma_1 \Gamma_2 \Gamma_3 \Gamma_4 \epsilon 
 = - \epsilon
}
then 
\eqn\entil{
\delta \Psi = - {1 \over 4} F^+_{ij} \Gamma_{ij} 
\epsilon + \zeta 
}
where $F^+$ is the selfdual part of $F$. We see that if 
\eqn\igen{
F^+ = -\omega^+
}
where $\omega^+$ is a constant then $\zeta$ can always be 
adjusted to cancel the first term. In other words if 
\eqn\igento{
F^+ = - \omega^+ 
} 
then all $\epsilon$ obeying eq.\endnuetnavn are preserved. 
So 8 supercharges are preserved. This works exactly as in the 
commutative case, which is a special case of this. The 
connections satisfying the instanton equation, eq.\igen , 
are the ones that preserve half the supersymmetry. 

Locally a connection has the form
\eqn\lokalform{
\nabla_i = \partial_i + A_i 
}
where $A_i$ is a matrix valued one-form. The field strength 
is 
\eqn\feltstyr{
F_{ij} = \partial_i A_j - \partial_j A_i 
         +A_i A_j - A_j A_i 
}
the difference between the commutative and noncommutative case 
is that the matrix entries are functions on respectively a 
commutative and a noncommutative space. Alternatively one 
can make the $A_i$ ordinary functions but then the 
product should be replaced with the $*$-product defined in 
eq.\stjernemultiplikation
\eqn\defins{
F_{ij} = \partial_i A_j - \partial_j A_i 
         +A_i *  A_j - A_j *  A_i 
}
We thus see that the instanton equations are being deformed 
by the noncommutativity.

Let us now discuss the possible singularities in the instanton 
 moduli space. Consider first the D0-D4 system in type IIA on 
$\bf R^{1,9}$ without any D2-branes and zero $B^{NS}$-field. 
A D0-brane alone preserves the supersymmetry
\eqn\nulalene{
\Gamma_0 \epsilon_L = \epsilon_R 
}
where $\epsilon_L$,$\epsilon_R$ is left and right chirality 
spinors of $SO(1,9)$. A D4-brane oriented along directions 
$1,2,3,4$ preserve 
\eqn\firealene{
\Gamma_0 \Gamma_1 \Gamma_2 \Gamma_3 \Gamma_4 
 \epsilon_L = \epsilon_R 
}
Each brane preserves 16 supercharges. If they are both 
present, separated from each other, they preserve 
8 supercharges, namely the $\epsilon_L$,$\epsilon_R$ that 
obey both equations, eq.\nulalene , eq.\firealene. 
This means there will be no force between a D0-brane and 
a D4-brane. One can move the D0-brane into the D4-brane where 
it can dissolve in a bound state. This looks like an 
instanton on the D4-brane worldvolume. The instanton 
moduli space has singularities coming from small instantons. 
The singularities in the instanton moduli space exactly reflects 
that there is a branch where the branes are separated.

Let us now turn on a constant $B^{NS}$-field along directions 
$1,2,3,4$. In other words $B^{NS}$ is a closed 2-form on $\bf R^4$. 
 The D0-brane still preserves 
\eqn\nulaleneto{
\Gamma_0 \epsilon_L = \epsilon_R 
}
The D4-brane preserves
\eqn\firealeneto{
{1 \over \sqrt{det(1+B_{ij})}} e^{-B_{ij} 
{\delta \over \delta \Gamma_i}{\delta \over \delta \Gamma_j}}
\Gamma_0 \Gamma_1 \Gamma_2 \Gamma_3 \Gamma_4 
 \epsilon_L = \epsilon_R 
}
where $\delta \over \delta \Gamma_i$ removes a factor of 
$\Gamma_i$, similarly to the way we differentiate Grassmann 
variables. This formula can be derived as follows. First there is 
a Lorentz frame where $B$ has the form 
\eqn\bmatrix{
B= \left( \matrix{ 0 & B_{12} & 0 & 0 \cr
             -B_{12} & 0      & 0 & 0 \cr
                   0 & 0 & 0 & B_{34} \cr
                   0 & 0 & -B_{34} & 0 \cr}\right)
}
Here it is easy to show, compactifying on a torus and using 
T-duality for instance, that the preserved supersymmetry is
\eqn\bevsusyvirker{
\Gamma_0 {(\Gamma_1 \Gamma_2 + B_{12}) \over \sqrt{1+B_{12}^2}}
{(\Gamma_3 \Gamma_4 + B_{34}) \over \sqrt{1+B_{34}^2}}
\epsilon_L = \epsilon_R
}
The formula above is just the Lorentzinvariant version of this. 
Let us check for which B-fields the D0-brane and D4-brane 
preserves supersymmetry when they are separated. We will do it 
in the special Lorentzframe, even though it does not make a 
difference. We need to solve eq.\nulaleneto and eq.\firealeneto . 
Combining them we get 
\eqn\kombiligning{
{(\Gamma_1 \Gamma_2 + B_{12}) \over \sqrt{1+B_{12}^2}}
{(\Gamma_3 \Gamma_4 + B_{34}) \over \sqrt{1+B_{34}^2}}
\epsilon_L = \epsilon_L
}
In other words the matrix on the lefthand side needs to have the 
eigenvalue +1. $\Gamma_1 \Gamma_2$ and $\Gamma_3 \Gamma_4$ have 
eigenvalues $\pm i$ and can be diagonalised simultaneously. It is 
easy to see that we need either 
\eqn\betingleser{
\Gamma_1 \Gamma_2 =i , \;\; \Gamma_3 \Gamma_4 = -i 
}
or
\eqn\betingleser{
\Gamma_1 \Gamma_2 =-i , \;\;  \Gamma_3 \Gamma_4 = i 
}
and 
\eqn\enafbet{
B_{12}=B_{34}
}
The Lorentz invariant version of the last equation is that 
$B^{NS}$ is selfdual. We have thus derived that if and only 
if $B^{NS}$ is selfdual can a separated D0-brane and D4-brane 
preserve supersymmetry. They preserve 8 supercharges in this case. 

What about a bound state of D0-branes and D4-branes. When $B^{NS}=0$ 
it exists as a BPS state. It has a description as 
instantons, as we discussed above. In other words it is a single 
particle state in a short multiplet. When we vary $B^{NS}$ the 
dimension of the representation can not jump. We thus conclude that the 
bound state always exists as a BPS state. This argument has been used 
 often in string dualities, in going from weak to strong coupling. 
Here we do not vary the coupling but the $B^{NS}$-field.

Suppose a D0-brane is bound to $N_4$ D4-branes in the presence of a 
$B^{NS}$-field. It can be described as an instanton in the  
noncommutative $U(N_4)$ theory. If $B^{NS}$ is not selfdual the 
D0-brane can not leave the D4-branes meaning that there is no small 
instanton. We thus see that the singularities in the instanton 
moduli space has been resolved. Here we argued from string theory. 
It would be interesting to confirm this picture directly from 
the instanton equations. A discussion of various 
aspects of noncommutative instantons can be found in 
\refs{\rNekSch , \rMicha, \rANS}.

This discussion was for the case of $\bf R^4$. It stays valid on 
$T^4$, since the preserved supersymmetry is identical on 
$\bf R^4$ and $T^4$. $T^4$ can be obtained from $\bf R^4$ by a 
periodic identification which does not change any of the above.
In chapter 5 we will return to noncommutative instantons.

\vfill\eject

\chapter{Twisted $(2,0)$ and Little-String Theories}

Now we will leave the realm of noncommutative geometry for a while. 
In this chapter we will study the compactification of the $(2,0)$
theory and the little-string theory on $\MS{1}$, $\MT{2}$ and $\MT{3}$.
The $(2,0)$-theory describes the low-energy modes coming from type-IIB
on an $A_{k-1}$ singularity \rWitCOM\ or, equivalently,
 $k$ 5-branes of M-theory \rStrOPN.
The little-string theory is the theory of $k$ type-II NS5-branes
decoupled from gravity \rSeiVBR.
In order to get an interesting low-energy question we will
twist the boundary conditions along $\MT{d}$ by elements of
the $Spin(5)$ (or $Spin(4)$ for the little-string theory) R-symmetry.
In this way we obtain new kinds of theories with 8 supersymmetries.
The aim of this paper is to find the low-energy description of
these theories. We will present an explicit construction in the
case $k=2$.
The construction for $k=2$ involves the moduli space of the
heterotic 5-brane wrapped on tori.

In certain limits we recover the known moduli spaces of
Super-Yang-Mills theories  with a massive adjoint hypermultiplet.
In the compactified little-string theories, examination of the moduli
space shows that for certain values of the external parameters 
there is a phase transition to a phase where little-strings condense.

The chapter is organized as follows.
In section (4.1) we explain our notation, present the problem
and discuss the parameters on which the compactifications depend.
In section (4.2) we present the general solution for $k=2$.
In section (4.3) we study in more detail
various limiting cases of the solution. In particular, we
study the limits where Super-Yang-Mills is obtained.
In subsection (4.3.2) we observe the phase transition.
In section (4.4) we explain the relation between the twist and
the mass of the adjoint hypermultiplets in the effective low-energy
description of Super-Yang-Mills.
In section (4.5) we discuss in more detail what it means to 
twist the little-string theories. We study what happens to the
twists after T-duality and suggest that the R-symmetry twists
are a special case of a more general twist.
We end with a discussion and open problems.


\section{The problem}
The problem that we are going to study is to find
the Seiberg-Witten curves of certain $\SUSY{2}$ theories in 3+1D
and to find the \hk\ moduli space of certain
$\SUSY{4}$ theories in 2+1D.
The $\SUSY{2}$ theories will be obtained by compactifying the
$(2,0)$ theory or, slightly generalizing,  the little-string theory,
on $\MT{2}$ with twisted R-symmetry
boundary conditions along the sides of
the torus.
The $\SUSY{4}$ theories in 2+1D are similarly obtained by
compactification on $\MT{3}$.
In this section we will describe the setting and the notation.

\subsection{Definitions}
Let us denote by $T(k)$ the $(2,0)$ low-energy theory of $k$ 5-branes
of M-theory \refs{\rWitCOM,\rStrOPN}.
We denote 
by $S_A(k)$ ($S_B(k)$) the theory of $k$ type-IIA (type-IIB)
NS5-branes in the limit
when the string coupling goes to zero keeping the string
tension fixed \rSeiVBR.
Compactified on a circle, these two theories are T-dual.
$T(k)$ is often called ``{\it the $(2,0)$ theory}'' and $S(k)$
is referred to as ``{\it the little-string theory}''.

\subsection{The $(2,0)$ theory and the little-string theories}
When $T(k)$ is compactified on $\MT{2}$ we get a 3+1D theory
which at low-energy becomes $k$ free vector multiplets (at
generic points in the moduli space).
The vector-multiplet moduli space is $(\MS{1}\times\MR{5})^k/S_k$
where the size of $\MS{1}$ is $A^{-1/2}$
and $A$ is the area of $\MT{2}$.
When we compactify $T(k)$ on $\MT{3}$ the low-energy
is (generically on the moduli space) given by a $\sigma$-model
on the \hk\ manifold $(\MT{3}\times\MR{5})^k/S_k$.
The $\MT{3}$ in the moduli space has the same
shape as the physical $\MT{3}$ but its volume is $V^{-1/2}$,
where $V$ is the volume of the physical $\MT{3}$.
(See \rSeiSTN\ for review.)
$S_A(k)$ has a low-energy
description given by 5+1D SYM and has a scale $M_s$. The scale
is related to the SYM coupling constant $M_s^{-1}$.
The parameters of the compactification
are now the metric on $\MT{3}$ and also the NSNS 2-form
on $\MT{3}$. The 2-form couples as a $\theta$-angle in the
effective 5+1D low-energy SYM, i.e.  as $\int B\wdg \trp{F\wdg F}$.
Together they parameterize 
\eqn\slfour{\eqalign{
& SO(3,3,\BZ)\backslash SO(3,3,\BR)/(SO(3)\times SO(3)) \cr
& = SL(4,\BZ)\backslash SL(4,\BR)/ SO(4).
}}
The moduli space is given by
$(\MT{4}\times\MR{4})^k/S_k$ where $\MT{4}$ has the shape
given by the point in $SL(4,\BZ)\backslash SL(4,\BR)/ SO(4)$
and has a fixed volume $M_s^2$.

We have to mention that the arguments of \rMS\  (see also \rGK)
show that the theories $S(k)$ are far more complicated
than the $T(k)$ theories, in the sense that they have a continuous
spectrum starting at energy around $M_s$ and this spectrum describes
graviton states propagating in a weakly coupled throat.
Below the scale $M_s$ there is a discrete spectrum (up to
the  effect of the $4k$ non-compact scalars).
Since there is a mass gap, one can still ask low-energy
questions, as we are doing.

\subsection{R-symmetry Wilson lines}
The compactifications discussed above have 16 supersymmetries
and therefore the moduli spaces obtained in 2+1D are flat
and only their global structure is interesting.
To get interesting metrics on the moduli space we need to break
the supersymmetry down by ${1\over 2}$.
This can be done as follows.
Suppose we identify a global symmetry of the $(2,0)$ theory.
When we compactify on $\MS{1}$ of radius $R$ and coordinate
$0\le x\le 2\pi R$, we can glue
the points $x=0$ and $x=2\pi R$ by adding a twist of the global
symmetry. When we compactify on $\MT{3}$ we can twist along all
3 directions so long as the twists commute. The global symmetry
of $T(k)$ is the $Spin(5)$ R-symmetry. Such a twist
has been recently used in \rWAdSII\ to break the supersymmetry
of the $(2,0)$ theory in compactifications.

When we compactify the little-string theory $S_A(k)$ ($S_B(k)$) it is not
immediately obvious that we can use such a twist because
the space-time interpretation is not unique. However, since
we can embed the twist as a geometrical twist in type-IIA,
the question is well defined. We will elaborate more on that point
in section (6).

Let us now take the $(2,0)$ theory $T(k)$ on $\MT{3}$ with
three commuting twists $g_1,g_2,g_3\in Spin(5)$ along $\MT{3}$.
The 16 super-charges of $T(k)$ transform as a space-time spinor
which also has indices in the $\rep{4}$ of $Spin(5)$.
The condition that 8 supersymmetries will be preserved
is the condition that $g_1,g_2,g_3$ preserve a two-dimensional
subspace of the representation $\rep{4}$
of $Spin(5)$. This becomes the following condition.
Take $SU(2)_B\times SU(2)_U = Spin(4)\subset Spin(5)$
and let $g_1,g_2,g_3$ be 3 commuting elements in the first
$SU(2)_B$ factor. This is the generic twist which preserves
$\SUSY{4}$ in 2+1D.
Similarly, for the little-string theory $S(k)$ the R-symmetry
is $Spin(4)$ and we need,
$$
g_1,g_2,g_3 \in SU(2)_B\subset SU(2)_B\times SU(2)_U = Spin(4).
$$
Since the $g_i$'s are commuting they can be taken inside
a $U(1)$ subgroup of $SU(2)_B$.
Then $g_i = e^{i\tw_i}\in U(1)\subset SU(2)_B$.
The subscripts $B$ and $U$ are short for ``broken''
and ``unbroken'' respectively.
We can now ask what is the low-energy description of $T(k)$,
$S_A(k)$ and $S_B(k)$ compactified, in turn,
 on $\MS{1}$, $\MT{2}$ and $\MT{3}$ with twists $\tw_i$.
The most general question is about $S(k)$ on $\MT{3}$ since
all others can be obtained by taking appropriate limits.
The low-energy description in 2+1D is a $\sigma$-model on a
$4(k-1)$-dimensional \hk\ manifold. We will always ignore the
decoupled ``center of mass''.
Furthermore, as will be elaborated in section (4),
in appropriate limits we obtain
3+1D or 2+1D $SU(k)$ SYM with a massive adjoint hypermultiplet.

What is the external parameter space?
The parameter space for the metric and $B$ fields on $\MT{3}$
is given by \slfour. The parameter space for conjugacy
classes of three commuting $SU(2)$ R-symmetry twists along
$\MT{3}$ is given by $\MHT{3}/\BZ_2$ where $\MHT{3}$ is 
the torus dual to $\MT{3}$
and $\BZ_2$ is the Weyl group of $SU(2)$.
However, with R-symmetry twists,
we can no longer divide by the full T-duality group
$SO(3,3,\BZ)$ (see the discussion in section (6)).
This means that the parameter space is a fibration of
$(\MHT{3}/\BZ_2)$ over 
$$
SL(3,\BZ)\backslash SO(3,3,\BR)/(SO(3)\times SO(3)).
$$

\subsection{Why is the problem not trivially solved by M-theory?}
Let us explain why we cannot just read off the
SW-curves and moduli spaces from M-theory.
To be specific, let us take the 6-dimensional non-compact space defined 
 as an $\MR{4}$-fibration over $\MT{2}$ with $Spin(4)$ twists
along the cycles of the $\MT{2}$. This is the geometric realization
of the R-symmetry twist, that we mentioned above (see section
(6) for a more detailed discussion).
M-theory compactified on this space preserves 16 supersymmetries if
the two twists $\tw_1,\tw_2$ are taken inside
$SU(2)_B\subset SU(2)_B\times SU(2)_U = Spin(4)$.
Let us wrap $k$ 5-branes on $\MT{2}$. Given the success of the method
described in \rWitFBR\ one may at first sight wonder whether
the classical moduli space of the $k$ 5-branes immediately gives the right
answer. The answer is negative. There is, in fact, a big difference
between the situation in \rWitFBR\ and ours.
The construction of \rWitFBR\ was used to solve certain QCD questions.
As explained there,
QCD is {\it not} the low-energy description of 5-branes in M-theory.
It is not even an approximate one. QCD is only a good approximation
in the region of moduli space where the 5-branes are close together
and the 11$^{th}$ dimension is very small. When this parameter was increased
the dynamics of the system is completely changed except for
the vacuum states (i.e. the moduli of the vector-multiplets).
This relied on the fact that the parameter that deforms the system
from close NS5-branes and D4-branes in type-IIA to M5-branes
decoupled from the vector-multiplet moduli space (similarly to the decoupling
in \rKV\ and \refs{\rBDS,\rSeiIRD}).
The classical result was correct for the M5-brane limit because 
all the relevant sizes were much larger than $M_{Pl}$ (the Planck scale).

In our case, not all the relevant sizes of the M5-brane configuration
are large. Let $A$ be the area of $\MT{2}$ and let $\Phi$
be the modulus of the tensor multiplet in 5+1D. $\Phi$ is related
to the separation $y$ between the 5-branes as $\Phi\sim M_p^3 y$.
The interesting region in moduli space is $\Phi A \sim 1$.
This region is $M_p^3 A y \sim 1$ and at least one of $y$ or $A$
cannot be made large.


\section{Solution}
In this section we will consider the theory $S_A(2)$ compactified on $\MT{3}$. 
We recall that $S_A(2)$ is the theory living on 2 coincident NS 5-branes in 
type IIA in the limit of vanishing string coupling with string scale,
$M_s$, kept fixed. The compactified theory has a moduli space of vacua 
which is a \hk\ manifold. The purpose of this section is to find this 
\hk\ manifold as a function of the parameters of the compactification. 
These parameters are described above. There is the IIA string scale,
$M_s$ (which is already a parameter in 6 dimensions).
There is the metric, $G^A_{ij}$ 
and NS-NS 2-form, $B^A_{ij}$, on the $\MT{3}$. Here A denotes the
underlying type IIA theory. Finally, there are the 
holonomies of the $Spin(4)$ R-symmetry around the 3 circles. The holonomies 
are taken inside an $SU(2)_B$ subgroup of $Spin(4)$ to preserve half of the 
supersymmetries. The 3 holonomies must commute and can thus be taken 
inside $U(1) \subset SU(2)_B$. We denoted the holonomies $e^{i\tw_1},
e^{i\tw_2},e^{i\tw_3}$, where $\tw_i$ is periodic with period 
 $2\pi$. Furthermore the Weyl group of $SU(2)_B$ relates $\tw_i$ to 
$-\tw_i$. These are the parameters of the theory.

The moduli space of vacua has real dimension 4, since we are dealing 
with 2 5-branes and we throw away the center of mass motion. We want 
to find the metric on this as a function of $M_s, G^A_{ij},B^A_{ij}$
and $\tw_i$.
Our strategy will be to start at the special point $\tw_i =0$ 
and then later understand how to do the general case. At $\tw_i =0$  
the theory actually has $\SUSY{8}$ supersymmetry in 3 dimensions
(like $\SUSY{4}$ in 4 dimensions). Here the moduli space is just the 
classical one. At the origin of the moduli space the low energy theory 
is an $SU(2)$, $\SUSY{8}$ theory. There are also heavy Kaluza-Klein modes 
with masses that go like multiples of $1 \over R_i$, where $R_i$ are 
the radii of the circles. In $\SUSY{4}$ language the multiplet is 
a vector-multiplet and an adjoint hypermultiplet. On the the moduli 
space of vacua $SU(2)$ is broken to $U(1)$. Dualizing the photon gives an 
extra scalar, so the vector-multiplet has 4 scalars. In the $\SUSY{8}$ 
theory the moduli space of vacua is 8 dimensional.
Four of the directions come 
from scalars in the hypermultiplet. These are lifted as soon as $\tw_i 
\neq 0$, because $\tw_i$ supply a mass to the hypermultiplet. We are 
really only interested in the 4 directions coming from scalars in the 
vector-multiplet. These 4 scalars are all compact. From the 5-brane point of 
view these scalars come about as follows. One of them is the relative 
position of the 5-branes on the 11$^{th}$ circle. The other 3 are the 
2-form living on the 5-brane with indices along the $\MT{3}$. These 4
 scalars are obviously compact. The Weyl group of the SU(2) gauge group 
changes the sign of all these. We thus see that the moduli space of 
vacua is $\MT{4} /\BZ_2$.
When we deform to $\tw_i\ne 0$, the moduli space remains compact.
The only compact 4 dimensional \hk\ manifolds are K3 and $\MT{4}$.
$\MT{4}/\BZ_2$ is topologically a K3 manifold with a 
singular metric. We thus conclude that for all parameters $G^A_{ij},B^A_{ij}, 
\tw_i$ the moduli space is topologically K3. We just need to find the 
\hk\ metric as a function of these parameters. 

Let us first recall the moduli space of \hk\ metrics on K3.
It is \rAspin,
$$
O(3,19,\BZ)\backslash O(3,19,\BR)/((O(3)\times O(19)) \times \MR{+}.
$$
$\MR{+}$ parameterizes the volume. This moduli space nicely 
coincides with the moduli space for Heterotic string theory on $\MT{3}$. 
This is a well-known consequence of the duality of M-theory on
K3 with heterotic on $\MT{3}$.
On the heterotic side the 
$\MR{+}$ denotes the dilaton. The space $O(3,19,\BR)/O(3) \times O(19)$ can 
be parameterized by the metric and NS-NS 2-form on the $\MT{3}$,
$G^H_{ij},B^H_{ij}$ and the Wilson lines around the 3 circles 
$V_1,V_2,V_3$. We will work with the $E_8 \times E_8$ Heterotic 
theory.
The reason for that will become clear in a moment.
There is a very nice way of obtaining the K3 on the M-theory side as 
a moduli space of vacua for a 3-dimensional $\SUSY{4}$ theory. 
This is the membrane of M-theory imbedded in $R^{1,6} \times 
K3$ with the world-volume along $R^{1,6}$ and at a point in K3. 
On the dual Heterotic side it corresponds to the 5-brane wrapped 
on $\MT{3}$ \rSeiIRD. This is thus the moduli space of the
$(1,0)$ little-string theory obtained from an NS5-brane
in the heterotic string by taking the coupling constant to zero \rSeiVBR.

Our aim can now be formulated as finding $G^H_{ij},B^H_{ij},V_1,V_2,V_3$ 
for given $G^A_{ij},B^A_{ij},\tw_i$.
According to the arguments of \rSeiHOL, the external parameters
can be combined into scalar components of auxiliary
vector-multiplets which are non-dynamical.
Supersymmetry then requires that the periods of the three 2-forms
which determine the \hk\ metric on the moduli space are linear
in these combinations of external parameters \rSWGDC.
To find the map subject to this restriction, we first examine $\tw_i=0$. 
We saw earlier that this was the $\SUSY{8}$ theory and the moduli 
space is $\MT{4} /\BZ_2$. Therefore, we can find the data of the $\MT{4}$ by 
classical analysis, starting from the $(1,0)$ tensor-multiplet 
living on the IIA 5-brane. (We have ignored the VEVs along 
the $(1,0)$ hypermultiplet direction.) The $(1,0)$ tensor-multiplet
is also the low-energy description of 
the $E_8 \times E_8$ Heterotic 5-brane and the scalar is 
compact since it corresponds to motion in the 11$^{th}$ direction, which 
is an interval. Let us compactify this theory on $\MT{3}$ with data 
$G^H_{ij},B^H_{ij},V_1,V_2,V_3$. To obtain the same moduli space of 
vacua as in the $S_A(2)$ case we need to set $G^H=G^A$,
$B^H=B^A$. What about $V_1,V_2,V_3$? The $S_A(2)$ theory had a $\MT{4} /\BZ_2$ 
as moduli space. $\MT{4} /\BZ_2$ has 16 $A_1$ singularities. This means 
that M-theory on this K3 has $SU(2)^{16}$ gauge symmetry. To achieve 
this we need very special Wilson lines. We can take $V_1$ to break 
$E_8 \times E_8$ to $SO(16) \times SO(16)$ and $V_2$ to break 
each SO(16) to $SO(8) \times SO(8)$ and $V_3$ to break each $SO(8)$
to $SO(4) \times SO(4)$. The unbroken symmetry group is thus 
$SO(4)^8 = SU(2)^{16}$ as desired.
These Wilson lines are unique up to $E_8\times E_8$ conjugation.
We can write down $V_1,V_2,V_3$ explicitly. The two $E_8$'s are treated 
symmetrically, so we restrict to one of them. Consider 
$\Gamma^8 \subset R^8$ where $\Gamma^8$ is the weight lattice of 
$E_8$. Recall that $\Gamma_8$ can be characterized as all sets 
$(a_1,\ldots,a_8)$ such that either all $a_i$ are half-integers
or all $a_i$ are integers. Furthermore $\sum a_i$ is even. A Wilson 
line around a circle can be specified by an element $V \in \MR{16}$ 
such that a ``state'' given by a weight vector $a$ is transformed as 
$e^{ia \cdot v}$\ on traversing the circle. In this notation
\eqn\wilsonlines{\eqalign{
V_1 &= (0,0, 0,0,0,0,0,1) \cr
V_2 &= (0,0,0,0,\half,\half,\half,\half) \cr
V_3 &= (0,0, \half,\half,0,0,\half,\half).
}}
Now we have the map in the case $\tw_i =0$. We will  make a  
proposal for the general case presently. The 16 singularities in $\MT{4} /\BZ_2$
 are due to an adjoint hypermultiplet becoming massless. When 
$\tw_i \neq 0$ the hypermultiplet is massive and we expect the 
singularities to disappear. Near the original singularities the 
theory now looks like pure $SU(2)$ SYM. This does not have any 
singularities \refs{\rSeiIRD,\rSWGDC}.
We thus see that the Wilson lines must change 
when we turn on $\tw_i$. We now make the following proposal. 
For nonzero $\tw_i$ we still have $G^H_{ij}=G^A_{ij}$,
$B^H_{ij}=B^A_{ij}$. The Wilson lines $W_1,W_2,W_3$ are taken to be,
$$
W_i = V_i + {\tw_i \over \pi}(\half,0,\half, 0,\half,0,\half,0),
$$
in the notation from above. This is the same as embedding 
$e^{\half i\tw_i}$ in the diagonal $SU(2) \subset SU(2)^{16} 
\subset E_8 \times E_8$. The coefficients of $\tw_i$ are 
chosen such that the period is $\tw_i \rightarrow \tw_i + 2\pi$. 

We do not have a proof of this proposal, but this certainly
satisfies the requirements of linearity in external
parameters, because this is also the moduli space of the compactified
$(1,0)$ little-string theory.
In the coming sections we will show that our proposal is consistent 
with string theory and field theory expectation.  

There is another very similar theory. This is the theory on 
2 coincident type IIB NS 5-branes in the limit of vanishing 
string coupling and fixed string mass. We call this theory 
$S_B(2)$. As soon as we compactify it on a circle it is T-dual 
to the theory studied above. When we compactify it on a $\MT{3}$ 
with R-symmetry twists we get a 3-dimensional theory with a 
K3 as the moduli space of vacua. Arguing exactly as in the IIA 
case we propose that this K3 is given in the same way as in 
the IIA case, except that we replace Heterotic $E_8 \times E_8$ 
with Heterotic SO(32). This is because the low energy description 
of the theory living on a IIB 5-brane is a gauge theory. The Heterotic 
SO(32) 5-brane is also described, at low energy, by a gauge theory. 
When we do the comparison at the point without an R-symmetry twist, 
the $\SUSY{8}$ point, the moduli spaces will automatically agree. 
This is analogous to the $\SUSY{8}$ point in the IIA case where 
we compared two tensor-multiplets. The T-duality between the IIA 
and IIB 5-brane theories on $\MT{3}$ fits very nicely with the 
T-duality between Heterotic SO(32) and Heterotic $E_8 \times 
E_8$ on $\MT{3}$ at the point $\tw_i=0$.
For $\tw_i\neq 0$, the R-symmetry twists do not remain R-symmetry
twists after T-duality.


\section{Limits}

Now that we have identified the moduli space of vacua for $S_A(2)$ 
compactified on $\MT{3}$ with arbitrary R-symmetry twists,
we can decompactify one or two of the circles to obtain 
the moduli space of vacua for $S_A(2)$ compactified to 
4 and 5 dimensions. Another limit is to take $M_s \rightarrow \infty$
in the $S_A(2)$ theory. This takes us to the $(2,0)$ theory, 
which we call $T(2)$. In this section we will consider these 
limits.

\subsection{Decompactification limits}

Let us first recall the correspondence between M-theory on K3 
and Heterotic $E_8 \times E_8$ on $\MT{3}$. M-theory on K3 has 
a Planck mass, $M_{Pl}$, and a moduli space 
$$
O(3,19,\BZ)\backslash O(3,19,\BR)/((O(3)\times O(19)) \times \MR{+}
$$
$\MR{+}$ denotes the volume of K3, $\Vol{K3}$. In Heterotic 
$E_8 \times E_8$ on $\MT{3}$ there is a string mass, $M_s$, and a 
moduli space, which is the same as for M-theory on K3. There 
is a 10-dimensional string coupling, $\lambda$. The $\MT{3}$ has 
a volume, $\Vol{\MT{3}}$, which is part of $O(3,19,\BR)/(O(3) \times O(19))$.
 Under the duality an M5-brane wrapped on K3 is mapped to the 
Heterotic string. Equating the tensions gives,
\eqn\volumen{
{M_{Pl}}^6\, \Vol{K3} = M_s^2.
}
Equating the 7-dimensional gravitational couplings gives,
\eqn\volum{
{M_{Pl}}^9\, \Vol{K3} = {M_s^8\Vol{\MT{3}} \over \lambda^2} 
}
We thus see, that the $\MR{+}$ on the Heterotic side is ${{\Vol{\MT{3}}}
\over {\lambda^2}}$, which of course is T-duality invariant. 
Eq.\volumen\ agrees with the fact that the volume of the moduli 
space of vacua of the Heterotic 5-brane is $M_s^2$ and 
${M_{Pl}}^6\,\Vol{K3}$ is the volume of the moduli space of the M 2-brane 
probe. We remember that scalar fields have dimension $\half$ in 
3 dimensions. A concrete way of tracing the duality between 
these two theories is to use T-duality from Heterotic $E_8 
\times E_8$ on $\MT{3}$ to Heterotic SO(32) on $\MT{3}$, and then 
S-duality to type-I on $\MT{3}$, then T-duality to type-IA on 
$\MT{3}$ which can be viewed as M-theory on K3.

Let us now consider the decompactification to 4 dimensions. 
This can be done by taking $\tw_3 =0$ and $R_3 \rightarrow 
\infty$. In this limit the K3 becomes elliptically fibered with the 
fiber shrinking. The area of the fiber $A$ is 
$$
{M_{Pl}}^3 A = {1 \over R_3}
$$
This can be seen by noting that a membrane wrapped on the fiber 
corresponds to momentum around the circle $R_3$ in the Heterotic 
theory. This limit of M-theory on an elliptically fibered K3 
is exactly what gives F-theory on this K3. The M2-brane probe 
becomes the D3-brane probe in F-theory on K3 \refs{\rSenFO,\rBDS}. Since the 
volume of K3 stays fixed and the fiber shrinks this means that the 
base grows. One might thus think that the moduli space seen by 
the D3-brane probe is infinite. However we should remember 
that a scalar field in 4 dimensions has dimension one, so 
we need a factor of the type-IIB string mass in the area of the 
moduli space.
Inserting this makes the area is up to a constant, $M_s^2$. 
This agrees with the expectation from $S_A(2)$ compactified on 
 $\MT{2}$. We can thus summarize our result for the 4-dimensional 
case. Take the theory $S_A(2)$ with mass scale $M_s$. Compactify it on 
a $\MT{2}$ with R-symmetry twists given by $\tw_1,\tw_2$. 
The $\MT{2}$ is specified by $G^A_{ij},B^A_{ij}$. The moduli space 
of vacua for this $\SUSY{2}$ theory in $D=4$ is the same as the moduli 
space of vacua for the $E_8 \times E_8$ Heterotic 5-brane wrapped 
on $\MT{2}$ with string mass, $M_s$, and a point in 
$O(18,2)/O(18) \times O(2)$ given as follows. The metric 
and 2-form on $\MT{2}$ is $G^A_{ij},B^A_{ij}$. The Wilson lines 
on $\MT{2}$ depend on $\tw_1, \tw_2$. In the case 
$\tw_i=0$ they are the essentially unique Wilson lines 
that break $E_8 \times E_8$ to $SO(8)^4$. For non-zero $\tw_i$ 
the Wilson lines are constructed as in the last section by embedding 
in a diagonal $SU(2)^{16} \subset SO(8)^4$. 

This wrapped 5-brane 
in the Heterotic theory is dual to the 3-brane probe in F-theory on 
the corresponding elliptic-fibered K3.
This K3 is the Seiberg-Witten curve for the 
moduli space. As in the 3 dimensional case, we are not 
saying that the compactified $S_A(2)$ theory is equal to the 
little-string theory on the Heterotic 5-brane, but just that the low-energy 
description is the same. It is obvious that they are not equal 
since the $S_A(2)$ theory has enhanced supersymmetry when $\tw_i=0$.

Decompactifying to 5 dimensions is now easy. The correspondence 
becomes the following. Consider the theory $S_A(2)$ compactified 
on $\MS{1}$ of radius R, string scale $M_s$ and R-symmetry twist $\tw$. 
This is a 5 dimensional theory with $\SUSY{1}$ supersymmetry. The 
coulomb branch is 1-dimensional. Topologically it is $\MS{1}/\BZ_2$. 
This moduli space is the same as the moduli space of the 
Heterotic $E_8 \times E_8$ 5-brane compactified on a circle with 
an $E_8 \times E_8$ Wilson line.
The Wilson line for one $E_8$ is,
$$
W = (0,0,0,0,0,0,0,1)
  + {\tw \over 2\pi} (\half,0,\half,0,\half,0,\half,0),
$$
and the same for the other $E_8$.

Completely analogous statements can be made for the type-IIB 5-brane 
theory, $S_B(2)$. Here the moduli space is given by the 5-brane in the 
Heterotic SO(32) theory. Let us describe this in detail for the case 
of 5 dimensions.
Consider $S_B(2)$ on a circle of radius $R$,
with R-symmetry twist $\tw$ and string scale $M_s$. The moduli space 
of vacua of this is the same as the 5-brane of SO(32) Heterotic 
string theory on a circle with radius R, string scale $M_s$ and SO(32) 
Wilson line
$$
W = (\underbrace{\half,\cdots,\half}_8,\,\underbrace{0,\cdots,0}_8)+
  {\tw \over \pi}(\underbrace{\half,0, \half,0,\cdots,\half,0}_{16})
$$
The string coupling $\lambda$ goes to zero to give a decoupled theory 
on the 5-brane.

There is a dual type-IA picture of the Heterotic theory. The 5-brane 
becomes a D4-brane living on an interval with 8-branes. The parameters 
of the type-IA theory are 
\eqn\etaparam{\eqalign{
M_s' &= {M_s \over \sqrt{\lambda}} \cr
R' &= {\lambda \over {M_s^2 R}} \cr
\lambda' &= {1 \over \sqrt{\lambda}M_s R}
}}
All quantities of interest to the D4-brane theory have a limit as 
$\lambda \rightarrow 0$. The positions of the D8-branes are given 
by the Wilson line.

\fig{The dual type-IA picture.}{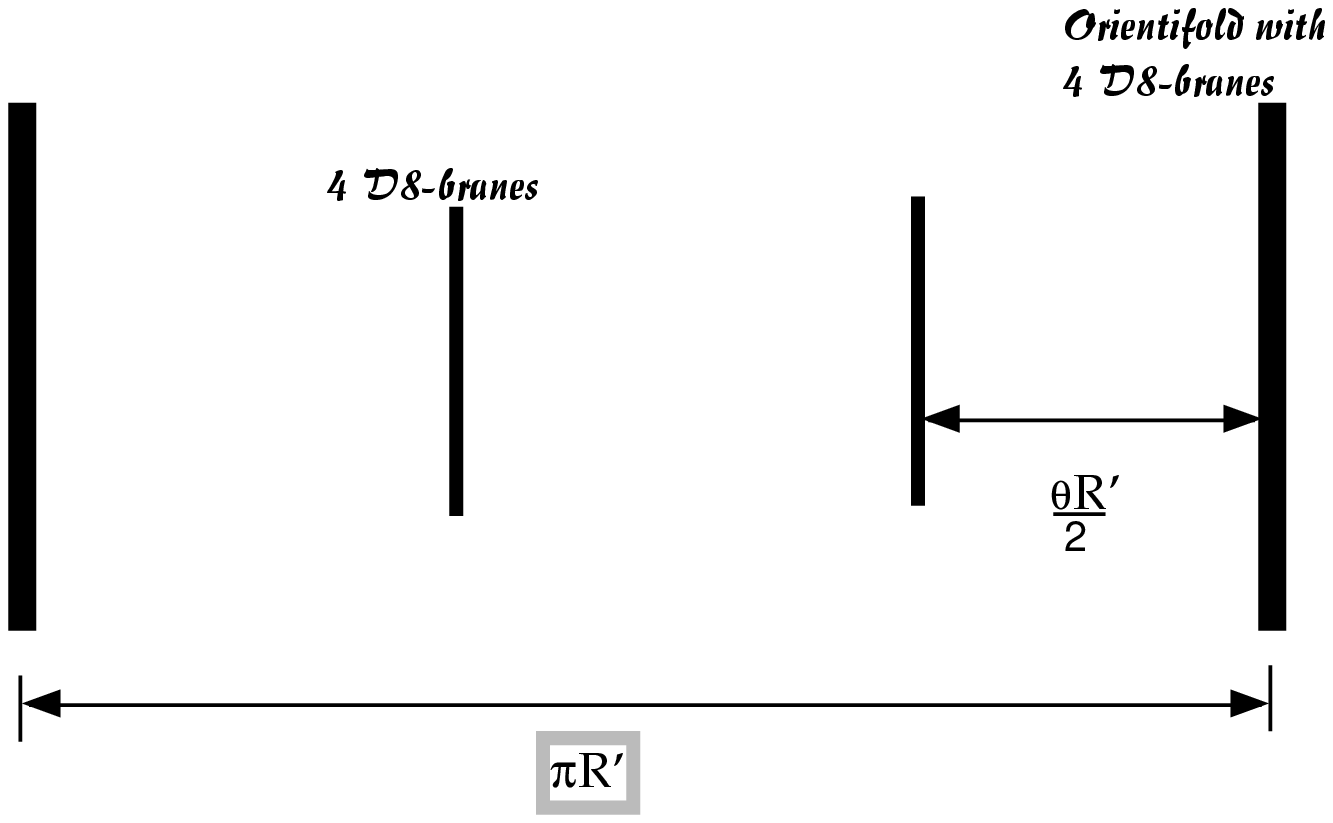} {5truein}

At each end there are 4 D8-branes. 
There are two more stacks of 4 D8-branes at distance ${\tw \over {2}}R'$ 
from each end. When $\tw = 2\pi$ the 8-branes reach 
the other end. This has to be the case since $\tw$ is periodic with 
period $2\pi$.  
We also remark that something interesting happens 
when $\tw = \pi$. Here 8 D8-branes are on top of each other. We will 
return to a discussion of this point later. There behavior in
$D=3,4$ is similar.

$S_A(2)$ compactified to 5 dimensions was described above by
mapping the R-symmetry twists to $E_8\times E_8$ Wilson lines.
For later purposes it will be more convenient to use the type-IA
dual description as the theory living on a D4-brane.
The chain of dualities going from Heterotic 
$E_8 \times E_8$ on $\MS{1}$ to type-IA is to first invoke 
T-duality from Heterotic $E_8 \times E_8$ to Heterotic $SO(32)$, 
and then proceed as above to reach type-IA. The parameters of the 
type-IA theory in terms of the parameters of the $S_A(2)$ theory 
become
\eqn\eotte{\eqalign{
M_s' &= { M_S \over \sqrt{\lambda}}(2({\tw^2 \over \pi^2} + 
          (M_s R)^2))^{1 \over 4} \cr
R' &= {\lam \over M_s^2 R}(2({\tw^2 \over \pi^2} + 
          (M_s R)^2))^{1 \over 2} \cr
\lam' &= {(2({\tw^2 \over \pi^2} + 
          (M_s R)^2))^{5 \over 4} \over \sqrt{\lam}M_s R}
}}
The configuration of D8-branes is as in the $S_B(2)$ case. 
At each end there are 4 D8-branes. At distance ${\tw' \over {2}}R'$ 
from the ends are 4 D8-branes. Here
\eqn\alfaprim{
\tw' = {\tw \over 2({\tw^2 \over \pi^2} + 
          (M_s R)^2)}.
}
We see an interesting effect here. When $\tw =0$, $\tw'$ is also zero. 
For small $\tw$,$\tw'$ is an increasing function. At $\tw = \pi M_s R$, 
$\tw'$ reaches its maximum and start to decrease.

The moduli spaces of all the $S_A(2)$ theories with all possible
$\tw$-twists occupy some subspace 
$$
\Modsp'\subset 
SO(1,17,\BZ)\backslash SO(1,17,\BR)/ SO(17).
$$
We wish to know what is the locus $\Modsp''$
which the $S_A(2)$ theories with the T-dual $\btw$-twists span.
In the above chain of dualities we started with the heterotic 
$E_8\times E_8$ 5-brane wrapped on $\MS{1}$. This represented 
$S_A(2)$ on a circle with a $\tw$-twist.
By definition, $S_A(2)$ with a $\btw$-twist is T-dual to
$S_B(2)$ with some $\tw$-twist and therefore corresponds to
a point in the moduli space of the $SO(32)$ 5-brane.
We have seen that the points on the $SO(32)$ which correspond
to our $S_B(2)$ theories map under heterotic T-duality
to the points on the $E_8\times E_8$ moduli space which correspond
to the $S_A(2)$ theories. Thus $\Modsp'$ and $\Modsp''$ are the
same locus. Nevertheless, $S_A(2)$ with
a $\tw$-twist is not equivalent to $S_A(2)$ with a $\btw$-twist.

\subsection{A peculiar phase transition} 
As we have explained above, when we compactify $S_B(2)$ with
a twist $\tw$ on $\MS{1}$ of radius $R$ we get a 4+1D theory
whose low-energy is the same as that of a D4-brane probe in
a configuration of D8-branes on an interval. In this configuration
there are 2 stacks of four coincident D8-branes. Whenever 
the D4-brane crosses the stack, a particle of $U(1)$ charge $2$ 
(coming from the adjoint of $SU(2)\supset U(1)$) becomes massless.
When the two stacks of D8-branes coincide we get
two massless hypermultiplets.
Since the low-energy description of a $U(1)$ with two massless
particles is weakly coupled, we can trust field-theory and the
conclusion is that there exists a Higgs phase
where the massless hyper-multiplets get a VEV and break the $U(1)$.

In the $S_B(2)$ case, this phase transition occurs for $\tw=\pi$
and for all values of $R$. In contrast, for $S_A(2)$ this only happens
for small enough $R$. We can see this from eq.\alfaprim.
The phase transition occurs
when $\tw' = \pi$ since this is where two stacks of D8-branes
coincide in the type-IA picture. 
This  has a real solution $\tw$, only if $M_s R \le {1 \over 4}$.
Such a bound is certainly to be expected for $S_A(2)$. The reason 
is that this phase transition happens when two NS 5-branes are 
on opposite points of the 11$^{th}$ circle. The 2 hypermultiplets that 
become massless originate from membranes stretched between the 
two 5-branes and wrapped on the compactified circle (the circle 
which takes us from a 6-dimensional theory to a 5-dimensional theory). 
The tension of the membrane gives a mass to these states. However 
there is also a contribution to the mass from the zero-point energy 
of the fields on the membrane. This contribution depends on $\tw$. 
For certain values of the parameters the zero-point energy can cancel 
the mass from the tension. This is how the hypermultiplet can become 
massless. Obviously the mass from the tension can not be canceled if 
the 11$^{th}$ circle is too big, or equivalently the compactified circle 
is too large. This is the reason for the above inequality.

\subsection{The $(2,0)$ limit}
Let us briefly consider another limit, namely the 
limit where $S_A(2)$ on $\MT{3}$ becomes $T(2)$ on $\MT{3}$. This happens 
when the 11$^{th}$ circle opens up. We see from eq.\volumen\ and eq.\volum\
that $\Vol{K3} \rightarrow \infty$, i.e. the moduli space becomes 
non-compact. This is as expected. Basically we just get half of the K3. 
The other half goes to infinity. In the $E_8 \times E_8$ Heterotic 5-brane
 picture it means that the distance between the ends of the world go to 
infinity and we only look at one end.

\subsection{Field theory limits} 
In this section we will compare the moduli spaces of vacua found in the other
sections with field theory results.  
At each point of the moduli spaces for the $T(k)$ and $S(k)$ theories, 
we can find a field theory description for the light modes. 
We are fortunate that such field theories in $D=3,4,5$ are known.  
The metric on the moduli space around the chosen point will be determined
by the light matter.  We are going to compare our exact metric with this
field theory expectation.  

Let us start with $S_B(2)$ compactified on $\MS{1}$. 
The effective field  theory
is that of the  $D4$-brane probe in type-IA.  
From $S_B(2)$ we have $SU(2)$ gauge theory with (1,1) supersymmetry.
(The field content of the (1,1) vector multiplet are a (1,0) vector multiplet
 and a (0,1) adjoint hypermultiplet.)
Upon compactification on $\MS{1}$ of radius
$R$ with a R-symmetry twist $\tw$, the moduli
space is parameterized by the sixth-component of the gauge field, $A_6$,
 in $U(1)\subset SU(2)$.   
The full R-symmetry of $S_B(2)$ theory is 
$$
SO(4)=SU(2)_U\otimes SU(2)_B
$$
which is broken down to $SU(2)_U$ by $e^{i\tw}\in U(1)\subset SU(2)_B$. 
We get in $5D$ an $SU(2)$ vector-multiplet
and massive adjoint vector-multiplets
(with masses ${n\over {R}}$ with $n\in\BZ_{\neq 0}$)
transforming non-trivially under $SU(2)_U$ R-symmetry.  
The boundary conditions on the two complex scalars in the adjoint
hypermultiplet are shifted by $\tw$:
\eqn\phtwst{
\phi_1(2\pi R) = e^{i\tw}\phi_1(0),\qquad
\phi_2(2\pi R) = e^{-i\tw}\phi_2(0).
}
This shifts the periodicity of the fields around the circle.  
The reduction also gives a tower
of adjoint hypermultiplets in $5D$ with masses,
$$
m^2 = {{(n+ {\tw\over {2\pi}})^2} \over {R^2}}, \qquad n\in \BZ.
$$  
For small $\tw>0$, we get one light adjoint hypermultiplet of mass 
$\tw \over {2 \pi R}$.  Now let us look at the moduli space around $A_6=0$.
From field theory it looks like $SU(2)$ theory with an adjoint hypermultiplet
of mass ${\tw\over {2\pi R}}$.
The gauge coupling is then given by \rSeiFIV,
$$
{1\over {g^2}} = b + c A_6, 
$$  
where  b and c are constants. 
The slope, c, changes when charged matter becomes massless. The change 
in the slope is proportional to the cube of the charge of the 
multiplet becoming massless. In $U(1) \subset SU(2)$ an 
adjoint field has components of charge $-2,0,+2$ under the U(1) in 
units where the $\rep{2}$ of $SU(2)$ has charge $\pm 1$.
This means that the change 
in the slope, c, is 8 times bigger for an adjoint hypermultiplet 
than for a fundamental. Let us calculate at what value of $A_6$ 
the charge 2 component of the adjoint hypermultiplet becomes massless. 
The holonomy around the circle is 
$$
\phi \rightarrow e^{-4\pi i R A_6}  \phi. 
$$
To cancel $e^{i\tw}$ we thus need
$$
A_6 = {\tw \over 4\pi R}.
$$
Let us now compare to the solution from the previous section. Here 
$\tw$ parameterizes the position of 4 D8-branes. For 
$\tw = 2\pi$ they reach the other end of the interval. In terms of 
$A_6$ the other end of the interval is at $1 \over 2R$, so the position 
of the 4 D8-branes is,
$$
A_6= {\tw \over 2\pi}\times {1 \over 2R}= { \tw \over 4\pi R}
$$
in exact agreement.
However the number of D8-branes is 4 and not 8 as naively expected 
from the discussion above. It seems like the change in slope
 is half of what should be expected from field theory. There is 
no discrepancy for a subtle reason. We compare, on one hand, 
 the U(1) low energy 
effective action for a D4-brane moving in an orientifold setting, 
with, on the other hand, a U(1) from a $SU(2)$ gauge theory. The 
U(1) on the D4-brane probe corresponds not to the 
$U(1) \subset SU(2)$ but to one of the U(1) factors in 
$U(1) \times U(1) \subset U(2)$.
The action for the diagonal $U(1)\subset U(2)$ is twice the action
for a single $U(1)$ factor. The normalization would contain
an extra $\sqrt{2}$ factor. Taking this factor of 2 into account
the change in the slope becomes 8 instead of 4.

Let us now consider the case of $S_A(2)$ on $\MT{3}$ with twists 
$\tw_1, \tw_2, \tw_3$. For simplicity the torus is taken 
to be rectangular with radii $R_1,R_2,R_3$ with $B_{ij}=0$. We will 
also take $\tw_i$ to be small.  We want 
to find the light fields. Finding the light fields in this case is 
not as easy as in the previous case, because $S_A(2)$ does not 
have a Lagrangian description. However we can figure out the result 
by first compactifying on a small $R_1$, with $\tw_1 =0$. Then the 
low energy description is a 5-dimensional $\SUSY{2}$ $SU(2)$ gauge 
theory. In $\SUSY{1}$ language it comprises a vector-multiplet and 
a hypermultiplet. Now we can compactify this on a second circle 
of radius $R_2 \gg R_1$. At scale $R_2$ the $SU(2)$ gauge theory is 
weakly coupled and we can perform a classical analysis to include the 
twists $\tw_2$. We get an $SU(2)$ gauge theory in $D=4$ with a 
hypermultiplet of mass $\tw_2 \over 2\pi R_2$. In $D=4$, $\SUSY{2}$ 
a hypermultiplet mass is complex. Since there is no distinction between 
direction 1 and 2 we expect a contribution $\tw_1 \over 2\pi R_1$
from direction 1. They have to combine into a complex mass
$$
m = {\tw_1 \over 2\pi R_1} +i {\tw_2 \over 2\pi R_2}.
$$
On compactifying down to 3 dimensions on $R_3$ (we assume
that $R_3> R_2$) there will similarly 
be a contribution $\tw_3 \over 2\pi R_3$. In $D=3$ a 
hypermultiplet mass consists of 3 real numbers that transform in the 
\rep{3} of $SO(3)$ \rSWGDC.
This $SO(3)$ is part of the R-symmetry group.
We thus 
conclude that the 3 real numbers are $\tw_i \over 2\pi R_i$. There 
is a region in moduli space where the theory looks like $\SUSY{4}$, 
$SU(2)$ gauge theory with an adjoint hypermultiplet with mass 
$m_i =  {{\tw_i} \over {2\pi R_i}}$.
As we have seen, this region is when $|\tw_i|\ll \pi$ and
when the mass scale set by the 2+1D SYM coupling constant (the smallest
of ${{R_1} \over {R_2 R_3}}$, ${{R_2}\over {R_1 R_3}}$ and 
  ${{R_3}\over {R_1 R_2}}$) is much smaller than the smallest
 compactification scale (the smallest of $R_1^{-1}$, $R_2^{-1}$
and $R_3^{-1}$). In our setting, $R_1 \ll R_2 < R_3$, this condition
is met. Note that if $|\tw_i|\ll \pi$ but $R_1\sim R_2\sim R_3$
are of the same order of magnitude, the correct approximation is
to start with the 2+1D CFT to which $\SUSY{8}$ 2+1D SYM flows
\refs{\rSavLen,\rTomNat} and deform it by the relevant operator
to which the mass deformation flows.
When $m_i=0$ we obtain a
$\SUSY{8}$ theory and the moduli space is $(\MR{3} \times \MS{1})/\BZ_2$. 
This has two $A_1$ singularities. When $m_i \neq 0$ these are blown 
up. From our solution in the previous section the sizes of the 
blow up can be read off as a function of $\tw_i$. This means 
that we have derived a formula for the size of the blow-up of 
the singularities in $D=3$, $\SUSY{4}$ $SU(2)$ gauge theory with a 
massive adjoint hypermultiplet.

We can do the same analysis in $D=4$. For $\tw_1 = \tw_2=0$ 
there are 4 singularities. Close to any one of them the system should 
be describable as an $\SUSY{2}$, $SU(2)$ gauge theory with an adjoint 
hypermultiplet. For small $\tw_i$ the mass of the hypermultiplet 
is 
$$
m = {\tw_1 \over 2\pi R_1} +i {\tw_2 \over 2\pi R_2}.
$$
We expect this to change the Seiberg-Witten curve. Our result 
also predicts how this goes. Our picture is that the 
Seiberg-Witten curve is the same as the D3-brane probe in F-theory
on the K3 as described earlier. For $\tw_i=0$ this has a description 
as a type-IIB orientifold 8 plane with 4 D7-branes on top making a 
$D_4$ singularity \refs{\rSenFO, \rBDS}. 
For non-zero $\tw_i$ two of the 7-branes move 
away, giving a $U(2) \times SO(4) = U(2) \times SU(2) \times 
SU(2)$ singularity. In a field theory setting this corresponds to the 
$SU(2)$ Seiberg-Witten theory with 4 fundamental hypermultiplets, 2 of 
them massless and 2 of them massive with equal mass. Our analysis 
thus predicts that this situation should have the same curve as 
the massive adjoint hypermultiplet. In the second Seiberg-Witten paper
\rSWII\ this was indeed found to be the case. In comparing the curves 
with the low energy effective action there is again a factor of 2 in 
the coupling constant $\tau$ because of a
difference in conventions between the adjoint and fundamental case. 
This is the same factor of 2 as explained in the 5-dimensional case above.



\section{Reduction of the twisted $(2,0)$ theory to 4+1D}
In this section we will study $T(2)$ on $\MS{1}$ with a twist $\tw$.
Neglecting the overall center-of-mass, the moduli space is
1-dimensional. The low-energy physics is a $U(1)$ vector-multiplet.
Let $\phi$ be the scalar partner of the vector field.
In this section we will study the BPS states in the theory.
There are two different regions in moduli space to consider.
Let $R$ be the radius of $\MS{1}$.
When $\phi R \ll 1$ we can use the effective 4+1D SYM Lagrangian.
We will show that for small $\tw$, the BPS states come from
the $W^\pm$ bosons and the charged states of a massive adjoint 
hyper-multiplet.
When $\phi R \gg 1$ we can identify the charged BPS states with
strings wound around $\MS{1}$.

The BPS masses in 4+1D are \rSeiFIV,
\eqn\bpsmas{
2\phi,\qquad m_0 + 2\phi,\qquad m_0 - 2\phi.
}
In the D4-brane and D8-brane picture, these come from
strings connecting the D4-brane to its image, and to the
two mirror D8-brane stacks. Here,
$$
m_0 = {\tw\over {2\pi R}}.
$$
This can be seen from eq.\eotte\ and eq.\alfaprim.
The states with mass $2\phi$ are vectors while those
with masses $2\phi\pm m_0$ are hyper-multiplets.

\subsection{Yang-Mills limit}
When $\tw=0$, the low-energy description of $T(2)$ on 
$\MS{1}$ is $SU(2)$ SYM with a coupling constant $g^2$ which
is proportional to $R$. As long as our energy scale is 
below the compactification scale $R^{-1}$, the coupling constant
is weak and the effective description is good.
When $|\tw|\ll 1$ it can be incorporated as a small perturbation
in the effective Lagrangian. It corresponds to
giving a bare mass of $m_0$ to the hyper-multiplet in the 
Lagrangian. After spontaneous breaking
of $SU(2)$ down to $U(1)$,
the masses in \bpsmas\ are easily calculated
in field theory. $2\phi$ is the mass of the $W^\pm$ bosons
while $2\phi\pm m_0$ come from the hypermultiplet.
The adjoint hypermultiplet also gives rise to a neutral multiplet
with a mass $m_0$.

\subsection{The large-tension limit}
Let us assume that $\phi R \gg 1$.
In this case, we can first reduce to the 5+1D low-energy of
a single $\SUSY{(2,0)}$ tensor multiplet and then reduce this 
tensor multiplet to 4+1D since the scale of the VEV $\phi$ is
much higher than the compactification scale.
In 4+1D, the neutral states come from the hypermultiplet in
5+1D with twists along $\MS{1}$ as in \phtwst.
The mass of these states is therefore (for small $\tw$),
$$
m = {{\tw}\over {2\pi R}}.
$$
The charged states come from quantization of the strings wrapped
on $\MS{1}$. 
Up to a correction proportional to ${{\a^2}\over {R^2}}$
(see \refs{\rSeiFIV,\rGMS,\rIMS}), the tension of the string
in 5+1D is $\Phi = \phi / 2 R$.
In the limit we are considering, $\Phi R^2\gg 1$, it is enough
to quantize only the low-energy excitations of the strings.
This is just as well, since the low-energy excitations are the only
things we know about these strings!
This means that our results are correct up to $O(1/\Phi R^2)$.
The low-energy description is given by a 1+1D $\SUSY{(4,4)}$ theory.
The VEV of the tensor multiplet of the 5+1D bulk breaks the $Spin(5)$
R-symmetry down to $Spin(4)$.
The 1+1D low-energy description of a string
contains 4 left-moving bosons and 4 right-moving bosons, 4 left-moving
fermions and 4 right-moving fermions.
The bosons are not-charged under the $Spin(5)$ R-symmetry.
The 8 fermions can be decomposed into representations,
of 
$$
(SU(2)_B\times SU(2)_U\times SU(2)'_1\times SU(2)'_2)_{SO(1,1)}
$$
Here $Spin(4)=SU(2)_B\times SU(2)_U$ is the unbroken
R-symmetry of the 5+1D theory,
$Spin(4)=SU(2)'_1\times SU(2)'_2$ is the subgroup of $Spin(5,1)$
of rotations transverse to the string and $SO(1,1)$ is the world-sheet
rotation group.
The fermions are in the 
$$
(\rep{2},\rep{1},\rep{2},\rep{1})_{+\half} +
(\rep{1},\rep{2},\rep{1},\rep{2})_{-\half} 
$$
with an added reality condition.
Under the embedding 
$$
U(1)\subset SU(2)_B\subset SU(2)_B\times SU(2)_U = Spin(4)\subset Spin(5),
$$
we find 2 left-moving fermions with charge $+1$ under $U(1)$,
2 left-moving fermions with charge $-1$ under $U(1)$,
and 4  right-moving fermions with charge $0$ under $U(1)$.
The boundary conditions on the charged fermions are twisted.
Quantization of this system gives low-lying vector-multiplets
and hyper-multiplets with masses,
$$
\Phi R,\qquad
{{\tw}\over {2\pi R}}\pm \Phi R.
$$
Recall that the derivation assumed that
$\Phi R^2 \gg 1$ and $|\tw| \ll \pi$.
This agrees with eq.\bpsmas.



\section{R-symmetry twists in the little-string theories}

For the $(2,0)$ theories, which are believed to have a local 
description, a twist by a global symmetry along a circle makes
perfect sense. For the little-string theories, the issue of
locality is more complicated and the meaning of an R-symmetry
twist has to be elaborated.
In this section we will describe the construction in more detail.
We will then see explicitly that T-duality of $S(k)$ does not preserve
the $\tw$-twists. Instead it maps them to T-dual ``$\btw$-twists''.
This raises the intriguing possibility to combine both kinds
of twists simultaneously.

\subsection{Geometrical realization}
One way to define an R-symmetry twist is to realize it geometrically
as follows.
We can start with $\MR{2,1}\times\MR{3}\times\MR{4}$
and mod out by a discrete $\BZ^3$ symmetry which is
generated by elements which act as a shift in $\MR{3}$ and
rotations in $\MR{4}$. We obtain $Z\times\MR{2,1}$
where $Z$ is an $\MR{4}$-fibration over $\MT{3}$.
Explicitly, we define  the 7-dimensional space
$$
Z_{\tw_1,\tw_2,\tw_3} = (\MR{3}\times\MC{2})/\BZ^3,
$$
where $\BZ^3$ is the freely acting group generated by,
\eqn\ygen{\eqalign{
s_1:&
(x_1,x_2,x_3,z_1,z_2)\rightarrow
(x_1 + 2\pi R_1, x_2, x_3,e^{i\tw_1}z_1, e^{-i\tw_1} z_2),\cr
s_2:&
(x_1,x_2,x_3,z_1,z_2)\rightarrow
(x_1, x_2 + 2\pi R_2, x_3, e^{i\tw_2}z_1, e^{-i\tw_2} z_2),\cr
s_3:&
(x_1,x_2,x_3,z_1,z_2)\rightarrow
(x_1, x_2, x_3+2\pi R_3, e^{i\tw_3}z_1, e^{-i\tw_3} z_2),\cr
}}
Here $(x_1,x_2,x_3)$ are coordinates on $\MR{3}$.
We can similarly define 
\eqn\yxdef{
Y_{\tw_1,\tw_2} = (\MR{2}\times\MC{2})/\BZ^2,\qquad
X_\tw = (\BR\times\MC{2})/\BZ.
}
The theory that we study in this paper, $S_A(k)$ on $\MT{3}$ with
a twist, can be obtained if we compactify type-IIA on 
$Z_{\tw_1,\tw_2,\tw_3}$, wrap $k$ NS5-branes on $\MT{3}$ and take
$\lam_s\rightarrow 0$ as in \rSeiVBR.
This shows that it makes sense to include an R-symmetry twist
in $S(k)$.

What is the meaning of these twists in terms of the theory $S(k)$
itself, without appealing to the underlying string-theory?
Let us first refine our terminology.
Let $p$ be a generic point in the parameter space
$$
\Modsp_A \equiv SO(3,3,\BZ)\backslash O(3,3,\BR)/(O(3)\times O(3)).
$$
We will denote the theory derived from $k$ type-IIA NS5-branes at
the type-IIA moduli space point $p\in\Modsp_A$ by $S_A(k;\,p)$.
Similarly there is an identical moduli space $\Modsp_B$ for type-IIB 
NS5-branes. We will denote the theory derived from $k$ type-IIB NS5-branes
at the type-IIB moduli space point $p\in\Modsp_B$ by $S_B(k;\,p)$.
T-duality implies that there is a map,
$$
T:\Modsp_A\rightarrow\Modsp_B,
$$
with $T^2=I$ such that $S_A(k,\,p) = S_B(k;\,T(p))$.
This map can be defined as follows.
Pick an element $v\in O(3,3,\BZ)$ with $\det v = -1$ (all such
elements are $SO(3,3,\BZ)$ conjugate to each other). For
$g\in O(3,3,\BR)$ which is a representative of a point in $p\in\Modsp_A$
take $v\circ g$ to be a representative of $T(p)\in\Modsp_B$.

A generic point $p'$ in the cover,
$$
SL(3,\BZ)\backslash O(3,3,\BR)/(O(3)\times O(3)),
$$
of the parameter space (note that we divided by $SL(3,\BZ)$ instead of
$SO(3,3,\BZ)$) will be called a {\it locality-frame}.
A generic point $p''$ in the cover,
$$
O(3,3,\BR)/(O(3)\times O(3))
$$
will be called a {\it coordinate-frame}.
There are the obvious maps,
$$
p''\rightarrow p'\rightarrow p.
$$
Now suppose that we are in a specific point $p\in\Modsp_A$, say,
and we fix a locality-frame $p'$ for $p$ and a coordinate-frame
$p''$ for $p'$. For a given $p''$ we can contemplate whether it makes
sense to define R-symmetry twists along the cycles of $\MT{3}$.
If they commute with each other,
an $SL(3,\BZ)$ transformation will permute the cycles and will act
on the twists in an obvious way. However, a full
$SO(3,3,\BZ)$ transformation takes one locality-frame to another
and an R-symmetry twist is not mapped back to an R-symmetry twist.

\subsection{The T-dual of an R-symmetry twist}
What does become of an R-symmetry twist after T-duality?
The effect of the R-symmetry twist is to make
a state which is R-charged have a fractional momentum,
because its boundary conditions are not periodic. The
momentum modulo $\BZ$ is related to the R-charge and the twist
in a linear way. Since T-duality replaces the momentum charge
with another $U(1)$ charge -- the winding number of little-strings,
one would deduce that after T-duality, R-charged states should have
fractional winding number.

To be more precise, let us take weakly coupled type-IIA on $X_\tw$
from \yxdef\ and perform T-duality.
Recall that,
$$
X_\tw = (\BR\times\MC{2})/\BZ,
$$
with $\BZ$ generated by,
\eqn\ys{
s: (x,z_1,z_2)\rightarrow (x+2\pi R, e^{i\tw} z_1, e^{-i\tw} z_2).
}
The world-sheet theory is the free type-IIA theory.
Let
\eqn\Xstrmodes{\eqalign{
X =& x+ w\sigma + p\tau 
+ \sum_{n\in\BZ_{\neq 0}} {{\a_{-n}}\over n} e^{i n (\tau-\sigma)}
+ \sum_{n\in\BZ_{\neq 0}} {{\widetilde{\a}_{-n}}\over n}
    e^{i n (\tau+\sigma)},\cr
Z_1 =&
 \sum_{s\in\BZ+\gamma_1} {{\zeta^{(1)}_{-s}}\over s} e^{i s (\tau-\sigma)}
+ \sum_{s\in\BZ+\gamma_1} {{\widetilde{\zeta}^{(1)}_{-s}}\over s}
      e^{i s (\tau+\sigma)},\cr
Z_2 =&
 \sum_{s\in\BZ+\gamma_2} {{\zeta^{(2)}_{-s}}\over s} e^{i s (\tau-\sigma)}
+ \sum_{s\in\BZ+\gamma_2} {{\widetilde{\zeta}^{(2)}_{-s}}\over s}
                  e^{i s (\tau+\sigma)},\cr
}}
$\gamma_{1,2}$ are real numbers which depends on the sector in 
a manner that we will write down below.
When $\gamma_i=0$, we need to add a piece $z_i + p_i \tau$ to $Z_i$.
$p_1,p_2$ are complex while $w,p$ are real.
Also $\a_{-n}^\dagger = \a_n$ and $\widetilde{\a}_{-n}^\dagger 
=\widetilde{\a}_n$.
Let $L$ be the total number of $\zeta^{(1)}$ creation
operators minus the total number of $\zeta^{(2)}$ creation operators
in a state. If some $\gamma_i=0$ we also need to add the rotation
generator $i(z_i p_i^\dagger - z_i^\dagger p_i)$.

\eqn\ndeftot{\eqalign{
L \equiv \sum_{s\in\BZ+\gamma_1} {1\over s}
          (\zeta^{(1)}_{-s})^\dagger \zeta^{(1)}_{-s}
        -\sum_{s\in\BZ+\gamma_2} {1\over s}
          (\zeta^{(2)}_{-s})^\dagger \zeta^{(2)}_{-s}
        + (\zeta \leftrightarrow \widetilde{\zeta}).
}}
Now we can determine which sectors are allowed.
First we require invariance under $s$ in \ys.
This is the world-sheet operator $e^{2\pi i p R - i\tw L}$, so we
require,
$$
 p R - {\tw\over {2\pi}} L\in \BZ.
$$
The sector twisted by $s^k$ has
$$
{w\over R} = k,\qquad \gamma = k{{\tw}\over {2\pi}}.
$$

What happens after T-duality?
In a world-sheet formulation, T-duality replaces $p$ with $w$ and
replaces $R$ with $R' = 1/R$. We now have the conditions
$$
{{w'}\over {R'}} - {\tw\over {2\pi}} L \in \BZ,\qquad
p' R' \in \BZ,\qquad \gamma = p' R' {{\tw}\over {2\pi}}.
$$

This suggests a more general twist, which can no longer be described
as modding out by a discrete symmetry.
This time we keep the sectors with
\eqn\abtwists{
p R - {{\tw}\over {2\pi}} L\in \BZ,\qquad
{w\over R} - {{\btw}\over {2\pi}} L\in \BZ,\qquad
2\pi \gamma = \tw {w\over R} + \eta p R.
}
We admit to not having checked that this is consistent with
modular invariance.
The following argument suggests that turning on both  $\tw$ and $\btw$
twists is consistent. For small $\tw$,
turning on a $\tw$-twists corresponds to making a small perturbation
with a certain operator to the Hamiltonian of $S(k)$.
An infinitesimal $\btw$-twist  also corresponds to a perturbation
but with another operator. Now we can make a small perturbation
with both a $\tw$-twist as well as a $\btw$-twist. They preserve
exactly the same supersymmetry. It could, however, happen that after
we turn on both $\tw$-twists and $\btw$-twists there is no longer
any super-symmetric vacuum. We do not know of any way to settle this
question.

\section{Discussion}

We have argued that the moduli space of vacua of $S_A(2)$ ($S_B(2)$)
compactified on $\MT{3}$ with 3 R-symmetry twists, $\tw_1,\tw_2,\tw_3$, 
is the same as the moduli space of vacua of the heterotic $E_8\times E_8$
($SO(32)$) $(1,0)$ NS5-brane theory compactified on the same $\MT{3}$
with Wilson lines given by an embedding of the twists in the gauge group.
We have also studied how T-duality of the little-string theory
acts on the R-symmetry $\tw$-twists. We have seen that they get mapped
to other types of twists ($\btw$-twists).
We have suggested that there exist theories with both kinds of twists
simultaneously.

Let us suggest a few questions for further research:
\item{1.}
Find an M-theoretic derivation of the moduli spaces, or perhaps
using compactification on a Calabi-Yau manifold.
\item{2.}
Study the BPS spectrum of the theories in 3+1D and 4+1D.
We have identified the moduli spaces of the twisted
$(2,0)$ theory with the moduli space of the compactified 
$E_8$ $(1,0)$ theory. However, these two theories
are not identical. It would be interesting to see
how this distinction is manifested in the multiplicities of
BPS states \refs{\rGTest,\rKMV,\rMNWI,\rMNWII,\rMNVW}.
\item{3.} 
Study the other phase where little-strings condense.

\vfill\eject


\chapter{Instantons on a Non-commutative $T^4$ from Twisted (2,0) 
        and Little-String Theories}
In this chapter we will continue the study of compactified 
$(2,0)$ and Little-string theories. The main result will be 
that their moduli spaces are equal to moduli spaces of 
instantons on noncommutative tori, which were discussed in 
chapter 3.
 
Starting with the work of \rSWI,
 the moduli-spaces of vacua have been found
for a large class of gauge theories with 8 super-charges in 3+1D
and in 2+1D. 
 These solutions were derived from string dualities in \rKV\ and
the works that followed.
String theory also suggested the existence of new theories 
in six dimensions \refs{\rWitCOM,\rSeiVBR}
(see also \refs{\rDVVQ-\rDVVS}).
Compactification of these theories to 3+1D reduces, in certain limits
of the external parameter spaces, to ordinary gauge theories.
As we will see, all the previously solved gauge theories with $\SUSY{2}$
supersymmetry and
$SU(N_1)\times\cdots\times  SU(N_r)$ gauge groups \rWitFBR\ can be
recovered at special limits of the external parameters of the 
compactification of the Little-string theories.
Let us recall that the Little-string theory is
 the world-volume theory on $k$ NS5-branes 
in type-IIA in the limit of vanishing string coupling keeping the string 
tension fixed \rSeiVBR. We denote this theory $S_A(k)$. It has 
$(2,0)$ supersymmetry. There is a similar theory coming from $k$
NS5-branes in type-IIB in the limit of vanishing string coupling 
keeping the string tension fixed.
We denote this theory $S_B(k)$. It has $(1,1)$ supersymmetry.
$S_A(k)$ and $S_B(k)$ are often referred to as the little-string theories.
They both have an inherent scale, $m_s$. In the limit $m_s \rightarrow 
\infty$, $S_A(k)$ becomes the theory on the world-volume of $k$ M5-branes
 -- the so called $(2,0)$ theory.

We will compactify these theories down to 3 dimensions. These theories        
have 16 super-charges, so if they are compactified on $\MT{3}$ the resulting 
theories will have $\SUSY{8}$ supersymmetry in three dimensions.
The low energy behavior of $\SUSY{8}$ theories is trivial.
Instead we want to study theories 
with $\SUSY{4}$ supersymmetry, i.e. 8 super-charges.
So we have to compactify 
in a way that breaks half the supersymmetry. We will do 
that as in \rCGK\ by introducing
 holonomies of the R-symmetry around the three circles in 
$\MT{3}$. To preserve half of the supersymmetries the holonomies were 
chosen inside a $SU(2)$ subgroup of the $Spin(4)$ R-symmetry group. 
The low energy behaviour of a $\SUSY{4}$ theory in $D=3$ is a 
sigma-model with the moduli-space of vacua as the target-space. 
So the low energy behaviour is given by the moduli-space 
of vacua and the its metric.

Let us start by identifying all the parameters of the 
compactification. Consider $S_A(k)$ compactified on $\MT{3}$. 
The scale of $S_A(k)$ is $m_s$, the string mass. The $\MT{3}$ 
is specified by a metric. For simplicity we will take it to 
be rectangular. It is easy to incorporate the more 
general case. Furthermore there can be a flux of the 2-form 
$B^{NS}$ field of type IIA through 2-cycles in the $\MT{3}$. 
 For simplicity we set $B^{NS}=0$. It is again not hard to incorporate
the more general case. Now we come to the most interesting 
parameters -- the twists. The R-symmetry group of $S_A(k)$ is 
$Spin(4)_R$, corresponding to transverse rotations. The twists are 
taken inside 
\eqn\introtwist{
U(1)_R \subset SU(2)_B \subset SU(2)_B \times SU(2)_U = 
Spin(4)_R 
}
This preserves 8 of the 16 super-charges. There is a twist, 
$\alpha_i$, along each of the 3 circles. The $\alpha_i$'s are 
periodic 
\eqn\periode{
\alpha_i \rightarrow \alpha_i +2 \pi,\qquad  i=1,2,3
}       
The twists can be described in the following way. States that are charged
under $U(1)_R$ receive a phase shift in traversing a circle.
In other words, momentum along the circle is shifted 
from $n \over R$ to ${n - {\alpha \over 2\pi} }\over R$.
By performing T-duality along all circles of the $\MT{3}$ 
we get $S_B(k)$ on another $\MT{3}$. Momentum has been exchanged 
with winding, so the T-dual of the twists has the following 
description. States that are charged under $U(1)_R$ have fractional 
winding numbers; $n - \alpha \over 2\pi$ instead of n. 
We call this kind of twist an ``$\eta$-twist.'' By combining these 
two types of twists we learn that the most general twist around 
a circle shifts both momentum and winding. In other words the 
$S_A(k)$ compactification on $\MT{3}$ depends on 6 parameters
\eqn\sekspar{
\alpha_i,\eta_i,\qquad    i=1,2,3,
}
where $\alpha_i$ shifts momentum and $\eta_i$ shifts winding. 
 The $\alpha_i$'s have a clear geometrical interpretation. In 
traversing the circle the transverse space is rotated. 
The $\eta_i$'s are harder to visualize. They are geometrical in the 
T-dual $S_B(k)$.       

We can actually generalize this system even more. Instead of 
$k$ NS5-branes we can consider $k$ NS5-branes on top of an
$A_{q-1}$ singularity. In other words the transverse space to the 
NS5-branes is $\MR{4} / Z_q$, where $Z_q$ is a subgroup of 
$U(1)_R$. $U(1)_R$ is still a symmetry of this space, so we can 
twist as before. These theories have 8 super-charges in 
6 dimensions. The $U(1)_R$ is a global symmetry
which commutes with super-charges.
The twists, therefore, do not break any more supersymmetry, so 
the compactified theory still has $\SUSY{4}$ in 3 dimensions.

Theories of branes on top of an ADE singularity have been 
studied in \refs{\rBluInt,\rIntNEW}. These 6 dimensional theories are,
loosely speaking, quiver gauge theories \rDM\ coupled 
to tensor theories or vice versa, depending on 
whether it is in type-IIA or type-IIB.

The 3 dimensional theory, obtained after compactification 
with twists, has a low energy description as a sigma-model
with a target-space, which is equal to the moduli-space 
of vacua. In this paper we will prove that the moduli-space 
of vacua is equal to the moduli-space of $k$ $U(q)$ instantons 
on a non-commutative $\MT{4}$. The non-commutativity is set        
by the 6 parameters $\alpha_i$ and $\eta_i$.
This generalizes the case of compactification without twists
where the moduli-space of the theories
turns out to be the moduli-space of ordinary instantons
\refs{\rIntNEW,\rGanSet}.

This result implies similar results for all the theories which 
are special cases of this. This includes firstly the $(2,0)$
theory which can be obtained from $S_A(k)$ by $m_s \rightarrow 
\infty$. Secondly, it includes all three-dimensional
$U(k)$ gauge theories with adjoint matter. 
By incorporating the $A_q$ singularity it 
also includes all gauge theories with group $U(k) \times 
\cdots \times U(k)$ and matter in $(k, \bar{k}, 1, \ldots ,1) + 
permutations$. By taking the gauge coupling
 to zero in some $U(k)$ we can get theories with the gauge group 
 being $U(k) \times \cdots \times U(k)$ with fundamental and bi-fundamental 
matter in various combinations with generic masses.

Our results imply that all these 2+1D gauge
theories have a moduli-space 
of vacua equal to the moduli-space of vacua of instantons on 
non-commutative $\MR{3} \times \MS{1}$. In the case of mass 
deformed $\SUSY{8}$ this result was derived earlier in \rKapSet.

By decompactifying one 
circle similar results hold for the moduli space of 
4 dimensional gauge theories on $\MR{3} \times \MS{1}$.       

We also find that for certain discrete values of the 
twists there are Higgs branches emanating from some locus of the 
Coulomb branch. We will identify these and calculate their 
dimensions. We will also calculate the existence of these 
branches from pure field theoretic arguments and find agreement 
in the structure of the Higgs branches. These branches 
generalize a branch found in \rCGK.

Moreover,
combining our results with the formulas in \rCGK\ for the special
case of $q=1$ and $k=2$, we get a prediction for the moduli-space
of two $U(1)$ instantons on a non-commutative $\MT{4}$.
This is a $K3$ (projecting out the center of mass)
and the exact point in moduli-space was given
in \rCGK\ as a function of the twists.

The organization of the chapter is as follows. In section (5.1)
we present the proof that the moduli space is equal to the 
moduli space of instantons on non-commutative $\MT{4}$. 
In section (5.2) we have a short review of the relevant aspects
 of non-commutative gauge 
theories. In section (5.3) we use this information about 
non-commutative gauge theory to make the claim about the 
moduli-space of non-commutative instantons precise and 
discuss some features of it. In section (5.4) we
describe the decompactification limit to 3+1D (compactification
of the 5+1D theories on $\MT{2}$ with twists).
In section (5.5) we present a more detailed geometrical formulation
of $\tw$-twists and especially $\etw$-twists.
We conclude with a summary of the results and possible further
direction.
\section{The solution}
\def\Step#1{\vskip 0.5cm {\it Step {#1}:}}

In this section, we derive the solution to the moduli space
of the twisted theory.
To construct the solution we will start with type-IIA on a space
$$
\MR{2,1}\times\MT{3}\times_{\vec{\tw}}\MR{4},
$$
where $\times_{\vec{\tw}}$ means that locally the space looks like
$\MR{2,1}\times\MT{3}\times\MR{4}$ but as we go around a cycle
of the $\MT{3}$ we have to twist the transverse space $\MR{4}$
by the appropriate element of $Spin(4)$ corresponding to the twist.
Now we take $k$ NS5-branes and let them stretch along
$\MR{2,1}\times\MT{3}$ and the origin of $\MR{4}$.
The question what is the low-energy effective action for this
system in the limit that the type-IIA string
coupling constant $\lam\rightarrow 0$.

As will be clear later on, it is easier to solve the problem if
we first replace the transverse $\MR{4}$ with another manifold $M_4$.
In the limit that the curvature of $M_4$ is small at the position
of the NS5-branes the switch from $\MR{4}$ to $M_4$ will not make
a big difference.
Moreover, we can argue that the quantum fluctuations
in the transverse position of the
NS5-brane are related to the fluctuations  of the scalars of $S_A(k)$ as,
$$
x\sim m_s^{-3}\lam\Phi,
$$
and for energy scales $m_s$, $\Phi$ is of the order of $m_s^2$.
In the limit $\lam\rightarrow 0$, the transverse fluctuations of
the NS5-brane go to zero and if the point in $M_4$ is smooth, it 
would seem that the dynamics of the NS5-brane will be the same
as on $\MR{4}$.
This argument should be taken with caution since the actual solitonic
solution of the NS5-brane has a cross-section of about $m_s$.
In any case, we will not have to rely on this argument.

The manifold $M_4$ that we will use is the Taub-NUT space.
The metric is,
\eqn\tnmet{
ds^2 = \TNS^2 U(dy - A_i dx^i)^2 + U^{-1} (d\vec{x})^2,\qquad
i=1\dots 3,\qquad 0\le y\le 2\pi.
}
where,
$$
U = \left(1 + {{\TNS}\over {2 | \vec{x} | }}\right)^{-1},
$$
and $A_i$ is the gauge field of a monopole centered at the origin.

The Taub-NUT space has the following desirable properties
(these properties were also used in \rWitNGT),
\item{(1)}
If we excise the origin, what remains is a circle fibration
over $\MR{3}-\{0\}$. Eqn\tnmet\ is written such that
$\vec{x}$ is the coordinate on this base $\MR{3}-\{0\}$.
 For $|\vec{x}|$ restricted to a constant, the fibration is
exactly the Hopf fibration of $\MS{3}$ over $\MS{2}$.

\item{(2)}
The origin $\vec{x} = 0$ is a smooth point.

\item{(3)}
As $|\vec{x}|\rightarrow\infty$ the radius of the fiber
becomes $\TNS$.

\item{(4)}
The space has a $U(1)$ isometry group that preserves the
origin $\vec{x}=0$. An element $g(\theta) = e^{i\theta}\in U(1)$
acts by $y\rightarrow y+\theta$.
It also acts on the tangent space $\MR{4}$
at the origin by embedding $e^{i\theta}$ inside
$$
U(1)\rightarrow SU(2)_L\rightarrow
(SU(2)_L\otimes SU(2)_R)/\BZ_2 = SO(4).
$$

Now that we have replaced the transverse $\MR{4}$ with a Taub-NUT space
we have $k$ NS5-branes on the space,
$$
\MR{2,1}\times\MT{3}\times_{\vec{\tw}} \TBNT(\rho).
$$
The $\tw$-twists are incorporated as follows.
As we go around a cycle of $\MT{3}$ we have to act on the fiber
$\TBNT(\rho)$ with $g(\tw_i)$ where $\tw_i$ is the appropriate
twist.
 In the limit $\rho\rightarrow\infty$,
$\TBNT(\rho)$ becomes $\MR{4}$
and the isometry $g(\tw_i)$ becomes the element in $SO(4)$ that
we have used for the twist.
The virtue of working with $\TBNT(\rho)$ instead of $\MR{4}$
is that at $\vec{x}=\infty$ the circle fiber becomes of
finite size which will help in subsequent dualities.

To generalize the construction to the case of $k$ NS5-branes
at an $A_{q-1}$ singularity, $\MR{4}/\BZ_q$,
we replace the transverse $\MR{4}/\BZ_q$ with a $q$-centered
Taub-NUT space, $\TBNT_q(\rho)$ with radius $\rho\rightarrow\infty$.
This space has similar properties,
\item{(1')}
If we excise the origin, what remains is a circle fibration
over $\MR{3}-\{0\}$.
 For $|\vec{x}|$ restricted to a constant, the fibration is
is a circle bundle over $\MS{2}$ with first Chern-class $c_1 = q$.

\item{(2')}
Near the origin $\vec{x} = 0$, $\TBNT_q$ looks like
$\MR{4}/\BZ_q$.

\item{(3')}
As $|\vec{x}|\rightarrow\infty$ the radius of the fiber
becomes $\TNS$.

\item{(4')}
The space has a $U(1)$ isometry group that preserves the
origin $\vec{x}=0$. An element $g(\theta) = e^{i\theta}\in U(1)$
 acts  at $\vec{x}=\infty$ by $y\rightarrow y+\theta$.
It also acts on the tangent space $\MR{4}/\BZ_q$
at the origin by embedding $e^{i\theta}$ inside
$$
U(1)\rightarrow SU(2)_L\rightarrow
(SU(2)_L\otimes SU(2)_R)/\BZ_2 = SO(4).
$$
Note that the discrete $\BZ_q$ by which we mod out is a subgroup
of the same $U(1)\subset SU(2)_L$ as well.

\subsection{Chains of Dualities}

We have seen that the twisted compactified little-string theories
can be realized as follows. Start with type-IIA on
$\MR{2,1}\times \MT{3}\times \TBNT_q$,
where the radii of $\MT{3}$ are $R_i$ (of the order of $m_s$)
and the radius of the fiber of
the Taub-NUT space is taken to be $\TNS$.
Put $k$ NS5-branes on $\MR{2,1}\times\MT{3}$ and study the limit,
$$
\lam\rightarrow 0,\qquad m_s\TNS\rightarrow\infty.
$$
In principle, we could probably settle on a constant $m_s\TNS$ as well,
since the transverse fluctuations of the NS5-brane are small.
However, the transverse size of the NS5-brane,
as a solitonic object, is of the order of $m_s^{-1}$.
Therefore, to be on the safe side, we take $m_s\TNS\rightarrow\infty$.
The technique for solving theories with 8 supersymmetries is \rKV\
to identify a parameter that decouples from the vector-multiplet
and such that at one limit of this parameter the theory
is described by gauge theory (or little-string theory, in our
case) and in another limit a dual description becomes weakly
coupled. In that second limit, the theory is no longer described
by the gauge theory but the vacuum structure remains the same
and is determined by the classical equations of motion.
This method was also applied in \refs{\rBDS,\rSeiIRD,\rWitFBR}.

In our case, to solve the problem we take the limit of strong
coupling keeping the Taub-NUT radius large.
\eqn\gvul{
\lam\rightarrow\infty,\qquad m_s\TNS\rightarrow \infty,
}
We will also require that $\lam (m_s\TNS)^{-3}\rightarrow\infty$.
We can think of $\TNS$ as being fixed but very large and
$\lam\rightarrow\infty$ much faster.
We will not show that this corresponds to a parameter that
is in a hyper-multiplet (and hence decouples from the vector-multiplets)
but this is the basic assumption.
Recall that in 2+1D hyper-multiplets and vector-multiplets
can be distinguished with the help of the
$U(1)_R\otimes SU(2)_U$ symmetry which is the unbroken subgroup
of \introtwist. The scalar fields of a  vector-multiplet are invariant
under $SU(2)_U$ while the scalar fields of a hyper-multiplet are in
the $\rep{2}$ (see \rSWGDC).
(The dilaton, which is a singlet,
is a quadratic expression in these fields.)
Similarly, the fermions of a hyper-multiplet 
are invariant under $SU(2)_U$
and the fermions of a vector-multiplet
are in the  $\rep{2}$.

The next step is to use string-dualities to convert the region
\gvul\ to a weakly coupled theory.

At this point we have 
$k$ NS5-branes in 
type-IIA on $\MR{2,1}\times \MT{3}\times \TBNT_q$
with string coupling $\lam$, string scale $m_s$,
$\MT{3}$-radii $R_i$, and twists $\tw_i$.
 For simplicity, we assume that $\MT{3}$ is of the form
$\MS{1}\times\MS{1}\times\MS{1}$ with no NS-NS 2-form fluxes.
Since $\lam\rightarrow\infty$ we view this as $k$ M5-branes in
M-theory on $\MR{2,1}\times \MT{3}\times\MS{1}\times \TBNT_q$.
Let $M_p$ be the 11-dimensional Planck scale.
The radius of $\MS{1}$ is, $R$. They are related according to,
$$
R = {\lam\over {m_s}},\qquad
M_p^3 = {{m_s^3}\over \lam}.
$$
The radius of $\TBNT_q$ is, $\rho$.

\Step{1}
Since, in the limit \gvul,
$$
M_p\rho = m_s \lam^{-1/3}\rho\rightarrow 0,
$$
we should view the fiber of the Taub-NUT
as the $11^{th}$ small dimension and convert to type-IIA
on $\MR{2,1}\times\MT{3}\times\MS{1}\times\MR{3}$.
We also have $k$ NS5-branes on $\MR{2,1}\times\MT{3}$
and $\TBNT_q$ became
$q$ D6-branes on $\MR{2,1}\times\MT{3}\times\MS{1}$.
The $\tw$-twists became RR 1-form Wilson lines along the
cycles of $\MT{3}$.
The string coupling constant is given by,
$$
\lam' = \lam^{-1/2}(m_s\TNS)^{3/2}\rightarrow 0.
$$
The new string scale is,
$$
M_s' = m_s^{3/2}\TNS^{1/2}\lam^{-1/2},
$$
and the radii of $\MT{3}$ satisfy,
$$
M_s' R_i = 
 m_s^{3/2}\TNS^{1/2}\lam^{-1/2} R_i\rightarrow 0.
$$
This means that we must perform T-duality on $\MT{3}$.

\Step{2}
After T-duality on $\MT{3}$ we obtain type-IIB on 
$\MR{2,1}\times\tilde{\MT{3}}\times\MS{1}\times\MR{3}$
with radii $\hat{R}_i$ which satisfy,
$$
M_s' \hat{R}_i = m_s^{-3/2}\TNS^{-1/2}\lam^{1/2} R_i^{-1}
       \rightarrow\infty.
$$
There are now
$k$ NS5-branes on $\MR{2,1}\times\tilde{\MT{3}}$
and $q$ D3-branes on $\MR{2,1}\times\MS{1}$.
At this point the $\tw$-twists became RR 2-form fluxes,
$$
\tw_i\epsilon_{ijk} = \int_{C_{jk}} B^{RR},\qquad i,j,k=1\dots 3,
$$
where  $C_{jk}$ is the 2-cycle made out of the $j^{th}$ and $k^{th}$
directions in $\MT{3}$.
The string coupling is now,
$$
\lam^{(2)} = {{\lam'}\over {m_s^{9/2}\TNS^{3/2}\lam^{-3/2} R_1 R_2 R_3}}
          = \lam m_s^{-3} R_1^{-1} R_2^{-1} R_3^{-1}
            \rightarrow\infty.
$$
This means that we must do S-duality.

\Step{3}
After S-duality we get type-IIB with $q$ D3-branes  and $k$ 
D5-branes in the same geometry.
The string coupling constant is now,
$$
\lam^{(3)} = \lam^{-1} m_s^3 R_1 R_2 R_3\rightarrow 0,
$$
and the string scale is,
$$
{M_s^{(3)}} 
= \lam^{-1}\TNS^{1/2}m_s^3 (R_1 R_2 R_3)^{1/2}.
$$
The radii satisfy,
$$
M_s^{(3)} \hat{R}_i 
 = \TNS^{-1/2}(R_1 R_2 R_3)^{1/2} R_i^{-1}
   \rightarrow 0,
$$
and the radius of $\MS{1}$ satisfies,
$$
M_s^{(3)} R = 
   m_s^2\TNS^{1/2}(R_1 R_2 R_3)^{1/2}\rightarrow \infty.
$$
At this point,
$$
\tw_i\epsilon_{ijk} = \int_{C_{jk}} B^{NSNS},\qquad i,j,k=1\dots 3.
$$
Since $M_s^{(3)}\hat{R}_i\rightarrow 0$, we must perform another
T-duality on $\MT{3}$. However, because of the NS-NS 2-form fluxes,
just as in \rCDS, another T-duality will not help.
Instead, let us do a T-duality on $\MS{1}$ which brings us to
the final setup of gauge theory on a non-commutative $\MT{4}$.

\Step{4}
After T-duality along $\MS{1}$ we get type-IIA with $k$ D6-branes
and $q$ D2-branes.
The string coupling is now,
$$
\lam^{(4)} = {{\lam^{(3)}}\over { M_s^{(3)} R }}
 =\lam^{-1}m_s\TNS^{-1/2}(R_1 R_2 R_3)^{1/2}\rightarrow 0,
$$
and $\hat{M}_s = M_s^{(3)}$.
The radii satisfy,
$$
\hat{M}_s\hat{R}_i =
  \TNS^{-1/2}(R_1 R_2 R_3)^{1/2} R_i^{-1}\rightarrow 0,
$$
and the radius of the $\MS{1}$ satisfies,
$$
\hat{M}_s \hat{R} = 
   m_s^{-2}\TNS^{-1/2}(R_1 R_2 R_3)^{-1/2}\rightarrow 0.
$$
At this point, the $\tw$-twists are still NS-NS 2-form fluxes.
We thus end up with a system of $k$ D6-branes on $\MT{4}\times\MR{2,1}$
and $q$ D2-branes which are points on $\MT{4}$. The radii of $\MT{4}$
are given, in terms of the 3 radii $R_i$ of the original
$\MT{3}$, as follows,
\eqn\hiluf{\eqalign{
\hat{R}_i &=\hat{M}_s^{-1}\TNS^{-1/2}
 (R_1 R_2 R_3)^{1/2}R_i^{-1},\qquad i=1,2,3,\cr
\hat{R}_4 &= \hat{M}_s^{-1}m_s^{-2}\TNS^{-1/2}(R_1 R_2 R_3)^{-1/2}.\cr
}}
Here $\hat{M}_s$ denotes the final type-IIA (with
the D2-branes and D6-branes) string scale.
The final string coupling constant is,
$$
\hat{\lam} = 
 \lam^{-1}m_s\TNS^{-1/2}(R_1 R_2 R_3)^{1/2}.
$$

Similarly, we can start with $S_A(k)$ with 3 $\etw$-twists.
By definition, this is $S_B(k)$ on the dual $\MT{3}$ with
3 $\tw$-twists.
We realize this in type-IIB on the background 
$\MR{2,1}\times\MT{3}\times \TBNT_q$ and $k$ NS5-branes
on $\MR{2,1}\times\MT{3}$.
As before, the fiber of the Taub-NUT space is denoted by $\TNS$.
We first perform S-duality to replace the NS5-branes with $k$ D5-branes.
At this point the $\etw$-twists are off-diagonal components of
the metric $g_{i9}$ with $i$ in the direction of $\MT{3}$ and
$9$ in the direction of the Taub-NUT fiber.
Then, we perform T-duality on the direction of $\TNS$ to obtain
type-IIA on $\MR{2,1}\times\MT{3}\times\MS{1}\times\MR{3}$
with $q$ NS5-branes
on $\MR{2,1}\times\MT{3}$ and $k$ D6-branes on 
$\MR{2,1}\times\MT{3}\times\MS{1}$.
The $\etw$-twists became NS-NS 2-form fluxes $B_{i4}$ where
$4$ is the direction of $\MS{1}$.
Then, we do T-duality on the
three directions of $\MT{3}$. We obtain $k$ D3-branes on
$\MR{2,1}\times\MS{1}$ and $q$ NS5-branes.
The $\etw$-twists are now off-diagonal components $g_{i4}$.
We then do another S-duality to get $k$ D3-branes and $q$
D5-branes and, finally, another T-duality on $\MT{3}$.
At this point we are back with $k$ D6-branes and $q$ D2-branes.
The $\etw$-fluxes are now NS-NS 2-form fluxes $B_{i4}$.

The moduli space is thus the same as the moduli space
of $q$ D2-branes  inside $k$ D6-branes on $\MT{4}$ with
NS-NS 2-form fluxes. In the case of $\tw$-twists,
these fluxes have both indices in the direction
of $\MT{3}\subset \MT{4}$. In the case of $\etw$-twists, the 
fluxes had one index in the direction of $\MT{3}$ and the other
index in the $4^{th}$ direction.
In the generic case, we have both $\tw$-twists and $\etw$-twists
simultaneously. The result is that the NS-NS 2-form flux
is nonzero for all 6 2-cycles of $\MT{4}$.
The string scale, string coupling, and
the parameters of the $\MT{4}$ are as calculated above.
We could in principle follow the chain of dualities above
with simultaneous $\tw$-twists and $\etw$-twists
but the intermediate steps would involve cumbersome 
non-linear expressions.

The moduli space of $q$ D2-branes inside $k$ D6-branes on $\MT{4}$
with NS-NS 2-form fluxes, and in the limit that the size of the
$\MT{4}$ vanishes,
was shown to be equivalent to the moduli space
of $k$ instantons of $U(q)$ gauge theory on a
non-commutative $\MT{4}$
\refs{\rDouglas-\rANS}.
It is likely that this result is true even for $\MT{4}$ of finite
size, because the size decouples by arguments as above.

In the next sections we will review the non-commutative geometry
and formulate a precise statement about the moduli space.

\section{Review of Noncommutative Gauge Theory}
In this section we will review the elements of non-commutative 
gauge theory which are relevant to our situation. 

Non-commutative gauge theory first entered string theory in 
\rCDS\  where it was shown to provide a matrix model for 
M-theory on a torus with the $C^{(3)}$ field turned on 
along the light-like circle. Subsequently, a lot of interesting 
work on this topic was done
\refs{\rDH-\rHVer}.
What we need here is not 
the connection to matrix theory but just the study of D-branes 
with a $B^{NS}$ fields turned on.

Consider type-IIA on $\MR{1,9-d} \times \MT{d}$ with $q$ D0-branes. 
The radii of $\MT{d}$ are called $R_i, i=1,..,d$, the string mass 
$m_s$ and the coupling $\lambda$. Furthermore let there be a 
constant $B^{NS}$ field along $\MT{d}$. Let 
\eqn\bfelter{
b_{ij} = \int_{ij} B^{NS},\qquad     i,j=1,\ldots,d
}
be the flux of $B^{NS}$ through the $\MT{2}$ spanned by 
directions $i,j$. The $b_{ij}$ are periodic with period 
$2 \pi$ due to the gauge invariance of $B^{NS}$. 

In \rCK\ this system was studied using the approach 
of \rWati. The result is that the 
low energy physics is described by a $d+1$ dimensional 
 $U(q)$ gauge theory on a dual torus, $\MHT{d} \times 
 \MR{0,1}$ with radii
\eqn\durad{
\tilde{R}_i = {1 \over m_s^2 R_i }
}
and gauge coupling 
\eqn\kob{
{1 \over g^2} = {m_s^{2d-3} R_1 \ldots R_d \over \lambda}.
}
The effect of $b_{ij}$ is to change the action. Every time 
two fields are being multiplied, the multiplication is with 
the $*$-product defined as,
\eqn\stjerne{\eqalign{
&(\phi^{(2)} * \phi^{(1)})(x) = \cr
& \left.
e^{-{{b_{ij}} \over {2 m_s^4 R_i R_j}}
(\partial^{(2)}_i \partial^{(1)}_j
-\partial^{(2)}_j \partial^{(1)}_i)}
    \phi^{(2)}(x_2)  \phi^{(1)}(x_1) \right|_{x^{(2)}=x^{(1)}=x},\cr
& \partial^{(a)}_i \equiv{\partial\over {\partial x^{(a)}_i}},\qquad a=1,2.
\cr
}}
The action is the usual gauge theory action just with this 
modification.

If there had been no $B^{NS}$-field the resulting $d+1$ dimensional 
gauge theory could have been obtained by performing 
T-duality along $\MT{d}$. The $q$ D0-branes would have turned into 
$q$ Dd-branes. The radii and gauge coupling of the $U(q)$ theory 
can be calculated in this way. The important point to remember 
is that the only change from having a $B^{NS}$-field is to 
change the product into eq.\stjerne.
The radii and gauge coupling are independent of $b_{ij}$. 
This result could not have been obtained by T-duality, 
since $B^{NS}$-fields change the formulas of T-duality and would 
have given other radii and gauge coupling.

There is another way of formulating this gauge theory. Instead of 
working with the $*$-product, eq.\stjerne, one can say that the 
torus $\MHT{d}$ is non-commutative. The algebra
 of functions on the torus is,$A$, is generated by $U_1, \ldots , 
U_d$ with relations 
\eqn\rela{
U_i U_j = U_j U_i e^{i b_{ij}}
}
The generalization of finite dimensional vector fields is 
finitely generated projective modules over $A$. Let $E$ be 
such a module. One can define connections,$\nabla$, and curvature 
$F_{ij}$ of this module \refs{\rCDS,\rANS}. One can define 
the Chern character of the module $E$
\eqn\chkar{
ch(E) = \sum_{k=0} {\hat{\tau} (F^k) \over (2\pi i )^k k!}
}
$\hat{\tau}$ is the trace on $End_A(E)$. $ch(E)$ 
can be regarded as an element in the cohomology, 
$H^{*}(\MT{d},{\bf C})$, of $\MT{d}$, the original 
torus. $ch(E)$ is not integral but there exists an 
integral cohomology class $\mu(E) \in H^{*}(\MT{d},{\bf C})$
 such that
\eqn\hele{
ch(E) = e^{{1 \over 2\pi}{\iota (b)}} \mu(E)
}
Here $\iota(b)$ denotes contraction with $b$ considered as an 
element of $H_{*}(\MT{d},{\bf C})$ \rANS. 

The mathematical fact that the module $E$ is 
characterized by integers is in 
exact agreement with our expectation from D-brane physics. 
Besides the $q$ D0-branes on $\MT{d}$ there could be any number 
of D2-branes, D4-branes , etc. wrapped on $\MT{d}$. These numbers 
are exactly given by $\mu(E)$.
$ch(E)$
measures the fact that D2-branes with $B^{NS}$-fields turned on 
have an effective D0-brane charge and the equivalent phenomena 
for other branes. Suppose for instance that only  $\mu_0$ and
$\mu_1$ are nonzero, then,
\eqn\eksempel{
ch_0 = \mu_0 + {b_{12} \over 2\pi} \mu_1,\qquad
ch_1 = \mu_1.
}
This equation reflects the fact that the number of D2-branes 
is unchanged by the presence of the $B^{NS}$-field but the 
number of D0-branes is shifted by the product of 
the number of D2-branes and the $B^{NS}$-field along the 
D2-branes.

\section{Noncommutative Instantons as the Moduli-space}
Let us now go back to our system of $q$ D2-branes inside 
$k$ D6-branes given above. They have a common $\MR{1,2}$. This 
is the space-time in which the 3 dimensional theory is living. 
The 3 dimensional theory has a low energy description as a sigma model 
with the moduli space of vacua as target space. 
The moduli space of vacua is a Hyper-k\"ahler manifold. The 
moduli space of vacua comes from the dynamics on the $\MT{4}$,  
which is the same as the dynamics of $q$ D0-branes in $k$ D4-branes on 
$\MT{4}$. The radii of the $\MT{4}$, 
$\hat{R}_1,\hat{R}_2,\hat{R}_3,\hat{R}_4$, and the string coupling
$\hat{\lam}$ and string scale $\hat{M}_s$
are given in terms of the parameters of 
the $S_A(k)$ compactification in \hiluf\ which we repeat here,
\eqn\otekhi{\eqalign{
\hat{R}_i &= m_s^{-3}\lam\TNS^{-1} R_i^{-1},\qquad i=1,2,3,\cr
\hat{R}_4 &= m_s^{-5}\lam\TNS^{-1}(R_1 R_2 R_3)^{-1},\cr
\hat{M}_s &= \lam^{-1}m_s^3\TNS^{1/2} (R_1 R_2 R_3)^{1/2},\cr
\hat{\lam} &=
  \lam^{-1}m_s\TNS^{-1/2}(R_1 R_2 R_3)^{1/2},\cr
}}
 Furthermore there is a $B^{NS}$-field turned on along $\MT{4}$,
\eqn\bfelter{\eqalign{
\int_{12} B^{NS} = \alpha_3, &\qquad
\int_{31} B^{NS} = \alpha_2, \cr
\int_{23} B^{NS} = \alpha_1, &\qquad
\int_{i4} B^{NS} = \eta_i,\qquad i=1,2,3.\cr
}}
but the vacuum structure of the vector-multiplets should be 
independent of $\TNS$ in this limit.

According to the above review of non-commutative geometry, 
the moduli space is equal to the moduli space of $k$ instantons 
in $U(q)$ gauge theory on a non-commutative torus, $\MHT{4}$,
with non-commutativity parameters equal to $\alpha_i$, $\eta_i$.
As explained above the radii and gauge coupling of this gauge theory 
are the same as if $\alpha_i = \eta_i =0$. Hence they can be found by 
T-duality on $\MT{4}$. By this T-duality one obtains $k$ D2-branes 
in $q$ D6-branes on $\MHT{4}$ of radii,
\eqn\radierto{
\tilde{R_1} = {\lam \over {m_s^3 R_2 R_3}},\,\,
\tilde{R_2} = {\lam \over {m_s^3 R_1 R_3}},\,\,
\tilde{R_3} = {\lam \over {m_s^3 R_1 R_2}},\,\,
\tilde{R_4} = {\lam \over m_s},
}
and string mass, $\tilde{m_s}$, and coupling, $\tilde{\lambda}$,
\eqn\vaerd{\eqalign{
\tilde{m_s} &=
\hat{M}_s = \lam^{-1}m_s^3\TNS^{1/2} (R_1 R_2 R_3)^{1/2},\cr
\tilde{\lambda} &= \lam^{-1} m_s^3 \TNS^{3/2} (R_1 R_2 R_3)^{1/2}.
}}
In the $U(q)$ theory, this gives a gauge coupling of,
\eqn\qcdkob{
{1 \over g^2} = {\tilde{m_s}^3 \over \tilde{\lambda}} =
\lam^{-2} m_s^6 R_1 R_2 R_3.
}
Observe that $\TNS$ has dropped out of the radii and 
the gauge coupling.

 What about the limit $\lam\rightarrow\infty$ and $m_s$ fixed.
To see that the moduli space 
of vacua is well defined in this limit we should remember that 
scalar fields in three dimensions have dimension $\half$, if 
we want a standard kinetic term. We can either view the moduli space 
of vacua from the $U(q)$ gauge theory point of view or from the 
$U(k)$ theory on the D2-branes. {}From the last point of view 
the moduli space is the Higgs branch. The action of the $U(k)$ 
theory has a term,
\eqn\led{
\half {1 \over \tilde{\lambda}\tilde{m_s}} \int d^3x 
(\partial_\mu (\tilde{m_s}^2 X^i))^2
}
We define $\Phi^i = \tilde{\lam}^{-1/2}\tilde{m_s}^{3/2} X^i$. 
This $\Phi$ has a standard kinetic term,
\eqn\ledto{
\half \int d^3x (\partial_\mu \Phi^i)^2
}
The radii of the $\Phi^i$ are $R(\Phi^i) = 
\tilde{\lam}^{-1/2}\tilde{m_s}^{3/2}\tilde{R^i}$.
\eqn\endrad{\eqalign{
R(\Phi^1) = \sqrt{R_1 \over R_2 R_3},\qquad
R(\Phi^2) =& \sqrt{R_2 \over R_1 R_3},\qquad
R(\Phi^3) = \sqrt{R_3 \over R_1 R_2} \cr
R(\Phi^4) =& m_s^2 \sqrt{R_1 R_2 R_3}.
}}
We see that the limit $\lam \rightarrow \infty$ exists.
This last discussion was really superfluous. Since $S_A(k)$ only 
depends on the combination $m_s^2$ and does not feel $\TNS$, 
this had to be true. For finite $m_s\TNS$,
it could even be true for the full theory, not just the 
 moduli space of vacua. The effect of the twists is just to 
deform the moduli space and so does not change the fact that 
the moduli space is independent of $\TNS$ and has a limit when 
$\lam\rightarrow\infty$, keeping $m_s$ fixed.

We can also see from \endrad\ what happens in the limit of the $(2,0)$
theory. For this limit we take $m_s\rightarrow\infty$.
We find that the $\MT{4}$ degenerates to $\MT{3}\times \BR$.

Let us now be more precise about the space of instantons on 
a non-commutative $\MT{4}$. For this sake we will temporarily neglect the
 uncompactified directions and think of our system as $q$ D0-branes and 
k D4-branes on $\MT{4}$.
 According to the review of non-commutative geometry above, 
this is described by a gauge theory on the dual $\MHT{4}$ 
with non-commutativity parameters, $b_{ij}$, equal to the twists. 
 By gauge theory we really mean a projective module, $E$, which 
is characterized by 
\eqn\karbun{
\mu(E) = H^*(\MT{4},{\BZ}).
}
$\mu(E)$ has components in dimensions 0,2 and 4. $\mu_0 = q$ is 
the number of D0-branes on $\MT{4}$. $(\mu_1)_{ij}$ is the number of 
D2-branes in the $\MT{2}$ in direction $(i,j)$ with $i,j=1,2,3,4$. 
$\mu_2 = k$ is the number of D4-branes. So far we have not specified 
the number of D2-branes. Since we are interested in the low energy 
dynamics we should take the number of D2-branes to minimize the 
total energy in the D0,D2,D4 brane system. When $b_{ij}=0$ this 
is done by setting $\mu_1=0$, i.e. no D2-branes. Let us turn on 
$b_{12}$, say. {}From the formula
\eqn\heleto{
ch(E) = e^{{1 \over 2\pi}{\iota (b)}} \mu(E)
}
we get 
\eqn\mini{
(ch_1)_{34}= (\mu_1)_{34} + {b_{12} \over 2\pi} \mu_2
           = (\mu_1)_{34} + {b_{12} \over 2\pi} k.
}
To minimize the energy, $(ch_1)_{34}$ should be minimized. We see 
that when $b_{12} > {1 \over 2k} 2\pi$ we can lower the energy 
by taking $(\mu_1)_{34} = -1$. This phenomena divides the space 
of $b_{ij}$ into ``Brillouin'' zones. Each zone is a six dimensional 
cube of length $2\pi \over k$ in each direction. Inside a zone 
the low energy physics is described by the gauge theory corresponding 
to a module with the $\mu(E)$ which minimizes the energy. In crossing 
the boundary between 2 zones, $\mu(E)$ jumps. 

We also see another interesting phenomena. Whenever ${b_{12} \over 2\pi}k$ 
 is an integer we have $(\mu_1)_{34} =- {b_{12} \over 2\pi}$ and 
hence $(ch_1)_{34} =0$. This means that $ch(E)$ is nonzero only in 
dimensions 0 and 4 (We are keeping all other components of $b_{ij}=0$. 
Only $b_{12}= n {2\pi \over k}$). This is exactly like the pure D0,D4 
system with no $B^{NS}$-field. This system has a phase where the 
D0-branes and D4-branes are separated. To reach this phase the 
system has to go through zero-size instantons. We thus conclude that 
whenever $b_{12}= n {2\pi \over k}$,$n \in {\BZ}$ there is another phase. 
Of course, there is nothing special about $b_{12}$. Similar statements 
could be made for the other 5 components of $b_{ij}$ and even for all 
of them simultaneously. The point is that for each center of 
the ``Brillouin'' zone there is another branch emanating from a 
locus on the Coulomb branch. It emanates from the points on the  
Coulomb branch where some instantons have shrunk to zero size. 
The other phase consists of the $k$ D4-branes with $-n$ D2-branes 
inside moving away from the $q$ D0-branes. Let us calculate the  
dimension of this branch. Suppose first $n=1$, so there are $k$
D4-branes with $-1$ D2-brane inside (equivalently 1 anti D2-brane). 
This system has a bound state. It is not marginally bound. The system 
has an 8 dimensional moduli space. To see this we should really 
remember that it is really $k$ D6-branes with $-1$ D4-brane. 
4 of the dimensions are $U(1)$ Wilson lines on the $\MT{4}$. 
They are center of mass coordinates and are always present. We are 
not interested in these. The other 4 are 3 transverse positions and 
the dual photon in 3 dimensions. We conclude that the other phase is 
4 dimensional. Furthermore it emanates from a point 
on the Coulomb branch, since all instantons have to shrink on top 
of each other. The only freedom is the point where they shrink, but 
that is a center of mass degree of freedom which we ignore.

Let us now take $n$ to be generic. Let $g= gcd(n,k)$. The system of 
$n$ D2-branes inside $k$ D4-branes can split into $g$ separate systems. 
The dimension is thus $8g-4$, subtracting the center of mass again. 
It emanates from the Coulomb branch on a locus of dimension $4g-4$.

The special case of $q=1,k=2$ was studied in detail in 
\rCGK. Here it was found that there was another phase of 
dimension 4 for $\alpha = \pi$. We see that this agrees exactly 
with what was found here. However we get a much clearer 
picture of the other branch. In the next section we will 
understand these branches from a field theory point of view.

\subsection{Phase Transitions from the Gauge Theory}
With generic twists (non-commutativity parameters), the moduli-space
that we obtain is smooth.
However, for special values of the twists the moduli space has ADE-type
singularities. We would now like to explain the origin
of some of these singularities.

$S_B(k)$ is a gauge theory at low energies. Let us study it with an 
$\alpha$-twist along one circle and no twist along the other 2 circles. 
 Since there is a circle without twist we can T-dualize
on that direction to $S_A(k)$, so 
these remarks apply to $S_A(k)$ as well. We want to reproduce the 
existence of other branches of the moduli space. 
 For a related discussion see \rNek.

The fields in 6 dimensions are a $U(k)$ vector-multiplet and an 
adjoint hypermultiplet. In 3 dimensions there is a tower of $U(k)$ 
vector-multiplets with masses $({n_1 \over R_1},
{n_2 \over R_2},{n_3 \over R_3})$, $n_i \in {\BZ}$ 
and a tower of adjoint hypermultiplets with masses
$({n_1 - {\alpha \over 2\pi} \over R_1},
{n_2 \over R_2},{n_3 \over R_3})$, $n_i \in {\BZ}$.
We remember that a mass in $\SUSY{4}$ theories in
 3 dimensions is specified by 3 numbers.
The moduli space is $4k$-dimensional including the center 
of mass degrees of freedom. On the Coulomb branch the $U(k)$ is 
broken to $U(1)^k$. Each adjoint hypermultiplet splits into 
$k^2$ hypermultiplets of the following charges. There are 
hypermultiplets with charge $(0,\ldots,0)$, and
there are $k$ hypermultiplets  with 
charges $(1,-1,\ldots,0)$ plus permutations. There 
is a total of $k(k-1)$ of these. 
Some of these hypermultiplets can become massless on the 
Coulomb branch. For that to happen we have to turn on a 
Wilson line, $A_1$, along the first circle and set the other
$3k$ moduli zero. $A_1$ has the form 
\eqn\matriks{
A_1 = \left(\matrix{ a_1 & 0 & \ldots & 0 &0 \cr
                       0 & a_2 & \ldots & 0 &0 \cr
                       \vdots & \vdots&\ddots& \vdots&\vdots \cr
                     0   & 0 & \ldots & a_{k-1} & 0 \cr 
                     0   & 0 & \ldots & 0 & a_{k} \cr} \right)
}
The tower of hypermultiplets is now as follows.
There are $k$ of charge $(0,\ldots,0)$ with mass 
$({n_1 - {\alpha \over 2\pi} \over R_1},
{n_2 \over R_2},{n_3 \over R_3})$ and for every $i\ne j$ there is
a hypermultiplet with charge
$(0,\ldots,1,\ldots,-1,\ldots,0)$ plus permutations
with the 1 on the $i^{th}$ place and the -1 on the $j^{th}$ place. 
It has a  mass $({n_1 - {\alpha \over 2\pi} + {a_i \over 2\pi} 
- {a_j \over 2\pi} \over R_1},
{n_2 \over R_2},{n_3 \over R_3})$.
The uncharged ones never become massless, as long as the twist is 
not a multiple of $2\pi$. The charged ones become massless if 
\eqn\masseloes{
n_1 - {\alpha \over 2\pi} + {a_i \over 2\pi} 
- {a_j \over 2\pi} = n_2 = n_3 =0.
}
Now it is easy to make some of them massless by choosing $A_1$ 
appropriately. However to have a Higgs branch we need to have 
non trivial solutions to the D-flatness equations. For 
hypermultiplets charged under a $U(1)^r$ group there should 
be at least $r+1$ of them to have a non trivial solution. We thus 
need to find a number of massless hypermultiplets which is bigger than 
the number of $U(1)$'s under which they are charged. No hypermultiplets 
are charged under the diagonal $U(1)$. Let us first find a situation 
of k massless hypermultiplets which are charged under $U(1)^{k-1}$.
The hypermultiplet of charge $(1,-1,0,\ldots,0)$ is massless 
if,
\eqn\maslet{
n_1=0,\qquad a_1-a_2= \alpha.
}
The one of charge $(0,1,-1,0,\ldots ,0)$ is massless if,
\eqn\maslto{
n_1=0,\qquad a_2-a_3= \alpha
}
and so on, up to the multiplet of charge $(0,\ldots,0,1,-1)$ 
which is massless if,
\eqn\masltre{
n_1=0,\qquad a_{k-1}-a_k= \alpha
}
This gives $k-1$ massless hypermultiplets. To have one more we need 
$(-1,0,\ldots,0,1)$ to be massless. This is the case if,
\eqn\maslfire{
{\alpha \over 2\pi} = {a_k-a_1 \over 2\pi} + n_1,
}
for some integer $n_1$. Now,
\eqn\maslfem{
a_k - a_1 = (a_k -a_{k-1}) + \ldots + (a_2 -a_1) = -(k-1)\alpha
}
so we need ${{k\alpha} \over {2\pi}}$ to be an integer. So for $\alpha 
= {2\pi \over k}$ we have another phase of dimension 4. The 
dimension is 4 because there are $k$ massless hypermultiplets
 each having 4 scalar fields and the D-flatness conditions 
remove $4(k-1)$ dimensions leaving 4 real dimensions. This phase 
 agree agrees exactly with the exact result from the previous section. 
We thus see that a naive field theory treatment, keeping all 
Kaluza-Klein modes, reproduces the result. 
This phase emanates 
from the Coulomb branch whenever $a_i = a_{i-1} = \alpha$ 
as we saw above. This fixes the $a_i$ up to an overall shift. 
The overall shift is the $U(1)$ part which we discard anyway. 
This shows that the other phase emanates from one
particular point on the Coulomb branch.
Note that the field theory treatment is justified when $M_s R_i\gg 1$.

More generally, let us take $\alpha = n {2\pi \over k}$ and 
$g = gcd(n,k)$. Now we can play the same game as above but within 
$g$ blocks of the $U(k)$ matrix of size $k \over g$. We thus get 
g sets of $k \over g$ massless fields. Each set is charged under 
a $U(1)^{{k \over g} -1}$ subgroup.
This gives a $4g$ dimensional phase 
emanating from a locus on the Coulomb branch. This locus has 
dimension $4g-4$. The $4g$ comes from the diagonal $U(1)$ in each 
of the $g$ blocks. The center of mass is subtracted again. 
This branch has a total dimension of $4g+4g-4 = 8g-4$. We again 
find agreement with the exact result described previously. 

The branches described above are the only ones coming from the 
naive field theory description besides the cases $\alpha = 2 \pi n$, 
$n \in {\BZ}$ which behave like $\alpha =0$.


\section{The 3+1D limit}
In this section we will explain how to obtain the 3+1D Seiberg-Witten
curves of the various theories
compactified on $\MT{2}$ with a twist. This time we only
have two independent
$\tw$-twists corresponding to the two cycles of $\MT{2}$.
The way to obtain the 3+1D SW curves is to start with the moduli
space of the theory compactified on $\MT{2}\times\MS{1}$ where
$\MS{1}$ is of radius $R$ and take the limit $R\rightarrow\infty$.
Let the 2+1D
hyper-K\"ahler moduli space be of dimension $4n$.
In the limit $R\rightarrow\infty$, it 
can be written as a fibration of $T^{2n}$ over a base of dimension $2n$.
In the decompactification limit the fiber $T^{2n}$ shrinks to zero.
We interpret it as the Jacobian variety of a Riemann surface of genus
$n$ which varies over the base. This will then be the Seiberg-Witten curve
(see \rSWGDC).
Starting with the Blum-Intriligator little-string theories
of $k$ NS5-branes at an $A_{q-1}$ singularity compactified on $\MT{2}$ 
with twists we can get, in appropriate limits, a 3+1D gauge theory
with,
$$
SU(k)_1\times\cdots\times SU(k)_q,
$$
and massive adjoint hyper-multiplets in consecutive $(k,\bar{k})$
representations.
The Seiberg-Witten curves for these models have been derived in \rWitFBR.
As we will show below, we can reproduce these curves by taking the
appropriate  decompactification limit of 
the moduli space of $k$ $U(q)$ instantons on the non-commutative $\MT{4}$.

To start, we will recall how the reduction of the untwisted compactified
Blum-Intriligator theories works.

\subsection{From instantons to quiver gauge theories}
When we set all the $\tw$-twists to zero we obtain the statement
that the Coulomb-branch moduli space of the theories of
$k$ NS5-branes on an $A_{q-1}$ singularity, compactified on 
$\MT{3}$ is the same as the moduli space of $k$ ordinary instantons
with a $U(q)$ gauge group on $\MT{4}$.
This result has already been established in \refs{\rIntNEW, \rGanSet}.
Suppose we compactified on $\MT{3} = \MT{2}\times \MS{1}$ and take
the radius of $\MS{1}$, $R\rightarrow\infty$.
It can be checked (see \endrad)
that the auxiliary $\MT{4}$ becomes a product
$\MT{2}_B\times \MT{2}_F$.
The complex structure of $\MT{2}_F$ and $\MT{2}_B$
are fixed as $R\rightarrow\infty$ while the area of $\MT{2}_B$ is
proportional to $R$ and the area of $\MT{2}_F$ is proportional to $R^{-1}$.
Now take a particular gauge configuration corresponding to an instanton
of $U(q)$ with instanton number $k$.
We can encode the information in the instanton as follows
(see \refs{\rFMW,\rBJPS}).
At a local point on the base, the gauge field reduces to two commuting 
$U(q)$ Wilson lines on the fiber. We can describe them uniquely
as $q$ points on the dual $\MHT{2}$ of the fiber.
These $q$ points vary over the base $B$. The instanton equations
imply that they span a holomorphic curve $\Sigma_g$ of genus 
$g = q k + 1$. $\Sigma_g$ is called the {\it ``spectral curve''}.
 To completely describe the instanton we also need to
describe a line bundle over $\Sigma_g$ which corresponds to a point
in the Jacobian of $\Sigma_g$ (recall that the Jacobian of a genus
$g$ curve is $T^g$). The line bundle is called the
{\it ``spectral-bundle''}.
Alternatively, we can represent the moduli space of $U(q)$ instantons
at instanton number $k$ on $B\times F$ as the moduli space of 
$q$ D6-branes wrapped on $B\times F$ with $k$ D2-branes.
The curves are obtained by T-duality along the two directions 
of $F$. We obtain a D4-brane wrapped on a curve $\Sigma_g$
of homology cycle $q\lbrack B\rbrack + k \lbrack F\rbrack$.
 The curve $\Sigma_g$ is the Seiberg-Witten curve
of the point in the moduli space. It intersects a generic fiber $F$
in $q$ points and a zero section of the base $B$ at $k$ points.
It is also easy to see that as the base $B$ decompactifies to
$\MS{1}\times \MR{1}$ we reproduce exactly the curves from the
brane construction of \rWitFBR\ for the quiver gauge theory.

\subsection{The role of the non-commutativity}
Now let us repeat the same procedure but with two non-commutativity
parameters $\tw_1$ and $\tw_2$. We can take $\tw_1$
to be along the first cycle of the base $B=\MT{2}$ and the first
cycle of the fiber $F=\MT{2}$ and we take $\tw_2$ to be along the
second cycle of the base $B$ and the first cycle of the fiber $F$.
The $\eta$-twists will similarly correspond to non-commutativity
along the second cycle of $B$ and one of the two cycles of $F$.

To translate this to the curve $\Sigma_g$ we take the system of
$q$ D6-branes and $k$ D2-branes and put in NSNS 2-form fluxes
according to the non-commutativity parameters. After T-duality
along $F$ The NSNS fluxes become components of the metric $G_{IJ}$.

As a result, we obtain a tilted $\MT{4}\equiv\MR{4}/\Lambda$,
where $\Lambda$ is a lattice spanned by the following vectors:
\eqn\ves{\eqalign{
\ve_1 &= (1,0,0,0),\cr
\ve_2 &= (\tau_1,\tau_2,0,0),\cr
\ve_3 &= (\tw_1 + \etw_1\tau_1,\etw_1\tau_2,\chi,0),\cr
\ve_4 &= (\tw_2 + \etw_2\tau_2,\etw_2\tau_2,\chi\rho_1,\chi\rho_2).\cr
}}
Here, $\tau\equiv \tau_1 + i\tau_2$ is the complex structure
of $\MT{2}_F$, $\rho\equiv \rho_1 + i\rho_2$ is the complex 
structure of $\MT{2}_B$,
and,
$$
\chi = m_s (\tau_2\rho_2)^{-1},
$$
so that the overall volume of the unit  cell will be $m_s^2$.
We will denote the coordinates in $\MR{4}$ by $(x_1,x_2,x_3,x_4)$.
The D2 and D6 branes became a single D4-brane in the homology class,
$$
\lbrack \Sigma\rbrack = 
q\lbrack B'\rbrack + k \lbrack F'\rbrack.
$$
Here,
\eqn\fpbp{\eqalign{
 F' &\equiv \{s \ve_1 + t\ve_2\, |\, 0\le s,t\le 2\pi\},\cr
B' &\equiv \{s \ve_3 + t\ve_4\, |\, 0\le s,t\le 2\pi\}.\cr
}}
are two faces of $\MT{4}$.
Similarly to \rWitFBR\ the D4-brane will find a minimal-area surface in
this homology class.
In the complex structure given by,
$$
z = x_1 + i x_2,\qquad w = x_3 + i x_4,
$$
the cohomology class $\omega\in H^2(\BZ)$ which is Poincar\`e dual
to $\lbrack\Sigma\rbrack$ will, generically, be a mixture of
$(1,1)$, $(0,2)$ and $(2,0)$ forms. However, it is always possible to
find a complex structure (with respect to the flat metric) for which
$\omega$ is entirely a $(1,1)$ form. In this complex structure
the $\MT{4}$ is ``algebraic'' (see p315 of \rGriHar).
Given the complex structure, it is possible to write down
the curve $\Sigma$ as the zero locus of a $\theta$-function on
$\MT{4}$. These $\theta$-functions are the sections
of the line-bundle corresponding to $\lbrack\Sigma\rbrack$
and depend on $k q$ parameters which are the moduli (see \rGriHar\
for further details).

It is easy to see that the ``elliptic-models'' of \rWitFBR\
are recovered in the special limit in which we get a gauge
theory with massive hyper-multiplets.
In this case $\tau\rightarrow\infty$ and
there are no $\eta$-twists. The fiber $F'$ is replaced
with a strip $\MS{1}\times\MR{1}$. The class $\lbrack\Sigma\rbrack$
is analytic (i.e. the class $\omega$ is a $(1,1)$ 2-form)
and the Seiberg-Witten curves of \rWitFBR\ are recovered.


\section{Another Look at the $\etw$-twists}

In this section, we write explicitly the solution for type-IIA
(or type-IIB) theory, with both $\tw$-twists and $\etw$-twists
turned on. 
These solutions should be interpreted as string world-sheet
$\sigma$-models with a $B$-field.

We will start with a Taub-NUT space without NS5-branes.
It is straightforward to define  the $\tw$-twist.
One starts with some given background, which
is a principal $U(1)$ bundle cross a torus $\MT{d}$. 
Locally, the $\tw$-twist is just the change of coordinate in the
$\MS{1}$ fiber of the Taub-NUT space,
 of the form $y\to y+\sum\tw_I\psi^I$.
$y$ is the coordinate on the circle (see \tnmet) and $\psi^I$
is the coordinate on $\MT{3}$ ($I=1,2,3$).
Since it is
just the change of variables, the string theory equations of
motion are trivially satisfied. But globally, this is not
a valid coordinate transformation, since $\tw_I\psi^I$ is
not a periodic function on $\MT{3}$ modulo $2\pi$. Therefore,
we get a different background --
we call it the $\tw$-twisted background.
As for $\etw$-twists, they are related to $\tw$-twists by 
T duality in $\MT{3}$.

We will construct the background with both $\tw$ and $\etw$ twists
turned on in the following way. We first consider the background
containing Taub-NUT space cross a three-torus, without any twists.
We introduce $\tw$-twists along the three-torus, with the parameters
$\etw_I$. Then, we make a T-duality transformation, and get a background
with $\etw$-twists. This new background is again a $U(1)$ bundle
cross a (dual) torus, and we now $\tw$-twist it. In this way,
we get a background with both $\tw$-twists and
$\etw$-twists.

Let us do it explicitly.
Start with $\MR{1,2}\times \TBNT(\rho)\times \MT{3}$. The metric is:
\eqn\metriczero{\eqalign{
ds^2=&\rho^2 U_{[\rho]}(|\vec{r}|) {\cal A}^2
     + U_{[\rho]}(|\vec{r}|)^{-1} (d\vec{r})^2 
     \cr &
+ g_{IJ} d\psi^I d\psi^J - dx_0^2+dx_1^2+dx_2^2,\cr
}}
where we have denoted 
\eqn\defUr{
\Ur{\rho}\equiv\left( 1+{\rho\over 2 |\vec{r}|}\right)^{-1}
}
and $\cal A$ is the connection one-form
${\cal A}=dy-\vec{A}\cdot d\vec{r}$.
Also, we turn on the following $B$ field:
\eqn\Btostart{
B=b_{IJ} d\psi^I\wedge d\psi^J
}
We wish to introduce $\tw$-twists with the parameter $\etw_I$.
As was explained above, this means just the change of variables
$y\to y-\etw_Id\psi^I$. This amounts to replacing ${\cal A}^2$
with $({\cal A}-\etw_Id\psi^I)^2$ in \metriczero.

Now we make three T-dualities. We do this by the standard 
technique of treating $V^I_\a\equiv \px{\a}\psi^I$ (where
$\a$ is a string world-sheet coordinate) as an independent
variable and inserting a Lagrange multiplier, $\tilde{\psi}_I$, for,
$\px{\lbrack\a}V_{\b\rbrack}^I$.
We get the following metric:
\eqn\metricone{\eqalign{
ds^2=&{\rho^2 \Ur{\rho} \over 1+(\etw,\etw) \rho^2 \Ur{\rho}}
         ({\cal A}-b^{IJ}\etw_I d\tilde{\psi}_J)^2 
      + \Ur{\rho}^{-1}(d\vec{r})^2
      \cr
&+ l_s^4\left(g^{IJ}- 
       {\rho^2 \Ur{\rho}\over 1+\rho^2 (\etw,\etw)\Ur{\rho}}
      \etw^I \etw^J \right) d\tilde{\psi}_I d\tilde{\psi}_J \cr           
 &     -dx_0^2 +dx_1^2 +dx_2^2, 
}}
with the notation,
$\etw^I=g^{IJ}\etw_J$, $(\etw,\etw)=\etw_I\etw^I$,
and $g^{IJ}+b^{IJ}$ is the matrix inverse to
$g_{IJ}+b_{IJ}$.
Also, we have the following $B$ field:
\eqn\bfield{
\matrix{
B=-{\rho^2 \Ur{\rho}\over 1+\rho^2 (\etw,\etw)\Ur{\rho}}
  \etw^I d\tilde{\psi}_I \wedge ({\cal A}-b^{JK}\etw_J d\tilde{\psi}_K)+
   b^{IJ}d\tp_I\wedge d\tp_J 
}}
Notice that 
\eqn\secondrho{
{\rho^2 \Ur{\rho} \over 1+(\etw,\etw) \rho^2 \Ur{\rho}}=
{\rho^2\over 1+(\etw,\etw)\rho^2} \Ur{{\rho\over 1+(\etw,\etw)\rho^2}}
}

If we start with a non-degenerate torus and a very small coupling constant,
then T-duality gives us back a very small coupling constant.

Now we $\tw$-twist this background. Again, $\tw$-twisting is just
a replacement,
$$
{\cal A}\to {\cal A}-\tw^Id\tp_I,
$$
in all the formulas for the metric and the $B$ field.
It is convenient
to absorb $b^{IJ}\etw_I d\tp_J$ into $\tw^I d\tp_I$. 
Then, the background fields are:

\eqn\both{\eqalign{
ds^2 =&
R^2(|\vec{r}|)({\cal A}-\tw^I d\tp_I)^2+
\Ur{\rho}(d\vec{r})^2 \cr 
 &+(dx^{\mu})^2+
l_s^4 G^{IJ}(|\vec{r}|)d\tp_I d\tp_J,
\cr
B =&
({\cal A}-\tw^I d\tp_I)\wedge B^J d\tp_J + B^{IJ}d\tp_I\wedge d\tp_J
}}
where
\eqn\defs{
\matrix{
R^2(|\vec{r}|)&=&{\rho^2\Ur{\rho}\over 1+(\etw,\etw)\rho^2\Ur{\rho}} 
\hfill\cr
G^{IJ}(|\vec{r}|)&=&g^{IJ}-
{\rho^2\Ur{\rho}\over 1+(\etw,\etw)\rho^2\Ur{\rho}}\etw^I\etw^J 
\hfill\cr
B^I(|\vec{r}|)
&=&{\rho^2\Ur{\rho}\over 1+(\etw,\etw)\rho^2\Ur{\rho}}g^{IJ}\etw_J
\hfill\cr
B^{IJ}&=&b^{IJ}\hfill
}}
Also, the dilaton is not constant.
Let $\lambda$ be the string coupling at $|\vec{r}|\to\infty$.
Then, the string coupling at finite $|\vec{r}|$ is:
\eqn\StringCoupling{
\lambda(|\vec{r}|)=\lambda\sqrt{1+\etw^2\rho^2\over 
1+\etw^2\rho^2\Ur{\rho}}
}


The metric \both\ is not, strictly speaking,  Hyper-K\"ahler.
Indeed, although it does have three complex structures, they are
not covariantly constant with respect to the standard 
covariant derivative. But they must be covariantly constant, if we modify
$\Gamma_{\mu\nu}^{\rho}$ with the torsion, proportional to $H=dB$.

We want to study the moduli space of the theory on the NS5-brane,
sitting at $\vec{r}=0$ in this background.
As we remarked in section (2), the NS5-brane has a size of $l_s$
and, although it is very heavy, it could affect the metric.
We will explore  this later in this section. For now, we will
assume that it is safe to forget about the NS5-brane.
To study the moduli space, we perform
the chain of dualities. It is most convenient to think of these
dualities as acting on the asymptotic
 ($\vr\to\infty$) values of the fields.
Therefore, we would like to discuss how the background fields near
the position of the NS5-brane ($\vr\to 0$) are related to the asymptotic
values of the fields at $\vr\to\infty$.

Let us look first at the geometry near the origin in $\MR{3}$.
 {}From \metricone\ and \secondrho\ we see that the geometry becomes flat
when the following two conditions are satisfied:
\eqn\twocond{
|\vec{r}|\ll\rho \;\;\;\; {\rm and} \;\;\;\;  
|\vec{r}|\ll{\rho\over 1+(\etw,\etw)\rho^2}
}
In this limit, we have just $\MR{1,6}\times \MT{3}$
with the metric
\eqn\metricloc{
ds^2=(dx^{\mu})^2 + 
|d(e^{i\tw^J\tp_J}z_1)|^2+
|d(e^{-i\tw^J\tp_J}z_2)|^2 + g^{IJ} d\tp_I d\tp_J
}
The $B$ field becomes:
\eqn\bfieldloc{
B= -\etw^I d\tp_I \wedge {\rm Im} (z_1^* dz_1+ z_2^* dz_2)+
b^{IJ}d\tp_I\wedge d\tp_J 
}
We wish to study the moduli space for the NS five-brane 
sitting at $\vec{r}=0$. Notice that the transversal 
fluctuations of this five-brane at energy scale $\simeq m_s^2$
have the characteristic size $\Delta X^{\perp}\simeq \lambda l_s$.
If we take $\rho\simeq l_s$ and general $\etw$, then both of the
inequalities \twocond\ are satisfied for 
$|\vec{r}|\equiv \Delta X^{\perp}$. This suggests that the parameter
$\rho\simeq l_s$ actually does not affect the moduli space.
The reason why it might be not true is that the transversal
size of the NS5-brane is, actually, of the order $l_s$. Therefore
the curvature of the background should, presumably, affect the physics
even in the limit $\lambda\to 0$. 
The answer we will get  shows that the moduli space
does not really depend on $\rho$.

Now let us look at the fields at infinity. They are given by the
formulae \both\ and \defs\ with $\vr=\infty$. We will denote
the limits of $R^2(\vr)$, $G^{IJ}(\vr)$
and $B^I(\vr)$ as $\vr\to\infty$
by $R^2$, $G^{IJ}$ and $B^I$. It is convenient to have
a dictionary relating the fields at $\vr=\infty$ with the fields
at $\vr=0$. Let us first summarize our notations.
We have already introduced the matrices $g_{IJ}$, $b_{IJ}$,
$g^{IJ}$ and $b^{IJ}$ satisfying:
$$ (g^{IJ}+m_s^2 b^{IJ})(g_{JK}+l_s^2 b_{JK})=\delta^I_K $$
We have also introduced $G^{IJ}$ and $B^{IJ}$ in \both.
Now, we define $G_{IJ}$, $B_{IJ}$, $g^{-1}_{IJ}$ and
$G^{-1}_{IJ}$ in the following way:
\eqn\newdefs{
(G_{IJ}+B_{IJ})(G^{JK}+B^{JK})=\delta_I^K,\;\;\;
g^{-1}_{IJ}g^{JK}=\delta_I^K,\;\;\;
G^{-1}_{IJ}G^{JK}=\delta_I^K
}
Then, we have the following dictionary, relating asymptotic
background to the local background:
\eqn\dictionary{\eqalign{
\rho^2=R^2+(B,B), &\qquad R^{-2}=\rho^{-2}+(\etw,\etw),\cr
g^{IJ}=G^{IJ}+R^{-2}B^IB^J, &\qquad
         G^{-1}_{IJ}=g^{-1}_{IJ}+\rho^2\etw_I\etw_J,\cr
\etw_I={R^{-2}G^{-1}_{IJ}B^J\over 1+R^{-2}(B,B)}, &\qquad
         B^I={\rho^2\over 1+\rho^2(\etw,\etw)} g^{IJ}\etw_J,\cr
B^{IJ} &=b^{IJ}.\cr
}}
The local value, $\lambda_0$,
of the string coupling is related to the asymptotic
value $\lambda$ by the formula which follows from \StringCoupling:
\eqn\Coupling{
\lambda_0^2=(1+(\etw,\etw)\rho^2)\lambda^2
}


\subsection{The chain of dualities.}

We start by replacing the Taub-NUT circle with the M-theory
circle. We get a D6-brane wrapped on $\MT{4}$, with the NS5-brane
on top of it. 

At this point it is useful that we remember how the fields
of type-IIA theory are related to the fields of M-theory.
M-theory on a $U(1)$ bundle is type-IIA on the base of this bundle.
Suppose that the action of $U(1)$ is associated to the vector field
$v$. 
The M-theory three-form $C_M$ splits as follows:
\eqn\splitC{
C_M=\pi^* A^{(3)}+ {\cal A}\wedge \pi^* B
}
Also, we choose some local trivialization, and define the connection
one-form $A^{(1)}$ on the base, $dA^{(1)}={\cal F}$
(${\cal F}$ is the curvature two-form on the base,
$d{\cal A}=\pi^*{\cal F}$).  It should be identified
with the RR one-form  $C^{(1)}$ of type-IIA. Also, $B$ should be identified 
with the $B$ field of type-IIA (this follows from its coupling to the
fundamental string). What is the relation between $A^{(3)}$ and the
Ramond-Ramond three-form $C^{(3)}$ of type-IIA? 
Let us remember the general formula for the couplings of the Ramond-Ramond 
fields to the D-brane \rDouglas:
\eqn\couplings{
S_{RR}=\int\mu_p C\wedge {\rm tr} e^{F-B}
}
 For example, for the D2 brane we get:
\eqn\Dthree{
S_{RR}=\mu_2 \int C^{(3)}-C^{(1)}\wedge (B-F)
}
Here $C^{(1)}$ should be identified with the connection
one-form, ${\cal A}=d\phi+C^{(1)}$.
We have to keep in mind that various forms participating in this formula
are, in general, subject to gauge transformations.
 For example, under the gauge transformation $C^{(1)}\to C^{(1)}-d\psi$
we should have $C^{(3)}\to C^{(3)}-d\psi\wedge B$ (this is needed for
the coupling \Dthree\ to be correctly defined). This suggests that
\eqn\whatisC{
C^{(3)}=A^{(3)}+C^{(1)}\wedge B
}
(that is, $C_M=\pi^* C^{(3)}+d\phi\wedge \pi^* B$.) 
We may derive how Ramond-Ramond fields transform under T duality 
from their coupling to D branes. It follows that 
$Ce^{-B}$ transforms as a spinor of $O(d,d,{\bf Z})$.
Notice that
\eqn\CemB{
Ce^{-B}=A^{(1)}+A^{(3)}+{\rm forms\;\; of \;\; higher \;\; rank}.
}

Let us return to our dualities.
We assume that the M Theory circle 
in our original configuration has radius $S=\lambda l_s$,
where $l_s$ is the string scale in the configuration we start with,
and $\lambda$ is the original coupling constant (which has to be very
small, if we want to get Little String Theory on NS5 brane).
The three-form of M Theory is read from \both:
\eqn\Mthreeform{
C_M=({\cal A}-\tw^Id\tp_I) \wedge
B^Jd\tp_J\wedge d\theta+  B^{IJ}d\tp_I\wedge d\tp_J \wedge d\theta
}
If we now treat  the Taub-NUT circle as the M-theory circle, we
get \splitC\ with 
$$
A^{(3)}= B^{IJ}d\tp_I\wedge d\tp_J \wedge d\theta,\qquad
B=B^Id\tp_I\wedge d\theta.
$$
(Notice that ${\cal A}-\tw^I d\tp_I$ is just the connection 1-form after 
$\tw$-twist.)

In the new type-IIA theory, obtained by compactifying M Theory
on the Taub-NUT circle, we have the following asymptotic
values of the background fields:
\eqn\bgnd{\eqalign{
ds^2&=S^2 d\theta^2 + l_s^4 G^{IJ} d\tp_I d\tp_J
       +d\vec{r}^2 +(dx^{\mu})^2,\cr
B&=B^I d\tp_I\wedge d\theta,\cr
Ce^{-B}&=\tw^I d\tp_I+
d\theta\wedge B^{IJ}d\tp_I\wedge d\tp_J.\cr
}}
(We have used \CemB\ to find $Ce^{-B}$ in type-IIA.)
The new string length is:
\eqn\newls{
l_1^2={S\over R}l_s^2 = \lambda_0 {l_s^3\over\rho}
}
and the new string coupling constant is:
\eqn\couplingone{
\lambda_1=\left({R\over l_s}\right)^{3/2}{1\over\sqrt{\lambda}}
}
Making three $T$ duality transformations along $\MT{3}$, we get:
\eqn\bgndone{\eqalign{
ds^2=&{l_s^2\lambda_0^2\over\rho^2}\left[\rho^2 d\theta^2 +   
      G^{-1}_{IJ} d\psi^I d\psi^J 
     +2 G^{-1}_{IJ} B^I d\psi^J d\theta \right] \cr 
    &  +d\vec{r}^2 + (dx^{\mu})^2 \cr
B^{RR}=&\tw^I\epsilon_{IJK}d\psi^J\wedge d\psi^K
+d\theta\wedge\epsilon_{IJK}B^{IJ}d\psi^K
\cr
B^{NS}=&0 \cr
}}
with the string coupling constant,
\eqn\couplingtwo{
\lambda_2={\lambda\over l_s^3\sqrt{\det G^{\cdot\cdot}}}.
}
The NS5-brane remains an NS5-brane, wrapped on $\MT{3}$, and 
D6-brane becomes D3-brane. It shares with NS5 the directions
of $\MR{1,2}$. 

Now we do S-duality, so that $B^{RR}$ becomes $B^{NS}$, and NS5 becomes D5.
Also, we get the new string coupling and the new string length:
\eqn\thirdstep{
\lambda_3={l_s^3\sqrt{\det G^{\cdot\cdot}}\over\lambda},
\;\;\;\;\;
l_3=\lambda_0\sqrt{(\det g^{-1}_{\cdot\cdot})^{1\over 2}\over\rho}
}

Then, doing T-duality along the circle parameterized by $\theta$.
We have now $D6$ brane wrapped on the four-torus, and the D2 brane
inside it, orthogonal to the torus.
We end up with the following string coupling  and string length,
\eqn\finalcoupling{
\lambda_4=
{l_s^2\over\lambda_0\sqrt{\rho(\det g^{-1}_{\cdot\cdot})^{1\over 2}}},
\;\;\;\;\;
l_4=\lambda_0\sqrt{(\det g^{-1}_{\cdot\cdot})^{1\over 2}\over\rho}
}
and the following metric and $B$ field,
\eqn\metricfin{\eqalign{
ds_4^2&={l_s^2\over\rho^2}\lambda_0^2\left[ l_s^{-4}
(\det g^{-1}_{\cdot\cdot})
(d\tilde{\theta}-\epsilon_{IJK}b^{IJ}d\psi^K)^2 
+ g^{-1}_{IJ}d\psi^Id\psi^J\right],\cr
B&=\tw^I\epsilon_{IJK} d\psi^J\wedge d\psi^K+
  \etw_I d\theta\wedge d\psi^I.
}}
Let us summarize.
We have started with $k$ NS5-branes sitting at the center
of the Taub-NUT space, string coupling $\lambda_0$ and
string length $l_s$. 
The background fields are given by the equations
\metricloc\ and \bfieldloc, they correspond to both $\tw$-twists
and $\etw$-twists present. By the chain of dualities, we have mapped this
configuration to $k$ D6 branes wrapped on $\MT{4}$, and one D2 brane,
the metric and the $B$ field given by \finalcoupling\ and \metricfin.  
Notice that the volume of $\MT{4}$ is ${l_s^2\over \rho^2}l_4^4$.
In the limit we are interested in ($\lambda_0\to 0$)
it remains finite in the string units (specified by $l_4$).
The shape of the torus does not depend on $\rho$.


\subsection{World-sheet T-duality in the limit $\TNS\rightarrow \infty$}
Let us now see what happens in the limit $\TNS\rightarrow \infty$.
The strategy will be to start with type-IIA string-theory on 
the purely geometrical background which realizes the $\tw$-twist.
We will then perform world-sheet T-duality on $\MS{1}$ to obtain
a nonlinear world-sheet $\sigma$-model. Finally, we will insert the
NS5-branes back.

To describe the geometrical background we choose,
$$
X_6,\dots,X_9,
$$
as the transverse coordinates (on which the R-symmetry $SO(4)$ acts).
These replace the coordinates $y$ and $\vec{r}$ of $\TBNT(\rho)$.
We will denote,
$$
Z_1 = X_6 + i X_7,\qquad Z_2 = X_8 - i X_9.
$$
The other coordinates will be denoted,
$$
X_0\dots X_5,
$$
where $X_5$ is periodic with period $2\pi$.
They are  the world-sheet fields corresponding to
$x_0,x_1,x_2,\psi_1,\psi_2,\psi_3$ from the previous section.
The bosonic part of the world-sheet action is,
$$
L_0 = \sum_{\u,\v=0}^4 \eta^{\u\v} \px{\a}X_\u\qx{\a}X_\v
 + R^2 \px{\a}X_5\qx{\a}X_5 
+ \sum_{i=1,2} \px{\a}\bZ_i\qx{\a}Z_i.
$$
Let us, for simplicity, twist only along $X_5$ ($=\psi_3$).
The twist implies that $Z_i$ are not single-valued but rather,
$$
W_i = Z_i e^{-i{\tw\over {2\pi}} X_5},\qquad i=1,2
$$
are single-valued.
The world-sheet Lagrangian now reads,
\eqn\nylign{\eqalign{
L_0 =& \sum_{\u,\v=0}^4 \eta^{\u\v} \px{\a}X_\u\qx{\a}X_\v
 + R^2 \px{\a}X_5\qx{\a}X_5 \cr 
&+ \sum_{j=1,2} |\px{\a}W_j + {{i\tw}\over {2\pi}}W_j\px{\a}X_5|^2.
}}
Next we perform T-duality by the standard technique of treating
$V_\a \equiv \px{\a}X_5$ as an independent field and inserting
a Lagrange multiplier $Y$ for $\px{\lbrack\a}V_{\b\rbrack}$.

The result is a world-sheet action corresponding to the metric and 
$B$-field,
\eqn\reswsac{\eqalign{
ds^2 =& \sum_{\u,\v=0}^4 \eta^{\u\v} dX_\u dX_\v
     + |dW_1|^2 + |dW_2|^2
\cr
     &+ {{dY^2 + \sum_j (i W_j d\bW_j - i\bW_j d W_j)^2}
       \over
       {R^2 + {{\tw^2}\over {4\pi^2}} (|W_1|^2 + |W_2|^2)}},
\cr
B_{\u\v}dx^\u\wdg dx^\v =&
{{dY\wdg\sum_j (i W_j d\bW_j - i\bW_j d W_j)}
       \over
       {R^2 + {{\tw^2}\over {4\pi^2}} (|W_1|^2 + |W_2|^2)}}.
}}

\subsection{Adding in the NS5-brane}
Now we repeat the same excercise with the NS5-brane metric.
In string units, the metric is,
\eqn\strengenh{\eqalign{
L_0 =& \sum_{\u,\v=0}^4 \eta^{\u\v} \px{\a}X_\u\qx{\a}X_\v
 + R^2 \px{\a}X_5\qx{\a}X_5  \cr
&+ {1\over {|Z_1|^2 + |Z_2|^2}}\sum_{i=1,2} \px{\a}\bZ_i\qx{\a}Z_i.
}}
The dilaton is given by,
$$
g_s^2 = {1\over {|Z_1|^2 + |Z_2|^2}},
$$
and the solution is to be trusted when $g_s\ll 1$.
(See discussion in \rABS.)
After T-duality we obtain,
\eqn\reswsac{\eqalign{
ds^2 =& \sum_{\u,\v=0}^4 \eta^{\u\v} dX_\u dX_\v
     + {{|dW_1|^2 + |dW_2|^2}\over {|W_1|^2 + |W_2|^2}}
\cr
     &+ {{dY^2 + {1\over {\| W\|^4}}
                 \sum_j (i W_j d\bW_j - i\bW_j d W_j)^2}
       \over
       {R^2 + {{\tw^2}\over {4\pi^2}}}},
\cr
B_{\u\v}dx^\u\wdg dx^\v =&
{{dY\wdg\sum_j (i W_j d\bW_j - i\bW_j d W_j)}
       \over
       {(R^2 + {{\tw^2}\over {4\pi^2}})\| W\|^2 }}.\cr
}}

This is to be trusted when,
$$
\| W\|^2 \equiv |W_1|^2 + |W_2|^2  \gg 1.
$$
We see that as $R\rightarrow 0$, the $Y$-direction stays of 
finite size ${{2\pi}\over {\tw}}$.

\subsection{Large radius limit}
An interesting question is
what is the low-energy description of $S_B(k)$
compactified on $\MS{1}$ of radius $R$
with a fixed $\eta$-twist in the limit $R\rightarrow\infty$.
Naively, one can argue as follows.
To perform an $\eta$-twist we have to go over the ``fundamental''
degrees of freedom of $S_B(k)$ (whatever  they are!) and
separate them according to their charge $Q$ under the $U(1)$ subgroup
of the R-symmetry and according to their momenta $n$ 
and winding $w$ along $\MS{1}$.
We then add $\eta Q R$ to the mass of this field.
In the limit $R\rightarrow\infty$ and for generic $\eta$, this will
push all the $Q$-charged fields to high energy and we will be left with
only the $Q$-neutral sector.
Thus, if we start with $\SUSY{(1,1)}$ $U(k)$ SYM in 5+1D, as the
effective low-energy description, the conclusion would be that
we are left with $\SUSY{(1,0)}$ $U(k)$ SYM.
This conclusion cannot be correct since the gluinos of the
$\SUSY{(1,0)}$ vector-multiplet are chiral and the theory has a 
local gauge anomaly.

One possibility is that there is no 5+1D limit.
 For this to be true we must show
that there are no BPS states corresponding
to light KK states. On the type-IIA side we must show that there are no
states made 
by strings wrapped on the T-dual $\MS{1}$ which would become light.
Perhaps, when the circle is small enough, they do not form bound states
any more?


\section{Conclusion}

Let us summarize the results:
\item{1.}
The moduli space of the little-string theories of $k$ NS5-branes
compactified on $\MT{3}$ with $Spin(4)$ R-symmetry $\tw$-twists
is equal to the moduli space of $k$ $U(1)$ instantons on a
non-commutative $\MT{4}$.
The shape of the $\MT{4}$ is determined by the shape and
size of the physical $\MT{3}$ and by the NSNS 2-form fluxes
along it. The non-commutativity parameters are determined from the values
of the twists.

\item{2.}
In principle, there are 6 non-commutativity parameters on $\MT{4}$.
They are determined from the 3 geometrical $\tw$-twists and
the 3 non-geometrical $\etw$-twists.
The moduli space depends only on the 3 self-dual combinations
of the non-commutativity parameters and hence only on
the sum of the $\etw$-twists and $\tw$-twists.

\item{3.}
Combining the result for $k=2$ with the result of \rCGK, we obtain
a concrete prediction for the moduli space of 2 $U(1)$ instantons
on a non-commutative $\MT{4}$. This 8-dimensional
moduli space is a resolution of $(\MT{4}\times\MT{4})/\BZ_2$
by blowing up the singular locus. It can also be described as
a $\MT{4}$ fibration over a $\BZ_2^4$ quotient of a particular $K3$. 
The fiber corresponds to the ``center-of-mass'' of
the NS5-branes and the structure group is
$\BZ_2^4$ acting as translations of the fiber.
The particular point in the moduli space
of hyper-K\"ahler metrics on the $K3$ was constructed in \rCGK\
as a function of the $\tw$-twists, i.e. the non-commutativity parameters.
This $K3$ turns out to have a $\BZ_2^4$ isometry.  The $K3$ can be
described by blowing up $\MT{4}/\BZ_2$ and the $\BZ_2^4$ acts
by permuting the exceptional divisors of the blow-up. Note that
this $\BZ_2^4$ does not act freely.

\item{4.}
Similarly,
the moduli space of the little-string theories of \rIntNEW\ of $k$
NS5-branes at an $A_{q-1}$ singularity, compactified on $\MT{3}$
with $\tw$-twists (twists in the global $U(1)$),
is equal to the moduli space of
$k$ $U(q)$ instantons on a non-commutative $\MT{4}$.

\item{5.}
We studied the phase transitions which occur at singular points
of the moduli space.

\item{6.}
If instead of the little-string theories we start with
the $(2,0)$ theory (or the SCFT theory of \rBluInt\ in item (4) above),
we obtain the moduli spaces of instantons on 
a non-commutative $\MT{3}\times\BR$.
The non-commutativity parameters are only along $\MT{3}$, which
is in accord with the fact that there are no $\etw$-twists for
this  problem.

\vskip 0.5cm
Let us conclude with 3 open problems:
\item{a.}
Generalize to other gauge groups, in particular to D-type and E-type
little-string theories.
\item{b.}
Generalize to NS5-branes at D-type or E-type singularities.
\item{c.}
Study the $\etw$-twists, in particular how they are described at
large compactification radii.

\vfill\eject
\listrefs


\vfill\eject


\bye